\documentclass[11pt,a4paper]{article}
\usepackage{jheppub}
\pdfoutput=1

\usepackage[utf8]{inputenc}
\usepackage{amsmath}
\usepackage{amsfonts}
\usepackage{bm}
\usepackage{tikz}
\usetikzlibrary{arrows}
\usetikzlibrary{shapes}
\usepackage[english]{babel}
\usepackage[autostyle]{csquotes}
\usepackage{braket}

\usepackage{graphics}
\usepackage{tkz-euclide}
\usepackage[toc,page]{appendix}
\usetikzlibrary{decorations.markings,arrows}
\usetikzlibrary{arrows,shapes,backgrounds}
\usetikzlibrary{decorations.pathreplacing,decorations.markings}
\usetikzlibrary{calc, trees, positioning, arrows, shapes, shapes.multipart, shadows, matrix, decorations.pathreplacing, decorations.pathmorphing}
\usetikzlibrary{shapes.misc}
\usepackage{hyperref}
\usepackage{bookmark}
\usepackage{verbatim}
 
\DeclareMathOperator{\Ker}{Ker}
\DeclareMathOperator{\Tr}{Tr}
\DeclareMathOperator{\SU}{SU}
\DeclareMathOperator{\su}{su}
\DeclareMathOperator{\USp}{USp}
\DeclareMathOperator{\Sp}{Sp}
\DeclareMathOperator{\SO}{SO}
\DeclareMathOperator{\U}{U}
\DeclareMathOperator{\tr}{tr}

\newcommand{\gf}{G}
\newcommand{\oi}{\mathcal{O}_i}
\newcommand{\bC}{\mathbb{C}}
\newcommand{\hf}{\mathcal{H}_f}
\newcommand{\q}{\mathsf{Q}}
\newcommand{\bOr}[1]{\bar{\mathcal{O}}_{(#1)}}
\newcommand{\define}{\mathrel{\mathop:}=}
\newcommand{\tnode}[1]{\overset{\scriptstyle#1}{\overset{{\displaystyle\circ}}{\scriptstyle\vert}}}

\newcommand{\tnoder}[1]{\overset{\scriptstyle#1}{\overset{{\displaystyle\textcolor{red}{\circ}}}{\scriptstyle\vert}}}

\newcommand{\ba}{\begin{equation}\begin{aligned}}
\newcommand{\ea}{\end{aligned}\end{equation}}
\newcommand{\eref}[1]{(\ref{#1})}
\def\be{\begin{equation}}
\def\ee{\end{equation}}
\def\cb{\mathcal{M}_C}
\def\3d{$3d~\mathcal N=4$}
\def\sl#1{\mathfrak{sl}( #1,\mathbb{C})}
\def\hb#1{\mathcal{H}\left( #1 \right)}
\def\cb#1{\mathcal{C}\left( #1 \right)}
\def\node#1#2{\overset{#1}{\underset{#2}{\circ}}}

\def\noder#1#2{\overset{#1}{\underset{#2}{\textcolor{red}{\circ}}}}

\def\snode#1#2{\overset{#1}{\underset{#2}{\scriptstyle\square}}}
\def\flav#1{\overset{\scriptstyle#1}{\overset{\square}{\scriptstyle\vert}}}

\preprint{Imperial/TP/18/AH/09}

\title{Minimally Unbalanced Quivers}
\author[]{Santiago Cabrera}
\author[]{, Amihay Hanany}
\author[]{and Anton Zajac}
\affiliation[]{Theoretical Physics, The Blackett Laboratory, Imperial College London\\ 
Prince Consort Road, London, SW7 2AZ United Kingdom}
\emailAdd{santiago.cabrera13@imperial.ac.uk}
\emailAdd{a.hanany@imperial.ac.uk}
\emailAdd{anton.zajac@imperial.ac.uk}

\abstract{We develop a classification of \emph{minimally unbalanced} $3d~\mathcal{N}=4$ quiver gauge theories. These gauge theories are important because the isometry group $G$ of their Coulomb branch contains a single factor, which is either a classical or an exceptional Lie group. Concurrently, this provides a classification of hyperk\"ahler cones with isometry group $G$ which are obtainable by Coulomb branch constructions. HyperK\"ahler cones such as Coulomb branches of $3d~\mathcal{N}=4$ quivers are indispensable tools for describing Higgs branches of different theories in various dimensions. In particular, they are used to describe Higgs branches of $5d~\mathcal{N}=1$ SQCD with gauge group $SU(N_c)$ and $6d~\mathcal N = (1,0)$ SQCD with gauge group $Sp(N_c)$ at the respective UV fixed points.}


\begin{document}

\maketitle

\section{Introduction}

The study of the vacuum structure of SQED and SQCD with eight supercharges \cite{SW94a,SW94b,APS96,IS96,HW96} constitutes a prodigious bridge between physics and mathematics. It continues the spirit of Dirac's vision on the modern role of theoretical physics \cite{Dirac31}; the study of the geometrical properties of the vacuum moduli space rewards the researcher with the discovery of new physical phenomena. In this note we apply this principle to study the relationship between the geometry of hyperk\"ahler\footnote{In this note we follow the terminology in \cite{B20} and by the notion of hyperk\"ahler we mean symplectic and holomorphic without any statements on a metric.} cones \cite{B70,Sl80,KP82,K90,CM93,N94,N15,BFN16,N16} and the physics of gauge theories with eight superchagers.\footnote{The recent work in \cite{Lindstrom:2018aoc} reviews the role of sigma models and supersymmetric theories on the quest for new geometrical spaces. In particular, the emphasis is given to construction of geometrical spaces with hyperk\"ahler structure. The present note should be understood as a complimentary effort: the action of the hyperk\"ahler quotient on an initial Lagrangian with hyperk\"ahler geometry is replaced by the utilization of dressed monopole operators in the Coulomb branch of \3d supersymmetric quiver gauge theory.} Let us start by remembering some aspects of Supersymmetric Quantum Electrodynamics.

The Higgs branch $\hf$ of SQED with eight supercharges, $n$ electrons, and finite gauge coupling $g$, can be computed classically \cite{APS96}. It does not depend on the number of spacetime dimensions. It is a hyperk\"ahler cone of complex dimension $2n-2$ and it possesses an isometry under the flavor group $\gf=SU(n)$. It is isomorphic to the reduced moduli space of one $A_{n-1}$ instanton on $\mathbb{C}^2$ \cite{IS96}. Recent developments by Namikawa \cite{N16} show that it is actually one of the simplest hyperk\"ahler cones with an $SU(n)$ isometry, in the sense that the set of generators of the cone is minimal.\footnote{By generators we mean the set of linearly independent holomorphic functions that generate the holomorphic ring of the hyperk\"ahler cone. Remember that in a SCFT with eight supercharges the generators of the holomorphic ring of a hyperk\"ahler branch of the moduli space (i.e. Higgs branch in any dimension or Coulomb branch in $3d$) are found to be in one-to-one correspondence with chiral operators which generate the corresponding chiral ring.} In particular, this Higgs branch can be described as the set of all $n\times n$ complex matrices such that:
\begin{equation}
	\hf = \{ M\in \mathbb{C}^{n\times n}|\tr (M)=0,\ M^2=0 \ {\rm and ~rank}(M)\leq 1 \}
\end{equation}
This set of matrices transforms under the adjoint representation of the $\mathfrak{sl}(n,\mathbb{C})$ algebra. Given a matrix $X_{(2,1^{n-2})}$ such that:
\begin{equation}
 X_{(2,1^{n-2})}\define\left(\begin{array}{ccccc}
 0&1&0&\dots&0\\
 0&0&0&\dots&0\\
 0&0&0&\dots&0\\
 \vdots&\vdots&\vdots&\ddots&\vdots\\
 0&0&0&\dots&0\\
 \end{array}\right)_{n\times n},
\end{equation}
where $X_{(2,1^{n-2})}$ is a block diagonal matrix, and $(2,1^{n-2})$ indicates the presence of one elementary Jordan normal block of size 2, and $n-2$ elementary Jordan normal blocks of size 1, it can be acted upon by elements of the group $PSL(n,\mathbb{C})$. An orbit $\mathcal{O}_{(2,1^{n-2})}$ is defined as:
\begin{equation}
	\mathcal{O}_{(2,1^{n-2})}\define \{S\cdot X\cdot S^{-1}| S\in PSL(n,\mathbb{C})\}
\end{equation}
such that $\hf$ is isomorphic to the orbit's closure:
\begin{equation}\label{eq:min}
	\hf = \bar{\mathcal{O}}_{(2,1^{n-2})}
\end{equation}
Note that $(2,1^{n-2})$ is a partition of the integer number $n$. For each partition $\lambda$ of $n$, there is a matrix $X_\lambda$ of Jordan normal form, and its orbit $\mathcal{O}_{\lambda}$ can be defined such that its closure, $\bar{\mathcal O}_{\lambda}$, is a hyperk\"ahler cone \cite{CM93}. The set of all such orbits is called the set of all \emph{nilpotent orbits} of $\mathfrak{sl}(n,\mathbb{C})$ and their closures are the simplest hyperk\"ahler cones that can be built that enjoy an isometry under the group $\gf=SU(n)$. Any other hyperk\"ahler cone with $SU(n)$ isometry that is not the closure of a nilpotent orbit of $\mathfrak{sl}(n,\mathbb C)$ has a non-minimal\footnote{This notion will be explained in more detail in the next section.} set of generators \cite{N16}.

Hyperk\"ahler cones whose isometry $\gf$ contains a single factor (f.i. $SU(n)$) can be classified according to grading of the generators with respect to their charge under $SU(2)_R$, with the set of nilpotent orbits of $Lie(G)$ in the simplest level of the classification. Accordingly, supersymmetric quantum field theories whose Higgs branch is a hyperk\"ahler cone posses a similar stratification.

Three dimensional $\mathcal{N}=4$ supersymmetric gauge theories also have Coulomb branches that are hyperk\"ahler cones. They are often related to hyperk\"ahler $3d$ Higgs branches via $3d$ \emph{mirror symmetry} \cite{IS96}. Hence, when their isometry group $\gf$ has a single factor, they also admit a similar classification. Moreover, Higgs branches of $5d~\mathcal N=1$ and $6d~\mathcal N=(1,0)$  SQCD have been found to have description in terms of $3d$ Coulomb branches when the gauge coupling $g$ is taken to infinity \cite{FHMZ17,Hanany:2018uhm,Cabrera:2018ann,Hanany:2018vph}. All of these recent developments suggest that a thorough classification of $3d~\mathcal N=4$ Coulomb branches is essential in order to carry out a systematic study of hyperk\"ahler Higgs branches in any dimension.


 Theories whose Higgs or Coulomb branches are closures of nilpotent orbits have seen an extensive analysis (see for example \cite{GW09,BTX10,CDT13,CHMZ14,CH16,CHZ17,Cabrera:2017njm,Hanany:2018uzt}). In this paper we present a classification of $3d \ \mathcal N=4$ gauge theories that have a common property: their Coulomb branch has an isometry group $\gf$ which has a single factor, but need not necessarily be a closure of a nilpotent orbit of $Lie(G)$. In order to adroitly produce such classification, we rely on recent advances in the study of \3d Coulomb branches, herein denoted by $\mathcal{C}$. For \3d theories that have an associated \emph{quiver}, the isometry group $\gf$ of $\mathcal C$ has a powerful connection with the structure of the quiver. We exploit this fact, and the current understanding of \emph{dressed monopole operators} \cite{Aharony:1997bx,Kapustin:1999ha,Borokhov:2002ib,Borokhov:2002cg,Borokhov:2003yu,GW09,CHZ13} on $\mathcal C$.

In section \eref{sec:gen} we illustrate the main ideas of the paper with several examples. Section \eref{sec:clas} presents the general method of classifying \3d quivers such that their Coulomb branch has isometry $\gf$, where $\gf$ is any Lie group, and its set of generators is minimally extended. Sections \eref{4} contains all cases where $\gf$ corresponds to a simply laced Dynkin diagram. Section \eref{5} collects all cases with $\gf$ that corresponds to a non-simply laced Dynkin diagram. In section \eref{6} we present an exotic extension of the classification. In particular, we consider minimally unbalanced quivers with $G$ corresponding to a simply laced Dynkin diagram but with the unbalanced node connected to the rest of the quiver via a non-simply laced edge. Section \eref{7} contains an exotic extension of the classification of minimally unbalanced quivers with $G$ that corresponds to a non-simply laced Dynkin diagram and with the unbalanced node connected by a non-simply laced edge.  Finally, section \eref{8} offers a brief summary of the work and possible future directions in the study of minimally unbalanced supersymmetric quiver gauge theories.

\section{A $3d$ Coulomb Branch with isometry $SU(n)$ and minimal set of generators}\label{sec:gen}

Let us start by remembering a very well known effect: the \emph{mirror symmetry} of \3d quiver gauge theories \cite{IS96}. Continuing with the example from the introduction:
\be \label{eq:mir}
	\hb{\node{\flav n}1} = \cb{\underbrace{\node{\flav 1}1-\node{}1-\dots-\node{}1-\node{\flav 1}1}_{n-1 \text{ nodes}}}
\ee
where $\hb{}$ denotes the Higgs branch of a the $3d$ quiver at finite coupling and $\cb{}$ denotes its Coulomb branch. Note, that the gauge groups are depicted by round nodes and the flavor groups by square nodes in the quiver. In this case, both sides of equation \eref{eq:mir} are equal to the the hyperk\"ahler cone in equation \eref{eq:min}, with \emph{highest weight generating function} \cite{Hanany:2014dia}: 
\be
	HWG(\mu_1,\dots,\mu_{n-1},t)=\frac{1}{1-\mu_1\mu_{n-1} t^2},
\ee
where the highest weight fugacities $\mu_1\mu_{n-1}$ signify that the generators of the holomorphic ring transform under the representation with highest weight $[1,0,0,\dots,0,0,1]$, i.e. the adjoint representation of the isometry group $SU(n)$. Let us focus on the RHS of equation \eref{eq:mir}. Note that the gauge nodes form the Dynkin diagram of $Lie(SU(n))$. Note also that all the gauge nodes are balanced.\footnote{They satisfy $N_f-2N_c=0$, where $N_c$ is the rank of the gauge node (or \emph{number of colors}) and $N_f$ is the sum over the ranks of adjacent nodes (or \emph{number of flavors}). In this particular example all gauge nodes have $N_c=1$ and $N_f = 1+1=2$.} As discussed in \cite{GW09}, a balanced node contributes to the holomorphic ring of the Coulomb branch with polynomials of degree\footnote{The degree of the polynomials is represented in the highest weight generating function by the power of the fugacity $t$.} $d=2$ (i.e. at the IR SCFT there are chiral operators $\mathcal O_i$ with conformal dimension $\Delta(\oi) = 1$). The set of all linearly independent holomorphic polynomials with degree $d=2$ transforms in the adjoint representation of the isometry group $\gf=SU(n)$ \cite{N16}. This means that the number of such polynomials in the example at hand is $n^2-1$. In \cite{N16} it is also shown that if there are only generators with degree $d=2$, then the space is the closure of a nilpotent orbit of $\mathfrak{sl}(n,\mathbb{C})=Lie(\gf)$. Hence:\\

\emph{Given an isometry $\gf$, the set of hyperk\"ahler cones with solely generators of degree $d=2$, which necessarily transform in the adjoint representation of $\gf$, is equivalent to the set of closures of nilpotent orbits of $Lie(\gf)$.} \\

As mentioned in the introduction, there is a one-to-one correspondence between the set of nilpotent orbits of $\sl n$ and the set of partitions of $n$, denoted by $\mathcal{P}(n)$. Therefore, there is a finite set of hyperk\"ahler cones with isometry $SU(n)$ and minimal number of generators (i.e. $n^2-1$ ) transforming in the adjoint representation. Each hyperk\"ahler cone corresponds to a different partition of $n$. Let us illustrate this using an example.

\subsection{$\gf=SU(3)$}

\paragraph{Minimal set of generators}

Let the isometry group be $\gf=SU(3)$. The set of partitions of $3$ is $\mathcal{P}(3)=\{(3),(2,1),(1,1,1)\}$.\footnote{In the following, we use exponential notation for partitions. For instance, partition $\{5,4,4,2,1,1,1\}$ is denoted by $\{5,4^2,2,1^3\}$.} There are three different hyperk\"ahler cones corresponding to $\bOr{3}$, $\bOr{2,1}$ and $\bOr{1^3}$, respectively. For each different nilpotent orbit closure there is a corresponding $3d~\mathcal N=4$ quiver \cite{GW09}:
\be	
	\begin{aligned}
		(3)&\rightarrow~ \node{\flav 3}{2}-\node{}1 \\
		(2,1)&\rightarrow ~\node{\flav 1}1-\node{\flav 1}1\\
		(1^3)&\rightarrow ~\node{}0-\node{}0
	\end{aligned}
\ee
such that
\be	\label{eq:nil3}
	\begin{aligned}
	\cb{\node{\flav 3}{2}-\node{}1}&= \bOr{3}\\
	\cb{\node{\flav 1}1-\node{\flav 1}1}&=\bOr{2,1}\\
	\cb{\node{}0-\node{}0}&=\bOr{1^3}
	\end{aligned}
\ee

The quiver with zero rank nodes has a trivial Coulomb branch. The remaining two quivers have Coulomb branches with highest weight generating functions:
\be
	HWG_{(3)}(\mu_1,\mu_{2},t)=\frac{1-\mu_1^3\mu_2^3 t^{12}}{(1-\mu_1\mu_{2} t^2)(1-\mu_1^3 t^6)(1-\mu_{2}^3 t^6)}
\ee
and
\be
	HWG_{(2,1)}(\mu_1,\mu_{2},t)=\frac{1}{1-\mu_1\mu_{2} t^2},
\ee

respectively. Both Coulomb branches are solely generated by holomorphic polynomials of degree $d=2$ in the adjoint representation of $\gf=SU(3)$, denoted by the term $\mu_1\mu_2t^2$ in the HWG. The degree of these polynomials, i.e. power of $t^d$, determines their spin $s=d/2$ under the $SU(2)_R$ (i.e. the R-symmetry). In this case the generators have spin $s=1$ (equivalently, the chiral ring associated with the Coulomb branch is generated by eight operators $\oi$ in the adjoint representation of $\gf=SU(3)$ with conformal dimension $\Delta(\oi) = 1$). Any other \3d Coulomb branch $\mathcal{C}$ with isometry $SU(3)$ is either isomorphic to $\mathcal{C}(\snode{}{3}-\node{}{2}-\node{}{1})$ or $\mathcal{C}(\snode{}1-\node{}1-\node{}1-\snode{}1)$, or has extra generators $\oi'$ with spin $s>1$ under $SU(2)_R$. In the latter case, the extra operators $\oi'$ have conformal dimension $\Delta(\oi')>1$.

\subsection{$\gf=SU(10)$}

Now consider the partition $\lambda = (2^5)\in \mathcal{P}(10)$. The quiver with Coulomb branch  $\bOr {2^5}\subset \sl {10}$ takes the form:
\be \label{eq:nil10}
	\cb {\node{}1-\node{}2-\node{}3-\node{}4-\node{\flav 2}5-\node{}4-\node{}3-\node{}2-\node{}1 } = \bOr{2^5}
\ee
where the round and square nodes denote gauge and flavor groups, respectively.\footnote{In the classification of this paper, all quivers contain unitary gauge nodes and no flavor nodes.} This Coulomb branch is minimally generated by operators $\mathcal O_i$ satisfying $\Delta (\mathcal O_i) = 1$, and transforming under the adjoint representation of $SU(10)$. The HWG reads:
\be 
	HWG(\mu_1,\dots,\mu_{9},t)=\prod_{i=1}^5\frac{1}{1-\mu_i\mu_{10-i} t^{2i}}.
\ee
\paragraph{Extension of the minimal set of generators} Let us consider the quiver in \eref{eq:muq10}
 
\be \label{eq:muq10}	\node{}1-\node{}2-\node{}3-\node{}4-\node{\tnode 2}5-\node{}4-\node{}3-\node{}2-\node{}1 
\ee
where the top gauge node is not balanced (i.e has non-zero \emph{excess}).\footnote{The \emph{excess} of a node is defined as $e\define N_f-2N_c$. If the node is \emph{balanced} its excess is zero $e=0$. The excess of the top node of the quiver in equation \eref{eq:muq10} is: $N_f-2N_c=5-2\times 2 = 1\neq 0$.} Written using the Plethystic exponential (PE) \cite{FHH07} the HWG reads \cite{HPsic17}:
\be
HWG(\mu_1,\dots,\mu_9,t)=PE[\mu_1\mu_9t^2 +\mu_5t^3 +(1+\mu_2\mu_8)t^4 +\mu_5t^5 +\mu_3\mu_7t^6 +\mu_4\mu_6t^8]
\ee
 The effect of the unbalanced node on the Coulomb branch is the appearance of new operators ${\mathcal{O}}'_i$ which are also the generators of the chiral ring.
Since the conformal dimension of the new operators is $\Delta( {\mathcal{O}}'_i) = 3/2$, they do not modify the global symmetry of the Coulomb branch (which is only determined by the operators $\mathcal O_i$ with $\Delta (\mathcal O_i)=1$). Therefore the Coulomb branch of \eref{eq:muq10} has an isometry group which contains a single factor $\gf = SU(10)$, but it is no longer a closure of a nilpotent orbit. This is an example of a theory that concerns the present work. In the next section we formally define the set of theories that share this property.

\section{A Classification: Minimally Unbalanced Quivers}\label{sec:clas}


This section provides the answer to the main question: Given a Lie group $\gf$, what is the set of \3d quivers such that their Coulomb branch $\mathcal C$ is generated by operators $\mathcal{O}_i$ with $\Delta(\mathcal O_i) = 1$ in the adjoint representation of $\gf$ (this set of operators is always present if $\gf$ is an isometry of $\mathcal C$) and an extra set of generators $\mathcal O'_i$ with $\Delta(\mathcal O'_i)>1$, such that $\gf$ remains the isometry of the Coulomb branch.
In order to address this question, let us employ the following claim, which results from the work on monopole operators on the $3d~\mathcal N=4$ Coulomb branch  \cite{Borokhov:2002cg,Borokhov:2002ib,GW09,CHZ13,CFHM14}: A gauge node of a \3d quiver determines the presence of operators $\mathcal O_i$ with $\Delta(\mathcal{O}_i)=1$ in the Coulomb branch in the following way:
\begin{itemize}
	\item If the node has \emph{excess} $e>0$, it contributes with a single Casimir operator $\phi_i$, such that $\Delta (\phi_i)=1$.
	\item A set of nodes with excess $e=0$, in the form of a Dynkin diagram of a Lie group $G$, contributes with a number of operators $\mathcal{O}_i$ (with $\Delta(\mathcal O_i)=1$) equal to the dimension of $G$. There will be one Casimir operator $\phi_i$ per gauge node. The remaining operators are \emph{bare monopole operators} $V_i$ that correspond to the different roots of the algebra $Lie(G)$.
	\item If the quiver has no flavor nodes, one Casimir operator with $\Delta(\mathcal \phi_i) = 1$ needs to be removed from the counting, corresponding to the adjoint representation of the decoupled $U(1)$ \emph{center of mass}.
\end{itemize}
Two different cases of \3d quivers with isometry $\gf$ on the Coulomb branch $\mathcal C$ can be readily identified employing this claim:
\begin{enumerate}
	\item \textbf{Nilpotent orbit's closure:} $\mathcal C = \bar{\mathcal{O}}_\lambda\subset Lie(\gf)$. All the generators of $\mathcal C$ have dimension $\Delta(\mathcal O_i) = 1$. The gauge nodes of the quiver form the Dynkin diagram of $Lie(\gf)$, for any classical or exceptional Lie group $\gf$. All gauge nodes of the quiver are balanced. Flavor nodes are added to ensure such balance condition. Moreover, the rank of the flavor nodes always follows the pattern of the \emph{weighted Dynkin diagram} \cite{CM93} of the corresponding nilpotent orbit $\mathcal O_\lambda$. This can realise nilpotent orbit's closures of height $ht(\mathcal O_\lambda) = 2$.\footnote{The \emph{height} of a nilpotent orbit is defined as in \cite[sec.~2]{Panyushev:1999on}. Note that for $\gf$ of $A$-type, this construction can be extended to nilpotent orbits of all heights $ht(\mathcal O_\lambda)$, where the flavor nodes are determined by the partition $\lambda$ of the nilpotent orbit. See \cite{GW09,HK16,CH16} for examples.} 
	\item \textbf{Minimally unbalanced quiver:} The gauge nodes of the quiver form a minimal extension of the Dynkin diagram of $Lie(\gf)$. By minimal extension we mean that there is a single extra gauge node, connected to the other gauge nodes that form the Dynkin diagram. There are no flavor nodes. All gauge nodes in the Dynkin diagram are balanced (with $e=0$). The the extra node is unbalanced, i.e. it has excess $e>0$.\footnote{In the following sections the cases with $e=-1$ are discussed separately from generic cases with $e<0$. If all nodes have $e=0$, the quiver forms an affine or twisted affine Dynkin diagram of the global symmetry $\gf$ and these cases are also discussed separately. We refrain from discussing the pathological case of the $A^{(2)}_2$ twisted affine Dynkin diagram.}
\end{enumerate}
Examples of the first case are the theories in equations \eref{eq:nil3} and \eref{eq:nil10}. Equation \eref{eq:muq10} is an example of the second case. Both cases have a number of generators $\mathcal O_i$, with $\Delta(\mathcal O_i)=1$, equal to the dimension of $\gf$. In all three examples the Coulomb branch has the same isometry $\gf$. The difference is that the first case has no extra generators of $\mathcal C$, while the second case has extra generators $\mathcal O_i'$ with $\Delta(\mathcal O'_i)>1$.
As mentioned before, $3d~\mathcal{N}=4$ quiver gauge theories whose Coulomb branches are closures of nilpotent orbits have already been extensively studied (note the recent progress for exceptional $\gf$ in \cite{Hanany:2017ooe}). In this note we present a classification of all minimally unbalanced quivers, for any classical Lie group $\gf$. We emphasize that all quivers presented in this paper are in the basic form such that the ranks are the lowest possible. Other theories can be obtained by multiplying the basic forms of the quivers by an integer number (this will not modify the isometry of the Coulomb branch).

\paragraph{Minimally unbalanced quivers}
We are in the position to present the general solution for finding all minimally unbalanced quivers with a Coulomb branch isometry $\gf$, where $\gf$ contains a single factor. The remaining sections of the paper contain the specific results for all the different types of Lie groups.
As the first step, consider a \3d quiver $\q$ with the shape of a particular Dynkin diagram and with an extra node attached to it in the simplest fashion.\footnote{Simplest fashion means that the extra node is attached by a simply laced edge to only one of the nodes of the balanced Dynkin diagram.} All nodes are $U(N_{i})$ gauge nodes, where $N_i$ is the number of colors of the $i$-th node. The nodes in the Dynkin diagram need to be balanced (with excess $e=0$). In order to impose the balancing condition one can remember the vectors $\vec{v}$ and $\vec{w}$ on Nakajima's quiver varieties \cite{N94}. In this case they are used slightly differently. Let $\vec v$ be the vector with the ranks of all the nodes of the part of the quiver that forms the (balanced) Dynkin diagram. Let $C$ be the corresponding Cartan matrix. Then, the vector $\vec{w}$ is defined as: 

\begin{equation}
	\vec{w} := C\cdot \vec{v}
\end{equation}

Note that $\vec w$ measures the excess in each of the nodes in $\vec v$ in the presence of no other nodes in the quiver. Now, one sets to zero the all components of $\vec w$ except of one. The non-zero component can be set to $k$. This corresponds to attaching an extra node of rank $k$ (the node that will become \emph{minimally unbalanced}) at the position of the non-zero element of $\vec w$ and simultaneously balancing all nodes in $\vec v$. After fixing the rank of the imbalanced node (node with $e\neq 0$) to $k$, the ranks of the balanced nodes are uniquely determined:\footnote{Note that the existence of the inverse of the Cartan matrix is guaranteed since we are dealing with finite-dimensional Lie algebras.}

\begin{equation}\label{eq:Cartan}
	\vec{v} = C^{-1}\cdot \vec{w}
\end{equation}

Finally, the value of $k$ can be chosen to be the smallest value such that all the other ranks are integer numbers. In the following sections we perform this computation for all different choices of the position of the non-zero component of $\vec w$. In this way, we obtain all possible minimally unbalanced quivers with a balanced subset of nodes corresponding to a certain Dynkin diagram.

\section{Simply Laced Minimally Unbalanced Quivers} \label{4}

We begin our classification of minimally unbalanced quiver gauge theories with Coulomb branch isometry $G$ that corresponds to a simply laced Dynkin diagram and the unbalanced node is also connected by a simply laced edge.

\subsection{$\gf $ of Type $A_n$}
Let us show one example of the approach described in the previous section. Choose $\gf = SU(9)$, with the Dynkin diagram of the form:
$$
\node{}{}-\node{}{}-\node{}{}-\node{}{}-\node{}{}-\node{}{}-\node{}{} -\node{}{}
$$
Let the Dynkin diagram be the balanced part of the \3d quiver $\q$. The vector $\vec v = (v_1,v_2,\dots,v_8)$ denotes the number of colors of each node:
\be
\node{}{v_1}-\node{}{v_2}-\node{}{v_3}-\node{}{v_4}-\node{}{v_5}-\node{}{v_6}-\node{}{v_7}-\node{}{v_8}
\ee
Let us attach an extra node with $k$ colors to the fourth node (which has the number of colors $v_4$). This determines $\vec w$:
\be 
	\vec w = (0,0,0,k,0,0,0,0)
\ee
The resulting quiver is:
\be
\q \define ~\node{}{v_1}-\node{}{v_2}-\node{}{v_3}-\node{\tnode k}{v_4}-\node{}{v_5}-\node{}{v_6}-\node{}{v_7}-\node{}{v_8}
\ee

Employing the Cartan matrix $C$ of $SU(9)$,
\be 	
	C\define \left(\begin{array}{cccccccc}
	2&-1&0&0&0&0&0&0\\
	-1&2&-1&0&0&0&0&0\\
	0&-1&2&-1&0&0&0&0\\
	0&0&-1&2&-1&0&0&0\\
	0&0&0&-1&2&-1&0&0\\
	0&0&0&0&-1&2&-1&0\\
	0&0&0&0&0&-1&2&-1\\
	0&0&0&0&0&0&-1&2\\
	\end{array}\right),
\ee
the imposed balancing condition on $v_i$:
\be 
	\vec v = C^{-1}\cdot \vec \omega
\ee
determines the ranks of the remaining nodes of the quiver. The resulting quiver takes the form:
\be
\q = ~\node{}{\frac{5k}9}-\node{}{\frac{10k}9}-\node{}{\frac{15k}9}-\node{\tnode k}{\frac{20k}9}-\node{}{\frac{16k}9}-\node{}{\frac{12k}9}-\node{}{\frac{8k}9}-\node{}{\frac{4k}9}
\ee
The ranks of the gauge groups are integer if $k=9p$, with $p\in \mathbb{N}$. All the nodes in the bottom row have excess $e=0$. The excess of the top gauge node with $k=9p$ colors is:
\be 
	e = 20p - 18p = 2p
\ee
The lowest value of $k$ such that all other ranks $v_i$ are positive integers is obtained for the choice $p=1$. Therefore the quiver of interest has the form:
\be 
	\q = ~\node{}5 - \node{}{10}-\node{}{15} -\node{\tnoder 9}{20}-\node{}{16}-\node{}{12}-\node{}{8}-\node{}{4}
\ee
where the top node (drawn red) has excess e = 2. Following \cite{GW09}, a bare monopole operator $V_i$ minimally charged under a node with excess $e$ has conformal dimension:
\be 
	\Delta(V_i)= (e+2)/2.
\ee
For $e=2$ there is a bare monopole operator $V$, only charged under the magnetic dual of the unbalanced node, with dimension $\Delta(V)=2$. This is part of the set of extra generators $\oi '$ of the Coulomb branch. In particular, we say that all the extra generators (with $\Delta(\oi')=2$) can be obtained by a procedure similar to that explained in \cite{GW09}, by turning on minimal charges of the balanced sector of the quiver. There is a total of $252$ such operators, transforming in the fourth antisymmetrization of the fundamental representation of $\gf=SU(9)$, denoted by Dynkin labels $[0,0,0,1,0,0,0,0]$, and its complex conjugate representation $[0,0,0,0,1,0,0,0]$.\footnote{Note that the appearance of the $252$ new operators can be read off the quiver since the unbalanced node is attached to the fourth Dynkin node, indicating that the extra generators transform in this representation and its complex conjugate representation.} \\

\begin{figure}
\center{
\begin{tikzpicture}[scale=0.70]
\draw (0.4,0) -- (0.8,0);
\draw (0,0) circle (0.4cm);
\draw (0,-0.8) node {\footnotesize{$\frac{ab}{s}$}};
\draw (1.6,0) -- (2,0);
\draw (1.2,0) node {\footnotesize{\dots}};
\draw(2.4,0) circle (0.4cm);
\draw (2.4,-0.8) node {\footnotesize{$\frac{3b}{s}$}};
\draw (-2.4,0) circle (0.4cm);
\draw (-2.4,-0.8) node {\footnotesize{$\frac{3a}{s}$}};
\draw (-2.8,0) -- (-3.2,0);
\draw (-3.6,0) circle (0.4cm);
\draw (-3.6,-0.8) node {\footnotesize{$\frac{2a}{s}$}};
\draw (-4,0) -- (-4.4,0);
\draw (-4.8,0) circle (0.4cm);
\draw (-4.8,-0.8) node {\footnotesize{$\frac{a}{s}$}};
\draw (2.8,0) -- (3.2,0);
\draw (3.6,0) circle (0.4cm);
\draw (3.6,-0.8) node {\footnotesize{$\frac{2b}{s}$}};
\draw (4,0) -- (4.4,0);
\draw (4.8,0) circle (0.4cm);
\draw (4.8,-0.8) node {\footnotesize{$\frac{b}{s}$}};
\draw (-0.4,0) -- (-0.8,0);
\draw (-1.2,0) node {\footnotesize{\dots}};
\draw (-1.6,0) -- (-2,0);
\draw (0,0.4) -- (0,0.8);
\draw (0,1.2)[red,fill=red!30] circle (0.4cm);
\draw (0,2) node {\footnotesize{$\frac{a+b}{s}$}};
\end{tikzpicture}
}
\caption{Generic quiver with $\gf = SU(n)$ global symmetry, with $n=a+b$, where $a,b\in \mathbb{N}$ and $s$ is the greatest common divisor of $a$ and $b$. The excess of the bottom nodes is $e=0$. The excess of the top node is $e=(ab-2a-2b)/s$. We are interested in the subset of quivers with $(ab-2a-2b)/s\neq0$. The quaternionic dimension of the Coulomb branch of the quiver can be expressed using parameters $a$ and $b$: $dim_{\mathbb{H}} = \frac{(ab+2)(a+b)}{2s}-1$.}
\label{fig:Aseries}
\end{figure}
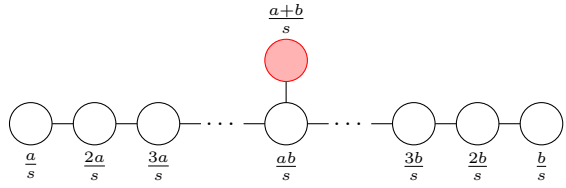

The various choices of $\vec w$ produce different quivers $\q$ where the extra node is attached either to the fourth, the third, the second or the first node in the Dynkin diagram of $\gf=SU(9)$.\footnote{The quiver in figure \ref{fig:Aseries} enjoys outer $\mathbb{Z}_2$ automorphism symmetry therefore other choices of $\vec w$ yield equivalent quivers to those already included.} In each case $k$ is chosen to be the smallest value such that the rest of the ranks are positive integers. The different results and the excess of the extra node are depicted in table \ref{tab:A8}. It is crucial to distinguish four cases based on the excess of the unbalanced node which in turn determines the presence of extra operators with various values of conformal dimension. Following the terminology of \cite{GW09}, we summarize the possible types of theories in table \ref{tab:goodbadugly}. This terminology is used throughout this paper.

Hence the first and the second row in table \ref{tab:A8} contain \emph{good} theories that are: unbalanced with positive excess and fully balanced, respectively. The third and fourth row in table \ref{tab:A8} contain \emph{bad} quivers that are both unbalanced with negative excess.

\begin{table}
	\centering
	\begin{tabular}{|c|c|c|}
	\hline
	Quiver & Excess & Type \\ \hline
	$\node{}5 - \node{}{10}-\node{}{15} -\node{\tnode 9}{20}-\node{}{16}-\node{}{12}-\node{}{8}-\node{}{4}$ & $2$ & Good\\ \hline
	$\node{}{2} - \node{}{4}-\node{\tnode 3}{6} -\node{}{5}-\node{}{4}-\node{}{3}-\node{}{2}-\node{}{1}$ & $0$ & Good\\ \hline
	$\node{}{7} - \node{\tnode 9}{14}-\node{}{12} -\node{}{10}-\node{}{8}-\node{}{6}-\node{}{4}-\node{}{2}$ & $-4$ & Bad\\ \hline
	$\node{\tnode 9}{8} - \node{}{7}-\node{}{6} -\node{}{5}-\node{}{4}-\node{}{3}-\node{}{2}-\node{}{1}$ & $-10$ & Bad\\ \hline
	\end{tabular}
	\caption{Quivers with a balanced $A_8$ subset and a single unbalanced node.}
	\label{tab:A8}
\end{table}

\begin{table}
	\centering
	\begin{tabular}{|c|c|c|c|}
	\hline
	\textbf{Excess} & \textbf{Type of the theory} & \textbf{$\Delta$ of extra generators} \\ \hline
	$e>0$& Good & $\Delta \geq 1$ \\ \hline
	$e=0$ & Good & $\Delta= 1$ \\ \hline
	$e=-1$ & Ugly &$\Delta > 0$ \\ \hline
	$e<-1$ & Bad & not applicable \\ \hline
	\end{tabular}
	\caption{Types of minimally unbalanced quiver gauge theories based on excess of the extra node.}
	\label{tab:goodbadugly}
\end{table}

\paragraph{General case} All quivers $\q$ that can be obtained with this procedure are summarized by a two parameter family, depicted in figure \ref{fig:Aseries}. The quiver in figure \ref{fig:Aseries} contains $a+b$ gauge nodes, of which $a+b-1$ are \emph{balanced}. The remaining unbalanced node (conveniently drawn red thorough this work) has excess:
\begin{align}\label{eq:e}
	e(a,b) = \frac{ab-2a-2b}{gcd(a,b)}
\end{align}
where $gcd(a,b)$ denotes the greatest common divisor of $a$ and $b$. 
For $e(a,b)>0$ the global symmetry of the Coulomb branch is $SU(n)$, where $n=a+b$, and one says that the quiver is \emph{minimally unbalanced} with positive excess. Therefore, the quiver in the first row in table \ref{tab:A8} with $e(4,5)=2$ corresponds to a good theory with positive excess. The quiver in the second row with $e(2,1)=0$ represents a good theory that is \emph{fully balanced} since all nodes have excess zero. The two bad theories with $e(a,b)=-4$ and $e(a,b)=-10$ are contained in the third and the fourth row of table \ref{tab:A8}, respectively. Minimally unbalanced quivers with the unbalanced node with excess $e=-1$ have either the entire or a part of the Coulomb branch freely generated.\footnote{See observation 3.1 in \cite{FG08}.}
Equation \eref{eq:e} defines a function:
\begin{equation}
	\begin{aligned}
		e: \mathbb{N}\times \mathbb{N}&\rightarrow \mathbb Z\\
		(a,b)&\mapsto e(a,b)
	\end{aligned}
\end{equation}
 that maps the two parameters of the family $a$ and $b$ to the excess of the top node of the corresponding quiver. This function can be visualized by defining a matrix $M$, with elements:
\be 
	M_{ab} = e(a,b)
\ee
Let $a$ and $b$ run from 1 to 16, then $M$ is $16\times 16$:
\be 
	M=\left(
\begin{array}{cccccccccccccccc} -3 & -4 & -5 & -6 & -7 & -8 & -9 & \bm{-10} & -11 & -12 & -13 & -14 & -15 & -16 & -17 & -18 \\
 -4 & -2 & -4 & -2 & -4 & -2 & \bm{-4} & -2 & -4 & -2 & -4 & -2 & -4 & -2 & -4 & -2 \\
 -5 & -4 & -1 & -2 & -1 & \bm{0} & 1 & 2 & 1 & 4 & 5 & 2 & 7 & 8 & 3 & 10 \\
 -6 & -2 & -2 & 0 & \bm{2} & 2 & 6 & 2 & 10 & 6 & 14 & 4 & 18 & 10 & 22 & 6 \\
 -7 & -4 & -1 & \bm{2} & 1 & 8 & 11 & 14 & 17 & 4 & 23 & 26 & 29 & 32 & 7 & 38 \\
 -8 & -2 & \bm{0} & 2 & 8 & 2 & 16 & 10 & 8 & 14 & 32 & 6 & 40 & 22 & 16 & 26 \\
 -9 & \bm{-4} & 1 & 6 & 11 & 16 & 3 & 26 & 31 & 36 & 41 & 46 & 51 & 8 & 61 & 66 \\
 \bm{-10} & -2 & 2 & 2 & 14 & 10 & 26 & 4 & 38 & 22 & 50 & 14 & 62 & 34 & 74 & 10 \\
 -11 & -4 & 1 & 10 & 17 & 8 & 31 & 38 & 5 & 52 & 59 & 22 & 73 & 80 & 29 & 94 \\
 -12 & -2 & 4 & 6 & 4 & 14 & 36 & 22 & 52 & 6 & 68 & 38 & 84 & 46 & 20 & 54 \\
 -13 & -4 & 5 & 14 & 23 & 32 & 41 & 50 & 59 & 68 & 7 & 86 & 95 & 104 & 113 & 122 \\
 -14 & -2 & 2 & 4 & 26 & 6 & 46 & 14 & 22 & 38 & 86 & 8 & 106 & 58 & 42 & 34 \\
 -15 & -4 & 7 & 18 & 29 & 40 & 51 & 62 & 73 & 84 & 95 & 106 & 9 & 128 & 139 & 150 \\
 -16 & -2 & 8 & 10 & 32 & 22 & 8 & 34 & 80 & 46 & 104 & 58 & 128 & 10 & 152 & 82 \\
 -17 & -4 & 3 & 22 & 7 & 16 & 61 & 74 & 29 & 20 & 113 & 42 & 139 & 152 & 11 & 178 \\
 -18 & -2 & 10 & 6 & 38 & 26 & 66 & 10 & 94 & 54 & 122 & 34 & 150 & 82 & 178 & 12 \\
\end{array}
\right)
\ee
The elements in bold are those that correspond to the quivers of length $a+b-1=8$, i.e. those in table \ref{tab:A8}. One can see that for a generic quiver the excess is positive. A theory with $a+b-1>8$ is \emph{bad} (negative excess) only if one of the two parameters is either 1 or 2. Furthermore, there are only three cases where the extra node is also balanced, i.e. excess $e(a,b)=0$. These are: $(a,b)=(3,6)$, $(a,b)=(4,4)$ and $(a,b)=(6,3)$. The first and last cases correspond to an enhancement of the global symmetry of the Coulomb branch from $SU(9)$ to $E_8$. The case $(a,b)=(4,4)$ sees a similar enhancement, this time from $SU(8)$ to $E_7$. The three cases with $e=-1$ are obtained for $(a,b)\in\{(3,3), (3,5), (5,3)\}$. For $(a,b)=(3,3)$ the greatest common divisor is $gcd(3,3)=3$, therefore, the quiver takes the form:
\be \label{eq:quiver12321w2on3}
	\q_{(3,3)} = ~\node{}1 -\node{}{2} -\node{\tnoder 2}{3}-\node{}{2}-\node{}{1}
\ee
and the Coulomb branch of this quiver is a freely generated variety (see 3.10 in \cite{Hanany:2018vph}):\footnote{Since the balanced sub-quiver corresponds to $A_5$ global symmetry, but $\mathbb{H}^{10}$ has isometry $Sp(10)$, we find an embedding: $SU(6) \hookleftarrow Sp(10)$. In particular, the pseudo-real fundamental rep of $Sp(10)$ projects to the pseudo-real $3^{rd}$ rank antisymmetric rep of $SU(6)$: $[1,0,0,0,0]_{Sp(10)} \hookrightarrow [0,0,1,0,0]_{SU(6)}$.}
\be
\mathcal{C}=\mathbb{H}^{10}.
\ee
For $(a,b)=(3,5)$ (or equivalently $(a,b)=(5,3)$) the quiver takes the form:
\be \label{eq:quiver3691215105w8on15}
	\q_{(3,5)} = ~\node{}3 -\node{}{6}  -\node{}{9} -\node{}{12}-\node{\tnoder 8}{15}-\node{}{10}-\node{}{5}
\ee
The quaternionic dimension of the Coulomb branch in \ref{eq:quiver3691215105w8on15} is $67$. The unbalanced node connects to the Dynkin node that corresponds to the $SU(8)$ representation with Dynkin labels $[0,0,0,0,1,0,0]$ and with dimension $56$. Drawing intuition from the quiver in \ref{eq:quiver12321w2on3} one would expect $112$ new operators transforming in the $[0,0,0,0,1,0,0]$ and its complex conjugate rep $[0,0,1,0,0,0,0]$. Although the excess is $e=-1$ (i.e. same as in freely generated \ref{eq:quiver12321w2on3}) the Coulomb branch of \ref{eq:quiver3691215105w8on15} seems to be more complicated (i.e has both a freely generated as well as a non-trivial part) and we leave its explicit computation for future study.


A formula for the HWG for minimally unbalanced $A$-type quivers with $a=b, s=1$ (i.e. with outer $\mathbb{Z}_2$ automorphism symmetry) is given by equation (23) in \cite{HPsic17}. Quivers of this type also show up in the study of Higgs branches of $5d$ $\mathcal{N}=1$ theories with $8$ supercharges \cite{FHMZ17}.


\subsection{$\gf $ of Type $D_n$}

Let us turn our focus to minimally unbalanced quivers with Coulomb branch isometry $\gf = SO(2n)$. The Dynkin diagram of $\mathfrak{so}(2n)$ is of the form:
$$ \node{}{}-\node{}{}-\node{}{}-\dots -\node{}{}-\node{\tnode{}}{}-\node{}{}$$
We find a two parameter family $a,n$, where $a$ is the position of the extra node starting from the left, and $n$ is the total number of \emph{balanced} nodes. Based on whether the unbalanced node attaches to one of the nodes on the main chain (i.e. $a<n-1$) or to one of the spinor nodes (i.e. $a=n$) we distinguish two categories with two further sub-categories:
\begin{itemize}
\item Unbalanced node attached to a node on the main chain:
\begin{itemize}
\item Unbalanced node of rank $1$ connects to an even node and the total number of balanced nodes is either even or odd. This family of quivers is contained in the first row in table \ref{tab:Dseries}.
\item Unbalanced node of rank $2$ connects to an odd node and the total number of balanced nodes is either even or odd. This family of quivers is depicted in the second row of table \ref{tab:Dseries}.
\end{itemize}
\item Unbalanced node attached to one of the spinor nodes:
\begin{itemize}
\item Unbalanced node is of rank $2$ and the total number of balanced nodes is even. This family of quivers is depicted in the third row in table \ref{tab:Dseries}.
\item Unbalanced node is of rank $4$ and the total number of balanced nodes is odd. This family of quivers is contained in the fourth row in table \ref{tab:Dseries}.
\end{itemize}
\end{itemize}
Note that the excess depends on a single parameter $a$. It is given by linear equations shown in the third column in table \ref{tab:Dseries}. 

\begin{table}
	\centering
	\begin{tabular}{|c|l|c|}
	\hline
	$a$ &  \multicolumn{1}{c|}{Quiver} & Excess \\ \hline
	$\begin{array}{c}
	a<n-1\\
	a=2m
	\end{array}$ &\begin{tikzpicture}[scale=0.70]

\draw (0.4,0) -- (0.8,0);
\draw (0,0) circle (0.4cm);
\draw (0,-0.8) node {\footnotesize{$2m$}};
\draw (1.6,0) -- (2,0);
\draw (1.2,0) node {\footnotesize{\dots}};
\draw (-2.4,0) circle (0.4cm);
\draw (-2.4,-0.8) node {\footnotesize{$3$}};
\draw (-2.8,0) -- (-3.2,0);
\draw (-3.6,0) circle (0.4cm);
\draw (-3.6,-0.8) node {\footnotesize{$2$}};
\draw (-4,0) -- (-4.4,0);
\draw (-4.8,0) circle (0.4cm);
\draw (-4.8,-0.8) node {\footnotesize{$1$}};
\draw (2.4,0) circle (0.4cm);
\draw (2.4,-0.8) node {\footnotesize{$2m$}};
\draw (-0.4,0) -- (-0.8,0);
\draw (-1.2,0) node {\footnotesize{$\dots$}};
\draw (-1.6,0) -- (-2,0);
\draw (0,0.4) -- (0,0.8);
\draw (0,1.2) [red,fill=red!30] circle (0.4cm);
\draw (0,2) node {\footnotesize{$1$}};
\draw [decorate,decoration={brace,amplitude=6pt}] (0.1,0.5) to (2.3,0.5);
\draw (1.4,1.1) node {\footnotesize{$n-a-1$}};
\draw (3.4,-1) circle (0.4cm);
\draw (4.26,-1) node {\footnotesize{$m$}};
\draw (2.7,-0.24) -- (3.1,-0.75);
\draw (3.4,1) circle (0.4cm);
\draw (4.26,1) node {\footnotesize{$m$}};
\draw (2.7,0.24) -- (3.1,0.75);

\end{tikzpicture} & $a-2$\\ \hline	
	$\begin{array}{c}
	a<n-1\\
	a=2m+1
	\end{array}$ &\begin{tikzpicture}[scale=0.70]

\draw (-4,0) -- (-4.4,0);
\draw (-4.8,0) circle (0.4cm);
\draw (-4.8,-0.8) node {\footnotesize{$2$}};
\draw (-2.8,0) -- (-3.2,0);
\draw (-3.6,0) circle (0.4cm);
\draw (-3.6,-0.8) node {\footnotesize{$4$}};
\draw (-1.6,0) -- (-2,0);
\draw (-2.4,0) circle (0.4cm);
\draw (-2.4,-0.8) node {\footnotesize{$6$}};
\draw (-0.4,0) -- (-0.8,0);
\draw (-1.2,0) node {\footnotesize{$\dots$}};
\draw (0,0) circle (0.4cm);
\draw (-0.2,-0.8) node {\footnotesize{$2(2m+1)$}};
\draw (0.4,0) -- (0.8,0);
\draw (1.2,0) node {\footnotesize{\dots}};
\draw (1.6,0) -- (2,0);
\draw (2.4,0) circle (0.4cm);
\draw (2.2,-0.8) node {\footnotesize{$2(2m+1)$}};
\draw (0,0.4) -- (0,0.8);
\draw (0,1.2) [red,fill=red!30] circle (0.4cm);
\draw (0,2) node {\footnotesize{$2$}};
\draw [decorate,decoration={brace,amplitude=6pt}] (0.1,0.5) to (2.3,0.5);
\draw (1.4,1.1) node {\footnotesize{$n-a-1$}};
\draw (3.8,-1) circle (0.4cm);
\draw (3.8,1) circle (0.4cm);
\draw (5.1,-1) node {\footnotesize{$2m+1$}};
\draw (5.1,1) node {\footnotesize{$2m+1$}};
\draw (2.76,-0.16) -- (3.5,-0.75);
\draw (2.76,0.16) -- (3.5,0.75);

\end{tikzpicture} & $2a-4$\\ \hline	
	$\begin{array}{c}
	a=n\\
	a=2m
	\end{array}$ &
\begin{tikzpicture}[scale=0.70]

\draw (0,0) circle (0.4cm);
\draw (-.1,-0.8) node {\footnotesize{$2m-2$}};
\draw (-2.4,0) circle (0.4cm);
\draw (-2.4,-0.8) node {\footnotesize{$3$}};
\draw (-2.8,0) -- (-3.2,0);
\draw (-3.6,0) circle (0.4cm);
\draw (-3.6,-0.8) node {\footnotesize{$2$}};
\draw (-4,0) -- (-4.4,0);
\draw (-4.8,0) circle (0.4cm);
\draw (-4.8,-0.8) node {\footnotesize{$1$}};
\draw (-0.4,0) -- (-0.8,0);
\draw (-1.2,0) node {\footnotesize{\dots}};
\draw (-1.6,0) -- (-2,0);
\draw (1.4,-1) circle (0.4cm);
\draw (0.34,-0.2) -- (1.1,-0.75);
\draw (1.4,1) circle (0.4cm);
\draw (0.34,0.2) -- (1.1,0.75);
\draw (1.4,-1.7) node {\footnotesize{$m$}};
\draw (2.6,1) node {\footnotesize{$m-1$}};
\draw (1.4,1.6) node {\footnotesize{$$}};
\draw (1.8,-1) -- (2.4,-1);
\draw (2.8,-1) [red,fill=red!30] circle (0.4cm);
\draw (2.8,-1.7) node {\footnotesize{$2$}};

\end{tikzpicture} & $\frac{a}{2}-4$\\ \hline	
	$\begin{array}{c}
	a=n\\
	a=2m+1
	\end{array}$ &
\begin{tikzpicture}[scale=0.70]

\draw (0,0) circle (0.4cm);
\draw (-0.2,-0.8) node {\footnotesize{$2(2m-1)$}};
\draw (-2.4,0) circle (0.4cm);
\draw (-2.4,-0.8) node {\footnotesize{$6$}};
\draw (-2.8,0) -- (-3.2,0);
\draw (-3.6,0) circle (0.4cm);
\draw (-3.6,-0.8) node {\footnotesize{$4$}};
\draw (-4,0) -- (-4.4,0);
\draw (-4.8,0) circle (0.4cm);
\draw (-4.8,-0.8) node {\footnotesize{$2$}};
\draw (-0.4,0) -- (-0.8,0);
\draw (-1.2,0) node {\footnotesize{\dots}};
\draw (-1.6,0) -- (-2,0);
\draw (1.4,-1) circle (0.4cm);
\draw (0.34,-0.2) -- (1.1,-0.75);
\draw (1.4,1) circle (0.4cm);
\draw (0.34,0.2) -- (1.1,0.75);
\draw (1.4,-1.7) node {\footnotesize{$2m+1$}};
\draw (2.6,1) node {\footnotesize{$2m-1$}};
\draw (1.4,1.6) node {\footnotesize{$$}};
\draw (1.8,-1) -- (2.4,-1);
\draw (2.8,-1) [red,fill=red!30] circle (0.4cm);
\draw (2.8,-1.7) node {\footnotesize{$4$}};

\end{tikzpicture} & $a-8$ \\ \hline	
	\end{tabular}
	\caption{Classification of minimally unbalanced quivers with $\gf = SO(2n)$.}
	\label{tab:Dseries}
\end{table}

\newpage

Note that we find the following special cases:
\begin{itemize}
\item In the first row:
\begin{itemize}
\item for $a=2$ the quiver has zero excess and corresponds to the reduced moduli space of one $D_n$ instanton on $\mathbb{C}^2$. 
\end{itemize}
\item In the third row:
\begin{itemize}
\item for $m=4$ one obtains the affine $E_8$ Dynkin diagram corresponding to the reduced moduli space of one $E_8$ instanton on $\mathbb{C}^2$. The Coulomb branch is denoted as $\mathcal{C}=\overline{min_{E_8}}$.
\end{itemize}
\item In the last row:
\begin{itemize}
\item for $m=3$ (or equivalently $n=7$) one obtains a peculiar quiver with $e=-1$ of the form:
\be \label{eq:peculiarDquiver}
\node{}2-\node{}4-\node{}6-\node{}8-\node{\tnode 5}{10}-\node{}7-\noder{}4 
\ee
Similarly as for the quiver in \ref{eq:quiver3691215105w8on15} the unbalanced node does not connect to a node corresponding to a pseudo-real representation. In the case of \ref{eq:peculiarDquiver}, the unbalanced node has negative excess $e=-1$ and it connects to a Dynkin node corresponding to the complex spinor representation with dimension $64$. The dimension of the Coulomb branch is $45$.\footnote{Recall that the quaternionic dimension of the Coulomb branch can be read off from the quiver as: $dim(\mathcal{C}) = \sum_i r_i -1$, where $r_i$ denotes the rank of a particular node.} Analogically with \ref{eq:quiver3691215105w8on15}, it seems that the Coulomb branch is rather complicated with both a freely generated as well as a non-trivial part. The explicit computation of $\mathcal{C}$ is left for future study.
\end{itemize}
\end{itemize}
The HWG for the case in the third row in table \ref{tab:Dseries} is given using $D_n$ highest weight fugacities by equation (26) in \cite{HPsic17}.

\subsection{$\gf$ of Type $E_n$}

Let us finally proceed by analyzing the last category of simply laced theories with $G$ of type $E_n$. All the different minimally unbalanced quivers with a certain $E$-type exceptional global symmetry can be written down explicitly. We report the excess of the unbalanced node in the second column of the classification tables. Note that all the quivers in the classification that are balanced (extra node drawn orange) are the affine Dynkin diagrams of the corresponding global symmetry, where the affine node has rank $1$. When such node is taken to be a flavor node, the quiver corresponds to both the closure of the minimal nilpotent orbit of $\mathfrak{e}_n$ algebra, and to the reduced moduli space of one $E_n$ instanton on $\mathbb{C}^2$ \cite{IS96,CHZ13}.
\subsubsection{$\gf$ of Type $E_6$}

For $\gf = E_6$ one explicitly writes down all the cases as displayed in table \ref{tab:E6series}. Note that there are only four distinct cases due to the $\mathbb{Z}_2$ outer automorphism of the $E_6$ Dynkin diagram. Also note that when the extra node happens to be balanced, it is drawn orange. This convention is used throughout the paper. The last row in table \ref{tab:E6series} is special (with excess $e=0$) and its Coulomb branch is the reduced moduli space of one $E_6$ instanton on $\mathbb{C}^2$ \cite{IS96,CHZ13}. The HWG for this quiver is given in terms of the $SU(6)\times SU(2)$ highest weight fugacities by equation (28) in \cite{HPsic17}.
\begin{table}
	\centering
	\begin{tabular}{|c|c|}
		\hline
		Quiver & Excess \\ \hline
		\begin{tikzpicture}[scale=0.70]

\draw (0.4,0) -- (0.8,0);
\draw (0,0) circle (0.4cm);
\draw (0,-0.8) node {\footnotesize{$6$}};
\draw (1.6,0) -- (2,0);
\draw (2.4,0) circle (0.4cm);
\draw (1.2,-0.8) node {\footnotesize{$4$}};
\draw(1.2,0) circle (0.4cm);
\draw (2.4,-0.8) node {\footnotesize{$2$}};
\draw (-2.4,0) circle (0.4cm);
\draw (-2.4,-0.8) node {\footnotesize{$4$}};
\draw (-0.4,0) -- (-0.8,0);
\draw (-1.2,0) circle (0.4cm);
\draw (-1.2,-0.8) node {\footnotesize{$5$}};
\draw (-1.6,0) -- (-2,0);
\draw (0,0.4) -- (0,0.8);
\draw (0,1.2) circle (0.4cm);
\draw (0,2) node {\footnotesize{$3$}};
\draw (-2.4,0.4) -- (-2.4,0.8);
\draw (-2.4,1.2)[red, fill=red!30] circle (0.4cm);
\draw (-2.4,2) node {\footnotesize{$3$}};
\end{tikzpicture} & $-2$\\ \hline
		
\begin{tikzpicture}[scale=0.70]

\draw (0.4,0) -- (0.8,0);
\draw (0,0) circle (0.4cm);
\draw (0,-0.8) node {\footnotesize{$12$}};
\draw (1.6,0) -- (2,0);
\draw (2.4,0) circle (0.4cm);
\draw (1.2,-0.8) node {\footnotesize{$8$}};
\draw(1.2,0) circle (0.4cm);
\draw (2.4,-0.8) node {\footnotesize{$4$}};
\draw (-2.4,0) circle (0.4cm);
\draw (-2.4,-0.8) node {\footnotesize{$5$}};
\draw (-0.4,0) -- (-0.8,0);
\draw (-1.2,0) circle (0.4cm);
\draw (-1.2,-0.8) node {\footnotesize{$10$}};
\draw (-1.6,0) -- (-2,0);
\draw (0,0.4) -- (0,0.8);
\draw (0,1.2) circle (0.4cm);
\draw (0,2) node {\footnotesize{$6$}};
\draw (-1.2,0.4) -- (-1.2,0.8);
\draw (-1.2,1.2) [red, fill=red!30] circle (0.4cm);
\draw (-1.2,2) node {\footnotesize{$3$}};

\end{tikzpicture} & $4$\\ \hline
		\begin{tikzpicture}[scale=0.70]

\draw (0.4,0) -- (0.8,0);
\draw (0,0) circle (0.4cm);
\draw (0,-0.8) node {\footnotesize{$6$}};
\draw (1.2,-0.8) node {\footnotesize{$4$}};
\draw (2.4,0) circle (0.4cm);
\draw (2.4,-0.8) node {\footnotesize{$2$}};
\draw(1.2,0) circle (0.4cm);
\draw (-2.4,0) circle (0.4cm);
\draw (-2.4,-0.8) node {\footnotesize{$2$}};
\draw (-0.4,0) -- (-0.8,0);
\draw (1.6,0) -- (2,0);
\draw (-1.2,0) circle (0.4cm);
\draw (-1.2,-0.8) node {\footnotesize{$4$}};
\draw (-1.6,0) -- (-2,0);
\draw (-0.26,0.3) -- (-0.65,0.83);
\draw (-0.8,1.2) [red, fill=red!30] circle (0.4cm);
\draw (-0.8,2) node {\footnotesize{$1$}};
\draw (0.26,0.3) -- (0.65,0.83);
\draw (0.8,1.2) circle (0.4cm);
\draw (0.8,2) node {\footnotesize{$3$}};

\end{tikzpicture} & $4$\\ \hline

\begin{tikzpicture}[scale=0.70]

\draw (0.4,0) -- (0.8,0);
\draw (0,0) circle (0.4cm);
\draw (0,-0.8) node {\footnotesize{$3$}};
\draw (1.6,0) -- (2,0);
\draw (2.4,0) circle (0.4cm);
\draw (1.2,-0.8) node {\footnotesize{$2$}};
\draw(1.2,0) circle (0.4cm);
\draw (2.4,-0.8) node {\footnotesize{$1$}};
\draw (-2.4,0) circle (0.4cm);
\draw (-2.4,-0.8) node {\footnotesize{$1$}};
\draw (-0.4,0) -- (-0.8,0);
\draw (-1.2,0) circle (0.4cm);
\draw (-1.2,-0.8) node {\footnotesize{$2$}};
\draw (-1.6,0) -- (-2,0);
\draw (0,0.4) -- (0,0.8);
\draw (0,1.2) circle (0.4cm);
\draw (0.8,1.2) node {\footnotesize{$2$}};
\draw (0,1.6) -- (0,2.0);
\draw (0,2.4) [orange, fill=orange!30] circle (0.4cm);
\draw (0.8,2.4) node {\footnotesize{$1$}};
\draw (0,3) node {\footnotesize{$$}};

\end{tikzpicture} & $0$\\ \hline
	\end{tabular}
	\caption{Minimally unbalanced quivers with $\gf=E_6$.}
	\label{tab:E6series}
\end{table}

\newpage

\subsubsection{$\gf$ of Type $E_7$}

Next, we consider minimally unbalanced quivers with $E_7$ global symmetry. One proceeds by attaching the unbalanced node from leftmost to the rightmost node. Due to the lack of any automorphism of the $E_7$ Dynkin diagram one has to exhaust all $7$ cases. The resulting minimally unbalanced quivers are collected in table \ref{tab:E7series}. The first row in table \ref{tab:E7series} is again a special case since its Coulomb branch is a reduced moduli space of one $E_7$ instanton on $\mathbb{C}^2$ \cite{IS96,Cremonesi:2015lsa}. The HWG for this theory is given in terms of $SU(8)$ highest weight fugacities, by equation (44) in \cite{HPsic17}. The last row of table \ref{tab:E7series} depicts a theory with excess $e=-1$. The Coulomb branch is freely generated:
\be
\mathcal{C}=\mathbb{H}^{28}
\ee
and we find the embedding $[1,0,0,0,0,0,0]_{E_7} \hookleftarrow [1,0,0,0,0,0,0]_{Sp(28)}$ of $E_7$ inside $Sp(28)$. Both of these $56$ dimensional representations are pseudo-real which is consistent with the expectation from $\mathbb{H}^{28}$. More generally, $\mathbb{H}^{n}$ is always generated by $2n$ generators that transform under the pseudo-real fundamental representation of $Sp(n)$.

\begin{table}
	\centering
		\begin{tabular}{|c|c|}
			\hline
			Quiver & Excess \\ \hline
			\begin{tikzpicture}[scale=0.70]

\draw (0.4,0) -- (0.8,0);
\draw (2.8,0) -- (3.2,0);
\draw (0,0) circle (0.4cm);
\draw (0,-0.8) node {\footnotesize{$4$}};
\draw (1.6,0) -- (2,0);
\draw (2.4,0) circle (0.4cm);
\draw (1.2,-0.8) node {\footnotesize{$3$}};
\draw (3.6,0) circle (0.4cm);
\draw (3.6,-0.8) node {\footnotesize{$1$}};
\draw (-2.4,1.2) [orange,fill=orange!30] circle (0.4cm);
\draw (-2.4,2) node {\footnotesize{$1$}};
\draw (-2.4,0.4) -- (-2.4,0.8);
\draw(1.2,0) circle (0.4cm);
\draw (2.4,-0.8) node {\footnotesize{$2$}};
\draw (-2.4,0) circle (0.4cm);
\draw (-2.4,-0.8) node {\footnotesize{$2$}};
\draw (-0.4,0) -- (-0.8,0);
\draw (-1.2,0) circle (0.4cm);
\draw (-1.2,-0.8) node {\footnotesize{$3$}};
\draw (-1.6,0) -- (-2,0);
\draw (0,0.4) -- (0,0.8);
\draw (0,1.2) circle (0.4cm);
\draw (0,2) node {\footnotesize{$2$}};

\end{tikzpicture} & 0\\ \hline
			 
\begin{tikzpicture}[scale=0.70]

\draw (0.4,0) -- (0.8,0);
\draw (2.8,0) -- (3.2,0);
\draw (0,0) circle (0.4cm);
\draw (0,-0.8) node {\footnotesize{$8$}};
\draw (1.6,0) -- (2,0);
\draw (2.4,0) circle (0.4cm);
\draw (1.2,-0.8) node {\footnotesize{$6$}};
\draw (3.6,0) circle (0.4cm);
\draw (3.6,-0.8) node {\footnotesize{$2$}};
\draw (-1.2,1.2) [red,fill=red!30] circle (0.4cm);
\draw (-1.2,2) node {\footnotesize{$1$}};
\draw (-1.2,0.4) -- (-1.2,0.8);
\draw(1.2,0) circle (0.4cm);
\draw (2.4,-0.8) node {\footnotesize{$4$}};
\draw (-2.4,0) circle (0.4cm);
\draw (-2.4,-0.8) node {\footnotesize{$3$}};
\draw (-0.4,0) -- (-0.8,0);
\draw (-1.2,0) circle (0.4cm);
\draw (-1.2,-0.8) node {\footnotesize{$6$}};
\draw (-1.6,0) -- (-2,0);
\draw (0,0.4) -- (0,0.8);
\draw (0,1.2) circle (0.4cm);
\draw (0,2) node {\footnotesize{$4$}};

\end{tikzpicture}& 4\\ \hline
			 
\begin{tikzpicture}[scale=0.70]

\draw (0.4,0) -- (0.8,0);
\draw (0,0) circle (0.4cm);
\draw (0,-0.8) node {\footnotesize{$12$}};
\draw (1.2,-0.8) node {\footnotesize{$9$}};
\draw (2.4,0) circle (0.4cm);
\draw (2.4,-0.8) node {\footnotesize{$6$}};
\draw(1.2,0) circle (0.4cm);
\draw (-2.4,0) circle (0.4cm);
\draw (-2.4,-0.8) node {\footnotesize{$4$}};
\draw (-0.4,0) -- (-0.8,0);
\draw (1.6,0) -- (2,0);
\draw (-1.2,0) circle (0.4cm);
\draw (-1.2,-0.8) node {\footnotesize{$8$}};
\draw (2.8,0) -- (3.2,0);
\draw (3.6,0) circle (0.4cm);
\draw (3.6,-0.8) node {\footnotesize{$3$}};
\draw (-1.6,0) -- (-2,0);
\draw (-0.26,0.3) -- (-0.65,0.82);
\draw (-0.8,1.2) [red, fill=red!30] circle (0.4cm);
\draw (-0.8,2) node {\footnotesize{$1$}};
\draw (0.26,0.3) -- (0.65,0.82);
\draw (0.8,1.2) circle (0.4cm);
\draw (0.8,2) node {\footnotesize{$6$}};

\end{tikzpicture}& 10\\ \hline
			
\begin{tikzpicture}[scale=0.70]

\draw (0.4,0) -- (0.8,0);
\draw (2.8,0) -- (3.2,0);
\draw (0,0) circle (0.4cm);
\draw (0,-0.8) node {\footnotesize{$12$}};
\draw (1.6,0) -- (2,0);
\draw (2.4,0) circle (0.4cm);
\draw (1.2,-0.8) node {\footnotesize{$9$}};
\draw (3.6,0) circle (0.4cm);
\draw (3.6,-0.8) node {\footnotesize{$3$}};
\draw (0,2.4) [red,fill=red!30] circle (0.4cm);
\draw (0.8,2.4) node {\footnotesize{$2$}};
\draw (0,1.6) -- (0,2);
\draw(1.2,0) circle (0.4cm);
\draw (2.4,-0.8) node {\footnotesize{$6$}};
\draw (-2.4,0) circle (0.4cm);
\draw (-2.4,-0.8) node {\footnotesize{$4$}};
\draw (-0.4,0) -- (-0.8,0);
\draw (-1.2,0) circle (0.4cm);
\draw (-1.2,-0.8) node {\footnotesize{$8$}};
\draw (-1.6,0) -- (-2,0);
\draw (0,0.4) -- (0,0.8);
\draw (0,1.2) circle (0.4cm);
\draw (0.8,1.2) node {\footnotesize{$7$}};
\draw (0,3) node {\footnotesize{$$}};

\end{tikzpicture} &3 \\ \hline
			 
\begin{tikzpicture}[scale=0.70]

\draw (0.4,0) -- (0.8,0);
\draw (2.8,0) -- (3.2,0);
\draw (0,0) circle (0.4cm);
\draw (0,-0.8) node {\footnotesize{$18$}};
\draw (1.6,0) -- (2,0);
\draw (2.4,0) circle (0.4cm);
\draw (1.2,-0.8) node {\footnotesize{$15$}};
\draw (3.6,0) circle (0.4cm);
\draw (3.6,-0.8) node {\footnotesize{$5$}};
\draw (1.2,1.2) [red,fill=red!30] circle (0.4cm);
\draw (1.2,2) node {\footnotesize{$1$}};
\draw (1.2,0.4) -- (1.2,0.8);
\draw(1.2,0) circle (0.4cm);
\draw (2.4,-0.8) node {\footnotesize{$10$}};
\draw (-2.4,0) circle (0.4cm);
\draw (-2.4,-0.8) node {\footnotesize{$6$}};
\draw (-0.4,0) -- (-0.8,0);
\draw (-1.2,0) circle (0.4cm);
\draw (-1.2,-0.8) node {\footnotesize{$12$}};
\draw (-1.6,0) -- (-2,0);
\draw (0,0.4) -- (0,0.8);
\draw (0,1.2) circle (0.4cm);
\draw (0,2) node {\footnotesize{$9$}};

\end{tikzpicture}& 13\\ \hline
			 
\begin{tikzpicture}[scale=0.70]

\draw (0.4,0) -- (0.8,0);
\draw (2.8,0) -- (3.2,0);
\draw (0,0) circle (0.4cm);
\draw (0,-0.8) node {\footnotesize{$6$}};
\draw (1.6,0) -- (2,0);
\draw (2.4,0) circle (0.4cm);
\draw (1.2,-0.8) node {\footnotesize{$5$}};
\draw (3.6,0) circle (0.4cm);
\draw (3.6,-0.8) node {\footnotesize{$2$}};
\draw (2.4,1.2) [red,fill=red!30] circle (0.4cm);
\draw (2.4,2) node {\footnotesize{$1$}};
\draw (2.4,0.4) -- (2.4,0.8);
\draw(1.2,0) circle (0.4cm);
\draw (2.4,-0.8) node {\footnotesize{$4$}};
\draw (-2.4,0) circle (0.4cm);
\draw (-2.4,-0.8) node {\footnotesize{$2$}};
\draw (-0.4,0) -- (-0.8,0);
\draw (-1.2,0) circle (0.4cm);
\draw (-1.2,-0.8) node {\footnotesize{$4$}};
\draw (-1.6,0) -- (-2,0);
\draw (0,0.4) -- (0,0.8);
\draw (0,1.2) circle (0.4cm);
\draw (0,2) node {\footnotesize{$3$}};

\end{tikzpicture}& 2\\ \hline
			 
\begin{tikzpicture}[scale=0.70]

\draw (0.4,0) -- (0.8,0);
\draw (2.8,0) -- (3.2,0);
\draw (0,0) circle (0.4cm);
\draw (0,-0.8) node {\footnotesize{$6$}};
\draw (1.6,0) -- (2,0);
\draw (2.4,0) circle (0.4cm);
\draw (1.2,-0.8) node {\footnotesize{$5$}};
\draw (3.6,0) circle (0.4cm);
\draw (3.6,-0.8) node {\footnotesize{$3$}};
\draw (3.6,1.2) [red,fill=red!30] circle (0.4cm);
\draw (3.6,2) node {\footnotesize{$2$}};
\draw (3.6,0.4) -- (3.6,0.8);
\draw(1.2,0) circle (0.4cm);
\draw (2.4,-0.8) node {\footnotesize{$4$}};
\draw (-2.4,0) circle (0.4cm);
\draw (-2.4,-0.8) node {\footnotesize{$2$}};
\draw (-0.4,0) -- (-0.8,0);
\draw (-1.2,0) circle (0.4cm);
\draw (-1.2,-0.8) node {\footnotesize{$4$}};
\draw (-1.6,0) -- (-2,0);
\draw (0,0.4) -- (0,0.8);
\draw (0,1.2) circle (0.4cm);
\draw (0,2) node {\footnotesize{$3$}};

\end{tikzpicture}& $-1$ \\ \hline
		\end{tabular}
	\caption{Minimally unbalanced quivers with $\gf = E_7$.}
	\label{tab:E7series}
\end{table}

\subsubsection{$\gf$ of Type $E_8$}

Finally, we present all minimally unbalanced theories with global symmetry $\gf = E_8$ in tables \ref{tab:E8series1} and \ref{tab:E8series2}. Again, we exhaust all eight distinct cases. The quiver in the shape of an affine $E_8$ Dynkin diagram in the last row of table \ref{tab:E8series2} is readily identified as the moduli space of one $E_8$ instanton on $\mathbb{C}^2$ \cite{IS96,Cremonesi:2015lsa}.  In terms of the $SO(16)$ highest weight fugacities, the HWG for this quiver is given by equation (142) in \cite{HPsic17}.

\begin{table}[htp]
	\centering
	\begin{tabular}{|c|c|}
		\hline
		Quiver & Excess \\ \hline
		 \begin{tikzpicture}[scale=0.70]

\draw (0.4,0) -- (0.8,0);
\draw (0,0) circle (0.4cm);
\draw (0,-0.8) node {\footnotesize{$8$}};
\draw (1.6,0) -- (2,0);
\draw (2.4,0) circle (0.4cm);
\draw (1.2,-0.8) node {\footnotesize{$6$}};
\draw(1.2,0) circle (0.4cm);
\draw (2.4,-0.8) node {\footnotesize{$4$}};
\draw (-2.4,0) circle (0.4cm);
\draw (-2.4,-0.8) node {\footnotesize{$7$}};
\draw (-2.8,0) -- (-3.2,0);
\draw (-3.6,0) circle (0.4cm);
\draw (-3.6,-0.8) node {\footnotesize{$4$}};
\draw (2.8,0) -- (3.2,0);
\draw (3.6,0) circle (0.4cm);
\draw (3.6,-0.8) node {\footnotesize{$2$}};
\draw (-0.4,0) -- (-0.8,0);
\draw (-1.2,0) circle (0.4cm);
\draw (-1.2,-0.8) node {\footnotesize{$10$}};
\draw (-1.6,0) -- (-2,0);
\draw (-1.2,0.4) -- (-1.2,0.8);
\draw (-1.2,1.2) circle (0.4cm);
\draw (-1.2,2) node {\footnotesize{$5$}};
\draw (-3.6,0.4) -- (-3.6,0.8);
\draw (-3.6,1.2) [red,fill=red!30] circle  (0.4cm);
\draw (-3.6,2) node {\footnotesize{$1$}};

\end{tikzpicture}&2 \\ \hline
		 
\begin{tikzpicture}[scale=0.70]

\draw (0.4,0) -- (0.8,0);
\draw (0,0) circle (0.4cm);
\draw (-3.6,-0.8) node {\footnotesize{$7$}};
\draw (-2.4,-0.8) node {\footnotesize{$14$}};
\draw (-1.2,-0.8) node {\footnotesize{$20$}};
\draw (0,-0.8) node     {\footnotesize{$16$}};
\draw (1.2,-0.8) node {\footnotesize{$12$}};
\draw (2.4,-0.8) node {\footnotesize{$8$}};
\draw (3.6,-0.8) node {\footnotesize{$4$}};
\draw (-1.2,2) node {\footnotesize{$10$}};
\draw (1.6,0) -- (2,0);
\draw (2.4,0) circle (0.4cm);
\draw(1.2,0) circle (0.4cm);
\draw (-2.4,0) circle (0.4cm);
\draw (-2.8,0) -- (-3.2,0);
\draw (-3.6,0) circle (0.4cm);
\draw (2.8,0) -- (3.2,0);
\draw (3.6,0) circle (0.4cm);
\draw (-0.4,0) -- (-0.8,0);
\draw (-1.2,0) circle (0.4cm);
\draw (-1.6,0) -- (-2,0);
\draw (-1.2,0.4) -- (-1.2,0.8);
\draw (-1.2,1.2) circle (0.4cm);
\draw (-2.4,0.4) -- (-2.4,0.8);
\draw (-2.4,1.2) [red,fill=red!30] circle  (0.4cm);
\draw (-2.4,2) node {\footnotesize{$1$}};

\end{tikzpicture}& 12 \\ \hline
		 
\begin{tikzpicture}[scale=0.70]

\draw (0.4,0) -- (0.8,0);
\draw (0,0) circle (0.4cm);
\draw (-3.6,-0.8) node {\footnotesize{$10$}};
\draw (-2.4,-0.8) node {\footnotesize{$20$}};
\draw (-1.2,-0.8) node {\footnotesize{$30$}};
\draw (0,-0.8) node     {\footnotesize{$24$}};
\draw (1.2,-0.8) node {\footnotesize{$18$}};
\draw (2.4,-0.8) node {\footnotesize{$12$}};
\draw (3.6,-0.8) node {\footnotesize{$6$}};
\draw (1.6,0) -- (2,0);
\draw (2.4,0) circle (0.4cm);
\draw(1.2,0) circle (0.4cm);
\draw (-2.4,0) circle (0.4cm);
\draw (-2.8,0) -- (-3.2,0);
\draw (-3.6,0) circle (0.4cm);
\draw (2.8,0) -- (3.2,0);
\draw (3.6,0) circle (0.4cm);
\draw (-0.4,0) -- (-0.8,0);
\draw (-1.6,0) -- (-2,0);
\draw (-1.2,0) circle (0.4cm);
\draw (-1.4,0.334) -- (-1.8,0.865);
\draw (-2,1.2) [red,fill=red!30] circle  (0.4cm);
\draw (-2,2) node {\footnotesize{$1$}};
\draw (-0.4,1.2) circle (0.4cm);
\draw (-0.4,2) node {\footnotesize{$15$}};
\draw (-1,0.334) -- (-0.6,0.865);

\end{tikzpicture}& 28 \\ \hline
		
\begin{tikzpicture}[scale=0.70]

\draw (0.4,0) -- (0.8,0);
\draw (0,0) circle (0.4cm);

\draw (-3.6,-0.8) node {\footnotesize{$5$}};
\draw (-2.4,-0.8) node {\footnotesize{$10$}};
\draw (-1.2,-0.8) node {\footnotesize{$15$}};
\draw (0,-0.8) node     {\footnotesize{$12$}};
\draw (1.2,-0.8) node {\footnotesize{$9$}};
\draw (2.4,-0.8) node {\footnotesize{$6$}};
\draw (3.6,-0.8) node {\footnotesize{$3$}};
\draw (-0.4,1.2) node {\footnotesize{$8$}};
\draw (1.6,0) -- (2,0);
\draw (2.4,0) circle (0.4cm);
\draw(1.2,0) circle (0.4cm);
\draw (-2.4,0) circle (0.4cm);
\draw (-2.8,0) -- (-3.2,0);
\draw (-3.6,0) circle (0.4cm);
\draw (2.8,0) -- (3.2,0);
\draw (3.6,0) circle (0.4cm);
\draw (-0.4,0) -- (-0.8,0);
\draw (-1.2,0) circle (0.4cm);
\draw (-1.6,0) -- (-2,0);
\draw (-1.2,0.4) -- (-1.2,0.8);
\draw (-1.2,1.2) circle (0.4cm);
\draw (-1.2,1.6) -- (-1.2,2);
\draw (-1.2,2.4) [red,fill=red!30] circle  (0.4cm);
\draw (-0.4,2.4) node {\footnotesize{$1$}};
\draw (0,3) node {\footnotesize{$$}};
		 
\end{tikzpicture}& 6 \\ \hline
	\end{tabular}
	\caption{First set of minimally unbalanced quivers with $\gf = E_8$.}
	\label{tab:E8series1}
\end{table}


\begin{table}[htp]
	\centering
	\begin{tabular}{|c|c|}
		\hline
		Quiver & Excess \\ \hline
		 \begin{tikzpicture}[scale=0.70]
		 
\draw (0.4,0) -- (0.8,0);
\draw (0,0) circle (0.4cm);
\draw (-3.6,-0.8) node {\footnotesize{$8$}};
\draw (-2.4,-0.8) node {\footnotesize{$16$}};
\draw (-1.2,-0.8) node {\footnotesize{$24$}};
\draw (0,-0.8) node     {\footnotesize{$20$}};
\draw (1.2,-0.8) node {\footnotesize{$15$}};
\draw (2.4,-0.8) node {\footnotesize{$10$}};
\draw (3.6,-0.8) node {\footnotesize{$5$}};
\draw (-1.2,2) node {\footnotesize{$12$}};
\draw (1.6,0) -- (2,0);
\draw (2.4,0) circle (0.4cm);
\draw(1.2,0) circle (0.4cm);
\draw (-2.4,0) circle (0.4cm);
\draw (-2.8,0) -- (-3.2,0);
\draw (-3.6,0) circle (0.4cm);
\draw (2.8,0) -- (3.2,0);
\draw (3.6,0) circle (0.4cm);
\draw (-0.4,0) -- (-0.8,0);
\draw (-1.2,0) circle (0.4cm);
\draw (-1.6,0) -- (-2,0);
\draw (-1.2,0.4) -- (-1.2,0.8);
\draw (-1.2,1.2) circle (0.4cm);
\draw (0,0.4) -- (0,0.8);
\draw (0,1.2) [red,fill=red!30] circle  (0.4cm);
\draw (0,2) node {\footnotesize{$1$}};

\end{tikzpicture}& 18 \\ \hline
		 
\begin{tikzpicture}[scale=0.70]

\draw (0.4,0) -- (0.8,0);
\draw (0,0) circle (0.4cm);
\draw (-3.6,-0.8) node {\footnotesize{$6$}};
\draw (-2.4,-0.8) node {\footnotesize{$12$}};
\draw (-1.2,-0.8) node {\footnotesize{$18$}};
\draw (0,-0.8) node     {\footnotesize{$15$}};
\draw (1.2,-0.8) node {\footnotesize{$12$}};
\draw (2.4,-0.8) node {\footnotesize{$8$}};
\draw (3.6,-0.8) node {\footnotesize{$4$}};
\draw (-1.2,2) node {\footnotesize{$9$}};
\draw (1.6,0) -- (2,0);
\draw (2.4,0) circle (0.4cm);
\draw(1.2,0) circle (0.4cm);
\draw (-2.4,0) circle (0.4cm);
\draw (-2.8,0) -- (-3.2,0);
\draw (-3.6,0) circle (0.4cm);
\draw (2.8,0) -- (3.2,0);
\draw (3.6,0) circle (0.4cm);
\draw (-0.4,0) -- (-0.8,0);
\draw (-1.2,0) circle (0.4cm);
\draw (-1.6,0) -- (-2,0);
\draw (-1.2,0.4) -- (-1.2,0.8);
\draw (-1.2,1.2) circle (0.4cm);
\draw (1.2,0.4) -- (1.2,0.8);
\draw (1.2,1.2) [red,fill=red!30] circle  (0.4cm);
\draw (1.2,2) node {\footnotesize{$1$}};

\end{tikzpicture}& 10\\ \hline
		 
\begin{tikzpicture}[scale=0.70]

\draw (0.4,0) -- (0.8,0);
\draw (0,0) circle (0.4cm);
\draw (-3.6,-0.8) node {\footnotesize{$4$}};
\draw (-2.4,-0.8) node {\footnotesize{$8$}};
\draw (-1.2,-0.8) node {\footnotesize{$12$}};
\draw (0,-0.8) node     {\footnotesize{$10$}};
\draw (1.2,-0.8) node {\footnotesize{$8$}};
\draw (2.4,-0.8) node {\footnotesize{$6$}};
\draw (3.6,-0.8) node {\footnotesize{$3$}};
\draw (-1.2,2) node {\footnotesize{$6$}};
\draw (1.6,0) -- (2,0);
\draw (2.4,0) circle (0.4cm);
\draw(1.2,0) circle (0.4cm);
\draw (-2.4,0) circle (0.4cm);
\draw (-2.8,0) -- (-3.2,0);
\draw (-3.6,0) circle (0.4cm);
\draw (2.8,0) -- (3.2,0);
\draw (3.6,0) circle (0.4cm);
\draw (-0.4,0) -- (-0.8,0);
\draw (-1.2,0) circle (0.4cm);
\draw (-1.6,0) -- (-2,0);
\draw (-1.2,0.4) -- (-1.2,0.8);
\draw (-1.2,1.2) circle (0.4cm);
\draw (2.4,0.4) -- (2.4,0.8);
\draw (2.4,1.2) [red,fill=red!30] circle  (0.4cm);
\draw (2.4,2) node {\footnotesize{$1$}};

\end{tikzpicture}& 4\\ \hline
		 
\begin{tikzpicture}[scale=0.70]

\draw (0.4,0) -- (0.8,0);
\draw (0,0) circle (0.4cm);
\draw (-3.6,-0.8) node {\footnotesize{$2$}};
\draw (-2.4,-0.8) node {\footnotesize{$4$}};
\draw (-1.2,-0.8) node {\footnotesize{$6$}};
\draw (0,-0.8) node     {\footnotesize{$5$}};
\draw (1.2,-0.8) node {\footnotesize{$4$}};
\draw (2.4,-0.8) node {\footnotesize{$3$}};
\draw (3.6,-0.8) node {\footnotesize{$2$}};
\draw (-1.2,2) node {\footnotesize{$3$}};
\draw (1.6,0) -- (2,0);
\draw (2.4,0) circle (0.4cm);
\draw(1.2,0) circle (0.4cm);
\draw (-2.4,0) circle (0.4cm);
\draw (-2.8,0) -- (-3.2,0);
\draw (-3.6,0) circle (0.4cm);
\draw (2.8,0) -- (3.2,0);
\draw (3.6,0) circle (0.4cm);
\draw (-0.4,0) -- (-0.8,0);
\draw (-1.2,0) circle (0.4cm);
\draw (-1.6,0) -- (-2,0);
\draw (-1.2,0.4) -- (-1.2,0.8);
\draw (-1.2,1.2) circle (0.4cm);
\draw (3.6,0.4) -- (3.6,0.8);
\draw (3.6,1.2) [orange,fill=orange!30] circle  (0.4cm);
\draw (3.6,2) node {\footnotesize{$1$}};

\end{tikzpicture}& 0 \\ \hline
	\end{tabular}
	\caption{Second set of minimally unbalanced quivers with $\gf = E_8$.}
	\label{tab:E8series2}
\end{table}


\section{Non-Simply Laced Minimally Unbalanced Quivers} \label{5}

In this section, we use the methods of section \ref{sec:clas} to classify all minimally unbalanced quivers with a global symmetry that corresponds to a non-simply laced Dynkin diagram and the unbalanced node connected to the rest of the quiver via a simply laced edge. By the sequel, this section contains minimally unbalanced quivers of $BCFG$-series. The excess of the unbalanced node is shown in a separate column in all classification tables.
 

\subsection{$\gf$ of Type $B_n$}

Analogically to $D$-type, minimally unbalanced quivers with $SO(2n+1)$ global symmetry divide into cases based on two parameters, $a$ and $n$, where $a$ is the position of the extra unbalanced node and $n$ is the total number of balanced nodes which are in the shape of a $B$-type Dynkin diagram. Similarly to $SO(2n)$, there are four different cases, collected in table \ref{tab:Bseries}. In the first case, the unbalanced node is attached to the $2m$-th node from the left, the rank of the unbalanced node is $1$ and the total number of balanced nodes is either even or odd. In the second case, the rank $2$ unbalanced node attaches to the $(2m+1)$-th node from the left and the total number of balanced nodes is either even or odd. The first two cases are contained in the first and the second row of table \ref{tab:Bseries}, respectively. When the unbalanced node attaches to the last \emph{spinor} node of the Dynkin diagram we distinguish two cases. Either the unbalanced node has rank $1$ and the total number of balanced nodes is even, or the unbalanced node is of rank $2$ and the total number of balanced nodes is odd. These two cases are shown is the third and fourth row in table \ref{tab:Bseries}, respectively. The excess is given in the last column of table \ref{tab:Bseries} in terms of $a$.\\

Lets discuss the special cases: 
\begin{itemize}
\item In the first row:
\begin{itemize}
\item for $a=2$ the quiver is the affine Dynkin diagram of $B_n$, the excess is zero, and we write $\mathcal{C}=\overline{min_{B_n}}$ (i.e. the Coulomb branch is the closure of the minimal nilpotent orbit of $SO(2n+1)$ or alternatively one says that is isomorphic to the reduced moduli space of one $B_n$ instanton on $\mathbb{C}^2$ \cite{CFHM14}).
\end{itemize}
\item In the third row:
\begin{itemize}
\item for $m=1$ one obtains the $C_2$ Dynkin diagram with an unbalanced node with $e=-1$. When the unbalanced node is ungauged we have $\mathcal{C}=\mathbb{H}^2$ (i.e. a freely generated Coulomb branch).
\item for $m=2$ one obtains the affine $F_4$ Dynkin diagram. Assuming the leftmost node is ungauged we have $\mathcal{C}=\overline{min_{F_4}}$ (see section 5.3).
\end{itemize}
\item In the fourth row:
\begin{itemize}
\item for $m=1$ one obtains the $B_3$ Dynkin diagram with an unbalanced node with excess $e=-1$ connected to the \emph{spinor} node. The spinor rep of $SO(7)$ is pseudo-real and therefore we expect $\mathcal{C}$ to be freely generated.
\end{itemize}
\end{itemize}

\begin{table}
	\centering
	\begin{tabular}{|c|l|c|}
	\hline
	a &  \multicolumn{1}{c|}{Quiver} & Excess \\ \hline
	$\begin{array}{c}
	a<n\\
	a=2m
	\end{array}$ &
\begin{tikzpicture}[scale=0.70]

\draw (0.4,0) -- (0.8,0);
\draw (0,0) circle (0.4cm);
\draw (0,-0.8) node {\footnotesize{$2m$}};
\draw (1.6,0) -- (2,0);
\draw (1.2,0) node {\footnotesize{\dots}};
\draw (-2.4,0) circle (0.4cm);
\draw (-2.4,-0.8) node {\footnotesize{$3$}};
\draw (-2.8,0) -- (-3.2,0);
\draw (-3.6,0) circle (0.4cm);
\draw (-3.6,-0.8) node {\footnotesize{$2$}};
\draw (-4,0) -- (-4.4,0);
\draw (-4.8,0) circle (0.4cm);
\draw (-4.8,-0.8) node {\footnotesize{$1$}};
\draw (2.8,0.08) -- (3.4,0.08);
\draw (2.8,-0.08) -- (3.4,-0.08);
\draw (3.2,0) -- (3,0.2);
\draw (3.2,0) -- (3,-0.2);
\draw (2.4,0) circle (0.4cm);
\draw (3.8,0) circle (0.4cm);
\draw (2.4,-0.8) node {\footnotesize{$2m$}};
\draw (3.8,-0.8) node {\footnotesize{$m$}};
\draw (-0.4,0) -- (-0.8,0);
\draw (-1.2,0) node {\footnotesize{\dots}};
\draw (-1.6,0) -- (-2,0);
\draw (0,0.4) -- (0,0.8);
\draw (0,1.2)[red,fill=red!30] circle (0.4cm);
\draw (0,2) node {\footnotesize{$1$}};
\draw [decorate,decoration={brace,amplitude=6pt}] (0.1,0.5) to (2.4,0.5);
\draw (1.38,1.1) node {\footnotesize{$n-a$}};
\end{tikzpicture} & $a-2$\\ \hline	
	$\begin{array}{c}
	a<n\\
	a=2m+1
	\end{array}$ &
\begin{tikzpicture}[scale=0.70]

\draw (0.4,0) -- (0.8,0);
\draw (0,0) circle (0.4cm);
\draw (0,-0.8) node {\footnotesize{$2(2m+1)$}};
\draw (1.6,0) -- (2,0);
\draw (1.2,0) node {\footnotesize{\dots}};
\draw (-2.4,0) circle (0.4cm);
\draw (-2.4,-0.8) node {\footnotesize{$6$}};
\draw (-2.8,0) -- (-3.2,0);
\draw (-3.6,0) circle (0.4cm);
\draw (-3.6,-0.8) node {\footnotesize{$4$}};
\draw (-4,0) -- (-4.4,0);
\draw (-4.8,0) circle (0.4cm);
\draw (-4.8,-0.8) node {\footnotesize{$2$}};
\draw (2.8,0.08) -- (4,0.08);
\draw (2.8,-0.08) -- (4,-0.08);
\draw (3.6,0) -- (3.4,0.2);
\draw (3.6,0) -- (3.4,-0.2);
\draw (2.4,0) circle (0.4cm);
\draw (4.4,0) circle (0.4cm);
\draw (2.4,-0.8) node {\footnotesize{$2(2m+1)$}};
\draw (4.4,-0.8) node {\footnotesize{$2m+1$}};
\draw (-0.4,0) -- (-0.8,0);
\draw (-1.2,0) node {\footnotesize{\dots}};
\draw (-1.6,0) -- (-2,0);
\draw (0,0.4) -- (0,0.8);
\draw (0,1.2)[red,fill=red!30] circle (0.4cm);
\draw (0,2) node {\footnotesize{$2$}};
\draw [decorate,decoration={brace,amplitude=6pt}] (0.1,0.5) to (2.4,0.5);
\draw (1.38,1.1) node {\footnotesize{$n-a$}};
\end{tikzpicture} & $2a-4$\\ \hline	
	$\begin{array}{c}
	a=n\\
	a=2m
	\end{array}$ &
\begin{tikzpicture}[scale=0.70]

\draw (0,0) circle (0.4cm);
\draw (0,-0.8) node {\footnotesize{$2m-1$}};
\draw (1.4,-0.8) node {\footnotesize{$m$}};
\draw (1.4,0) circle (0.4cm);
\draw (2.6,-0.8) node {\footnotesize{$1$}};
\draw (-2.4,0)  circle (0.4cm);
\draw (-2.4,-0.8) node {\footnotesize{$3$}};
\draw (-2.8,0) -- (-3.2,0);
\draw (-3.6,0) circle (0.4cm);
\draw (-3.6,-0.8) node {\footnotesize{$2$}};
\draw (-4,0) -- (-4.4,0);
\draw (-4.8,0) circle (0.4cm);
\draw (-4.8,-0.8) node {\footnotesize{$1$}};
\draw (2.6,0) [red,fill=red!30] circle (0.4cm);
\draw (0.4,0.08) -- (1,0.08);
\draw (0.4,-0.08) -- (1,-0.08);
\draw (0.8,0) -- (0.65,0.20);
\draw (0.8,0) -- (0.65,-0.20);
\draw (-0.4,0) -- (-0.8,0);
\draw (1.8,0) -- (2.2,0);
\draw (-1.2,0) node {\footnotesize{\dots}};
\draw (-1.6,0) -- (-2,0);
\draw [decorate,decoration={brace,amplitude=6pt}] (-4.7,0.5) to (-0.1,0.5);
\draw (-2.4,1.15) node {\footnotesize{$n-1=$ odd number}};

\end{tikzpicture} & $\frac{a}{2}-2$\\ \hline	
	$\begin{array}{c}
	a=n\\
	a=2m+1
	\end{array}$ &
\begin{tikzpicture}[scale=0.70]

\draw (0,0) circle (0.4cm);
\draw (-0.1,-0.8) node {\footnotesize{$2(2m)$}};
\draw (1.4,-0.8) node {\footnotesize{$2m+1$}};
\draw (1.4,0) circle (0.4cm);
\draw (2.6,-0.8) node {\footnotesize{$2$}};
\draw (-2.4,0)  circle (0.4cm);
\draw (-2.4,-0.8) node {\footnotesize{$6$}};
\draw (-2.8,0) -- (-3.2,0);
\draw (-3.6,0) circle (0.4cm);
\draw (-3.6,-0.8) node {\footnotesize{$4$}};
\draw (-4,0) -- (-4.4,0);
\draw (-4.8,0) circle (0.4cm);
\draw (-4.8,-0.8) node {\footnotesize{$2$}};
\draw (2.6,0) [red,fill=red!30] circle (0.4cm);
\draw (0.4,0.08) -- (1,0.08);
\draw (0.4,-0.08) -- (1,-0.08);
\draw (0.8,0) -- (0.65,0.2);
\draw (0.8,0) -- (0.65,-0.2);
\draw (-0.4,0) -- (-0.8,0);
\draw (1.8,0) -- (2.2,0);
\draw (-1.2,0) node {\footnotesize{\dots}};
\draw (-1.6,0) -- (-2,0);
\draw [decorate,decoration={brace,amplitude=6pt}] (-4.7,0.5) to (-0.1,0.5);
\draw (-2.4,1.15) node {\footnotesize{$n-1=$ even number}};
\end{tikzpicture} &$a-4$\\ \hline	
	\end{tabular}
	\caption{Classification of minimally unbalanced quivers with $\gf = SO(2n+1)$.}
	\label{tab:Bseries}
\end{table}


\subsection{$\gf$ of Type $C_n$}

Next, we classify all minimally unbalanced quivers with $Sp(n)$ global symmetry. We employ the same two parameters $n$, number of balanced nodes in form of the C-type Dynkin diagram, and $a$, position of the unbalanced node from the left. There are two different cases based on the position of the unbalanced node, see table \ref{tab:Cseries}. In the first row of table  \ref{tab:Cseries} we show the case where the unbalanced node is not attached to the rightmost balanced node ($a<n$). If the unbalanced node is connected to the rightmost node ($a=n$), it is of rank $2$ and the resulting quiver is depicted in the second row of table \ref{tab:Cseries}. In both cases the total number of balanced nodes is either even or odd.\\

Let us now look at the special cases: 
\begin{itemize}
\item In the first row: for $a=2$ the excess is zero and the quiver is the $A_{2k-1}^{(2)}$ twisted affine Dynkin diagram, where $2k-1=n+1$.
\item In the second row:
\begin{itemize}
\item for $a=3$ one obtains the $C_3$ Dynkin diagram with an unbalanced node connected to the rightmost node and with excess $e=-1$. Given the value of excess and the fact that the unbalanced node connects to the pseudo-real representation denoted by $[0,0,1]_{C_3}$ one expects that the Coulomb branch is freely generated, namely $\mathcal{C}=\mathbb{H}^7$. Indeed, this claim is confirmed by a computation.
\item for $m=4$ one obtains the affine $F_4$ Dynkin diagram with zero excess. Under the assumption that the rank $1$ node is ungauged one has $\mathcal{C}=\overline{min_{F_4}}$.
\end{itemize}
\end{itemize}

\begin{table}
	\centering
	\begin{tabular}{|c|l|c|}
	\hline
	a &  \multicolumn{1}{c|}{Quiver} & Excess \\ \hline
	$a<n$ &
\begin{tikzpicture}[scale=0.70]

\draw (0.4,0) -- (0.8,0);
\draw (0,0) circle (0.4cm);
\draw (0,-0.8) node {\footnotesize{$a$}};
\draw (1.6,0) -- (2,0);
\draw (1.2,0) node {\footnotesize{\dots}};
\draw (-2.4,0) circle (0.4cm);
\draw (-2.4,-0.8) node {\footnotesize{$3$}};
\draw (-2.8,0) -- (-3.2,0);
\draw (-3.6,0) circle (0.4cm);
\draw (-3.6,-0.8) node {\footnotesize{$2$}};
\draw (-4,0) -- (-4.4,0);
\draw (-4.8,0) circle (0.4cm);
\draw (-4.8,-0.8) node {\footnotesize{$1$}};
\draw (2.8,0.08) -- (3.4,0.08);
\draw (2.8,-0.08) -- (3.4,-0.08);
\draw (3,0) -- (3.2,0.2);
\draw (3,0) -- (3.2,-0.2);
\draw (2.4,0) circle (0.4cm);
\draw (3.8,0) circle (0.4cm);
\draw (2.4,-0.8) node {\footnotesize{$a$}};
\draw (3.8,-0.8) node {\footnotesize{$a$}};
\draw (-0.4,0) -- (-0.8,0);
\draw (-1.2,0) node {\footnotesize{\dots}};
\draw (-1.6,0) -- (-2,0);
\draw (0,0.4) -- (0,0.8);
\draw (0,1.2) [red,fill=red!30] circle (0.4cm);
\draw (0,2) node {\footnotesize{$1$}};
\draw [decorate,decoration={brace,amplitude=6pt}] (0.1,0.5) to (2.4,0.5);
\draw (1.36,1.1) node {\footnotesize{$n-a$}};
\end{tikzpicture} & $a-2$\\ \hline	

	$a=n$ &\begin{tikzpicture}[scale=0.70]

\draw (2,0) -- (2.8,0);
\draw (0,0) circle (0.4cm);
\draw (0,-0.8) node {\footnotesize{$a-1$}};
\draw (1.6,0) circle (0.4cm);
\draw (1.6,-0.8) node {\footnotesize{$a$}};
\draw (-2.4,0) circle (0.4cm);
\draw (-2.4,-0.8) node {\footnotesize{$3$}};
\draw (-2.8,0) -- (-3.2,0);
\draw (-3.6,0) circle (0.4cm);
\draw (-3.6,-0.8) node {\footnotesize{$2$}};
\draw (-4,0) -- (-4.4,0);
\draw (-4.8,0) circle (0.4cm);
\draw (-4.8,-0.8) node {\footnotesize{$1$}};
\draw (2.8,0) [red,fill=red!30] circle (0.4cm);
\draw (2.8,-0.8) node {\footnotesize{$2$}};
\draw (-0.4,0) -- (-0.8,0);
\draw (-1.2,0) node {\footnotesize{\dots}};
\draw (-1.6,0) -- (-2,0);
\draw (0.4,0.08) -- (1.2,0.08);
\draw (0.4,-0.08) -- (1.2,-0.08);
\draw (0.7,0) -- (0.9,0.2);
\draw (0.7,0) -- (0.9,-0.2);
\draw (0,1) node {\footnotesize{$$}};

\end{tikzpicture} & $a-4$\\ \hline	

	\end{tabular}
	\caption{Classification of minimally unbalanced quivers with $\gf = Sp(n)$.}
	\label{tab:Cseries}
\end{table}

\subsection{$\gf$ of Type $F_4$}

Let us now classify all minimally unbalanced quivers with $F_4$ global symmetry on their Coulomb branch. Starting form the Dynkin diagram of $F_4$ and employing the methods of section \ref{sec:clas} one obtains the four cases depicted in table \ref{tab:Fseries}. Note that the first and the last row of table \ref{tab:Fseries} contain quivers that are balanced. The Coulomb branch of the quiver in the first row is the reduced moduli space of one $F_4$ instanton on $\mathbb{C}^2$ \cite{CFHM14}. The Coulomb branch of the quiver in the last row is the closure of the next to minimal nilpotent orbit of $\mathfrak{f}_4$ algebra \cite{Hanany:2017ooe}.

\begin{table}
	\centering
	\begin{tabular}{|c|c|}
	\hline
	Quiver & Excess \\ \hline
	
\begin{tikzpicture}[scale=0.70]

\draw (0,0) circle (0.4cm);
\draw (0,-0.8) node {\footnotesize{$2$}};
\draw (1.2,-0.8) node {\footnotesize{$1$}};
\draw (1.2,0) circle (0.4cm);
\draw (-2.4,0)  circle (0.4cm);
\draw (-2.4,-0.8) node {\footnotesize{$2$}};
\draw (-1.2,-0.8) node {\footnotesize{$3$}};
\draw (-0.8,0.08) -- (-0.4,0.08);
\draw (-0.8,-0.08) -- (-0.4,-0.08);
\draw (-0.7,0.2) -- (-0.5,0);
\draw (-0.7,-0.2) -- (-0.5,0);
\draw (-2.4,1.2) [orange,fill=orange!30] circle (0.4cm);
\draw (-2.4,0.4) -- (-2.4,0.8);
\draw (-2.4,2) node {\footnotesize{$1$}};
\draw (0.4,0) -- (0.8,0);
\draw (-1.2,0) circle (0.4cm);
\draw (-1.6,0) -- (-2,0);

\end{tikzpicture} &0 \\ \hline
	 
\begin{tikzpicture}[scale=0.70]

\draw (0,0) circle (0.4cm);
\draw (0,-0.8) node {\footnotesize{$4$}};
\draw (1.2,-0.8) node {\footnotesize{$2$}};
\draw (1.2,0) circle (0.4cm);
\draw (-2.4,0)  circle (0.4cm);
\draw (-2.4,-0.8) node {\footnotesize{$3$}};
\draw (-1.2,-0.8) node {\footnotesize{$6$}};
\draw (-1.2,2) node {\footnotesize{$1$}};
\draw (-0.8,0.08) -- (-0.4,0.08);
\draw (-0.8,-0.08) -- (-0.4,-0.08);
\draw (-0.7,0.2) -- (-0.5,0);
\draw (-0.7,-0.2) -- (-0.5,0);
\draw (-1.2,1.2) [red,fill=red!30] circle (0.4cm);
\draw (0.4,0) -- (0.8,0);
\draw (-1.2,0) circle (0.4cm);
\draw (-1.2,0.4) -- (-1.2,0.8);
\draw (-1.6,0) -- (-2,0);

\end{tikzpicture} & 4\\ \hline
	 
\begin{tikzpicture}[scale=0.70]

\draw (0,0) circle (0.4cm);
\draw (0,-0.8) node {\footnotesize{$6$}};
\draw (1.2,-0.8) node {\footnotesize{$3$}};
\draw (1.2,0) circle (0.4cm);
\draw (-2.4,0)  circle (0.4cm);
\draw (-2.4,-0.8) node {\footnotesize{$4$}};
\draw (-1.2,-0.8) node {\footnotesize{$8$}};
\draw (0,2) node {\footnotesize{$1$}};
\draw (-0.8,0.08) -- (-0.4,0.08);
\draw (-0.8,-0.08) -- (-0.4,-0.08);
\draw (-0.7,0.2) -- (-0.5,0);
\draw (-0.7,-0.2) -- (-0.5,0);
\draw (0,1.2) [red,fill=red!30] circle (0.4cm);
\draw (0.4,0) -- (0.8,0);
\draw (-1.2,0) circle (0.4cm);
\draw (0,0.4) -- (0,0.8);
\draw (-1.6,0) -- (-2,0);

\end{tikzpicture} & 4\\ \hline
	 
\begin{tikzpicture}[scale=0.70]

\draw (0,0) circle (0.4cm);
\draw (0,-0.8) node {\footnotesize{$3$}};
\draw (1.2,-0.8) node {\footnotesize{$2$}};
\draw (1.2,0) circle (0.4cm);
\draw (-2.4,0)  circle (0.4cm);
\draw (-2.4,-0.8) node {\footnotesize{$2$}};
\draw (-1.2,-0.8) node {\footnotesize{$4$}};
\draw (-0.8,0.08) -- (-0.4,0.08);
\draw (-0.8,-0.08) -- (-0.4,-0.08);
\draw (-0.7,0.2) -- (-0.5,0);
\draw (-0.7,-0.2) -- (-0.5,0);
\draw (1.2,1.2) [orange,fill=orange!30] circle (0.4cm);
\draw (1.2,0.4) -- (1.2,0.8);
\draw (1.2,2) node {\footnotesize{$1$}};
\draw (-1.2,0) circle (0.4cm);
\draw (-1.6,0) -- (-2,0);
\draw (0.4,0) -- (0.8,0);

\end{tikzpicture} & 0\\ \hline
	\end{tabular}
	\caption{Classification of minimally unbalanced quivers with $\gf = F_4$.}
	\label{tab:Fseries}
\end{table}


\subsection{$\gf$ of Type $G_2$}

Finally, applying the methods of section \ref{sec:clas} to a $G_2$ Dynkin diagram yields two cases of minimally unbalanced quiver theories with $G_2$ global symmetry on their Coulomb branch. The two quivers are depicted in table \ref{tab:Gseries}. Note that both cases are balanced and the first row contains the reduced moduli space of one $G_2$ instanton on $\mathbb{C}^2$ \cite{CFHM14}. 
\begin{table}
	\centering
	\begin{tabular}{|c|c|}
		\hline
		Quiver & Excess \\ \hline  
\begin{tikzpicture}[scale=0.70]
\draw (0,0) circle (0.4cm);
\draw (0,-0.8) node {\footnotesize{$1$}};
\draw (-1.2,-0.8) node {\footnotesize{$2$}};
\draw (-1.2,1.2) [orange,fill=orange!30] circle (0.4cm);
\draw (-1.2,2) node {\footnotesize{$1$}};
\draw (-1.2,0.4) -- (-1.2,0.8);
\draw (-0.8,0.08) -- (-0.4,0.08);
\draw (-0.8,-0.08) -- (-0.4,-0.08);
\draw (-0.7,0.2) -- (-0.5,0);
\draw (-0.7,-0.2) -- (-0.5,0);
\draw (-0.4,0) -- (-0.8,0);
\draw (-1.2,0) circle (0.4cm);
\end{tikzpicture} & 0\\ \hline
		
\begin{tikzpicture}[scale=0.70]
\draw (0,0) circle (0.4cm);
\draw (0,-0.8) node {\footnotesize{$2$}};
\draw (-1.2,-0.8) node {\footnotesize{$3$}};
\draw (0,0.4) -- (0,0.8);
\draw (0,1.2) [orange,fill=orange!30] circle (0.4cm);
\draw (0,2) node {\footnotesize{$1$}};
\draw (-0.8,0.08) -- (-0.4,0.08);
\draw (-0.8,-0.08) -- (-0.4,-0.08);
\draw (-0.7,0.2) -- (-0.5,0);
\draw (-0.7,-0.2) -- (-0.5,0);
\draw (-0.4,0) -- (-0.8,0);
\draw (-1.2,0) circle (0.4cm);
\end{tikzpicture} & 0\\ \hline
	\end{tabular}
	\caption{Classification of minimally unbalanced quivers with $\gf = G_2$.}
		\label{tab:Gseries}
\end{table}



\section{Simply Laced Minimally Unbalanced Quivers with Unbalanced Node connected by a Non-Simply Laced Edge} \label{6}

In this section, we classify all minimally unbalanced quivers with the unbalanced node connected to the rest of the quiver with a double or triple laced edge, such that the balanced subset of nodes forms a finite simply laced Dynkin diagram. We term this part of the classification \emph{exotic} since it is the first time these quivers appear and the systematic study of the field theoretic aspects of these quiver gauge theories is yet to be done. The first level of distinction in the following classification is based on the Dynkin type of the balanced subset of nodes. Further levels of distinction are the following:
\begin{itemize}
\item Type of the non-simply laced edge: 
\begin{itemize}
\item Double Edge
\item Triple Edge
\end{itemize}
\item Position of the unbalanced node
\item Direction of the non-simply laced edge with respect to the unbalanced node
\begin{itemize}
\item Outwards from the unbalanced node
\item Inwards to the unbalanced node
\end{itemize}
\end{itemize}
Let us begin the exotic classification with $A$-type minimally unbalanced quivers with the unbalanced node connected by a double or triple laced edge.

\subsection{Exotic Minimally Unbalanced Quivers with $G$ of Type $A_n$}

Tables \ref{tab:AseriesExotic1} and \ref{tab:AseriesExotic2} collect all exotic minimally unbalanced quivers of $A_n$ type. Table \ref{tab:AseriesExotic1} contains quivers where the unbalanced node attaches via a double laced edge. Table \ref{tab:AseriesExotic2} depicts quivers with the unbalanced node attached by a triple laced edge.

\begin{table}
	\centering
	\begin{tabular}{|c|c|c|}
	\hline
$(a+b)/s$ divisible by 2 & Quiver & Excess \\ \hline

N.A. &\begin{tikzpicture}[scale=0.70]
\draw (0.4,0) -- (0.8,0);
\draw (0,0) circle (0.4cm);
\draw (0,-0.8) node {\footnotesize{$\frac{ab}{s}$}};
\draw (1.6,0) -- (2,0);
\draw (2.4,-0.8) node {\footnotesize{$\frac{3b}{s}$}};
\draw(2.4,0) circle (0.4cm);
\draw (1.2,0) node {\footnotesize{\dots}};
\draw (-2.4,0) circle (0.4cm);
\draw (-2.4,-0.8) node {\footnotesize{$\frac{3a}{s}$}};
\draw (-2.8,0) -- (-3.2,0);
\draw (-3.6,0) circle (0.4cm);
\draw (-3.6,-0.8) node {\footnotesize{$\frac{2a}{s}$}};
\draw (-4,0) -- (-4.4,0);
\draw (-4.8,0) circle (0.4cm);
\draw (-4.8,-0.8) node {\footnotesize{$\frac{a}{s}$}};
\draw (2.8,0) -- (3.2,0);
\draw (3.6,0) circle (0.4cm);
\draw (3.6,-0.8) node {\footnotesize{$\frac{2b}{s}$}};
\draw (4,0) -- (4.4,0);
\draw (4.8,0) circle (0.4cm);
\draw (4.8,-0.8) node {\footnotesize{$\frac{b}{s}$}};
\draw (-0.4,0) -- (-0.8,0);
\draw (-1.2,0) node {\footnotesize{\dots}};
\draw (-1.6,0) -- (-2,0);
\draw (0,1.2)[red,fill=red!30] circle (0.4cm);
\draw (0,2) node {\footnotesize{$\frac{a+b}{s}$}};
\draw (-0.08,0.4) -- (-0.08,0.8);
\draw (0.08,0.4) -- (0.08,0.8);
\draw (0,0.5) -- (-0.2,0.7);
\draw (0,0.5) -- (0.2,0.7);
\end{tikzpicture} & $\frac{2ab-2(a+b)}{s}$\\ \hline	
	
	Yes &\begin{tikzpicture}[scale=0.70]
\draw (0.4,0) -- (0.8,0);
\draw (0,0) circle (0.4cm);
\draw (0,-0.8) node {\footnotesize{$\frac{ab}{s}$}};
\draw (1.6,0) -- (2,0);
\draw (1.2,0) node {\footnotesize{\dots}};
\draw(2.4,0) circle (0.4cm);
\draw (2.4,-0.8) node {\footnotesize{$\frac{3b}{s}$}};
\draw (-2.4,0) circle (0.4cm);
\draw (-2.4,-0.8) node {\footnotesize{$\frac{3a}{s}$}};
\draw (-2.8,0) -- (-3.2,0);
\draw (-3.6,0) circle (0.4cm);
\draw (-3.6,-0.8) node {\footnotesize{$\frac{2a}{s}$}};
\draw (-4,0) -- (-4.4,0);
\draw (-4.8,0) circle (0.4cm);
\draw (-4.8,-0.8) node {\footnotesize{$\frac{a}{s}$}};
\draw (2.8,0) -- (3.2,0);
\draw (3.6,0) circle (0.4cm);
\draw (3.6,-0.8) node {\footnotesize{$\frac{2b}{s}$}};
\draw (4,0) -- (4.4,0);
\draw (4.8,0) circle (0.4cm);
\draw (4.8,-0.8) node {\footnotesize{$\frac{b}{s}$}};
\draw (-0.4,0) -- (-0.8,0);
\draw (-1.2,0) node {\footnotesize{\dots}};
\draw (-1.6,0) -- (-2,0);
\draw (0,1.2)[red,fill=red!30] circle (0.4cm);
\draw (0,2) node {\footnotesize{$\frac{a+b}{2s}$}};
\draw (-0.08,0.4) -- (-0.08,0.8);
\draw (0.08,0.4) -- (0.08,0.8);
\draw (0,0.7) -- (-0.2,0.5);
\draw (0,0.7) -- (0.2,0.5);
\end{tikzpicture}

& $\frac{ab-(a+b)}{s}$\\ \hline	
	
	No &\begin{tikzpicture}[scale=0.70]
\draw (0.4,0) -- (0.8,0);
\draw (0,0) circle (0.4cm);
\draw (0,-0.8) node {\footnotesize{$\frac{2ab}{s}$}};
\draw (1.6,0) -- (2,0);
\draw (1.2,0) node {\footnotesize{\dots}};
\draw(2.4,0) circle (0.4cm);
\draw (2.4,-0.8) node {\footnotesize{$\frac{6b}{s}$}};
\draw (-2.4,0) circle (0.4cm);
\draw (-2.4,-0.8) node {\footnotesize{$\frac{6a}{s}$}};
\draw (-2.8,0) -- (-3.2,0);
\draw (-3.6,0) circle (0.4cm);
\draw (-3.6,-0.8) node {\footnotesize{$\frac{4a}{s}$}};
\draw (-4,0) -- (-4.4,0);
\draw (-4.8,0) circle (0.4cm);
\draw (-4.8,-0.8) node {\footnotesize{$\frac{2a}{s}$}};
\draw (2.8,0) -- (3.2,0);
\draw (3.6,0) circle (0.4cm);
\draw (3.6,-0.8) node {\footnotesize{$\frac{4b}{s}$}};
\draw (4,0) -- (4.4,0);
\draw (4.8,0) circle (0.4cm);
\draw (4.8,-0.8) node {\footnotesize{$\frac{2b}{s}$}};
\draw (-0.4,0) -- (-0.8,0);
\draw (-1.2,0) node {\footnotesize{\dots}};
\draw (-1.6,0) -- (-2,0);
\draw (0,1.2)[red,fill=red!30] circle (0.4cm);
\draw (0,2) node {\footnotesize{$\frac{a+b}{s}$}};
\draw (-0.08,0.4) -- (-0.08,0.8);
\draw (0.08,0.4) -- (0.08,0.8);
\draw (0,0.7) -- (-0.2,0.5);
\draw (0,0.7) -- (0.2,0.5);
\end{tikzpicture} & $\frac{2ab-2(a+b)}{s}$ \\ \hline	

	\end{tabular}
	\caption{Exotic minimally unbalanced quivers with $G=SU(n)$, $n=a+b$ and a double  laced edge. $s$ is the greatest common divisor of $a$ and $b$.}
	\label{tab:AseriesExotic1}
\end{table}

\begin{table}
	\centering
	\begin{tabular}{|c|c|c|}
	\hline
$(a+b)/s$ divisible by 3 & Quiver & Excess \\ \hline

N.A. & \begin{tikzpicture}[scale=0.70]
\draw (0.4,0) -- (0.8,0);
\draw (0,0) circle (0.4cm);
\draw (0,-0.8) node {\footnotesize{$\frac{ab}{s}$}};
\draw (1.6,0) -- (2,0);
\draw (2.4,-0.8) node {\footnotesize{$\frac{3b}{s}$}};
\draw(2.4,0) circle (0.4cm);
\draw (1.2,0) node {\footnotesize{\dots}};
\draw (-2.4,0) circle (0.4cm);
\draw (-2.4,-0.8) node {\footnotesize{$\frac{3a}{s}$}};
\draw (-2.8,0) -- (-3.2,0);
\draw (-3.6,0) circle (0.4cm);
\draw (-3.6,-0.8) node {\footnotesize{$\frac{2a}{s}$}};
\draw (-4,0) -- (-4.4,0);
\draw (-4.8,0) circle (0.4cm);
\draw (-4.8,-0.8) node {\footnotesize{$\frac{a}{s}$}};
\draw (2.8,0) -- (3.2,0);
\draw (3.6,0) circle (0.4cm);
\draw (3.6,-0.8) node {\footnotesize{$\frac{2b}{s}$}};
\draw (4,0) -- (4.4,0);
\draw (4.8,0) circle (0.4cm);
\draw (4.8,-0.8) node {\footnotesize{$\frac{b}{s}$}};
\draw (-0.4,0) -- (-0.8,0);
\draw (-1.2,0) node {\footnotesize{\dots}};
\draw (-1.6,0) -- (-2,0);
\draw (0,1.2)[red,fill=red!30] circle (0.4cm);
\draw (0,2) node {\footnotesize{$\frac{a+b}{s}$}};
\draw (0,0.4) -- (0,0.8);
\draw (-0.08,0.4) -- (-0.08,0.8);
\draw (0.08,0.4) -- (0.08,0.8);
\draw (0,0.5) -- (-0.2,0.7);
\draw (0,0.5) -- (0.2,0.7);
\end{tikzpicture} & $\frac{3ab-2(a+b)}{s}$\\ \hline	

Yes & \begin{tikzpicture}[scale=0.70]
\draw (0.4,0) -- (0.8,0);
\draw (0,0) circle (0.4cm);
\draw (0,-0.8) node {\footnotesize{$\frac{ab}{s}$}};
\draw (1.6,0) -- (2,0);
\draw (1.2,0) node {\footnotesize{$\dots$}};
\draw (-2.4,-0.8) node {\footnotesize{$\frac{3a}{s}$}};
\draw(2.4,0) circle (0.4cm);
\draw (2.4,-0.8) node {\footnotesize{$\frac{3b}{s}$}};
\draw (-2.4,0) circle (0.4cm);
\draw (-2.8,0) -- (-3.2,0);
\draw (-3.6,0) circle (0.4cm);
\draw (-3.6,-0.8) node {\footnotesize{$\frac{2a}{s}$}};
\draw (-4,0) -- (-4.4,0);
\draw (-4.8,0) circle (0.4cm);
\draw (-4.8,-0.8) node {\footnotesize{$\frac{a}{s}$}};
\draw (2.8,0) -- (3.2,0);
\draw (3.6,0) circle (0.4cm);
\draw (3.6,-0.8) node {\footnotesize{$\frac{2b}{s}$}};
\draw (4,0) -- (4.4,0);
\draw (4.8,0) circle (0.4cm);
\draw (4.8,-0.8) node {\footnotesize{$\frac{b}{s}$}};
\draw (-0.4,0) -- (-0.8,0);
\draw (-1.2,0) node {\footnotesize{$\dots$}};
\draw (-1.6,0) -- (-2,0);
\draw (0,1.2)[red,fill=red!30] circle (0.4cm);
\draw (0,2) node {\footnotesize{$\frac{a+b}{3s}$}};
\draw (0,0.4) -- (0,0.8);
\draw (-0.08,0.4) -- (-0.08,0.8);
\draw (0.08,0.4) -- (0.08,0.8);
\draw (0,0.7) -- (-0.2,0.5);
\draw (0,0.7) -- (0.2,0.5);
\end{tikzpicture} & $\frac{3ab-2(a+b)}{3s}$ \\ \hline	

 No & \begin{tikzpicture}[scale=0.70]
\draw (0.4,0) -- (0.8,0);
\draw (0,0) circle (0.4cm);
\draw (0,-0.8) node {\footnotesize{$\frac{3ab}{s}$}};
\draw (1.6,0) -- (2,0);
\draw (1.2,0) node {\footnotesize{$\dots$}};
\draw (-2.4,-0.8) node {\footnotesize{$\frac{9a}{s}$}};
\draw(2.4,0) circle (0.4cm);
\draw (2.4,-0.8) node {\footnotesize{$\frac{9b}{s}$}};
\draw (-2.4,0) circle (0.4cm);
\draw (-2.8,0) -- (-3.2,0);
\draw (-3.6,0) circle (0.4cm);
\draw (-3.6,-0.8) node {\footnotesize{$\frac{6a}{s}$}};
\draw (-4,0) -- (-4.4,0);
\draw (-4.8,0) circle (0.4cm);
\draw (-4.8,-0.8) node {\footnotesize{$\frac{3a}{s}$}};
\draw (2.8,0) -- (3.2,0);
\draw (3.6,0) circle (0.4cm);
\draw (3.6,-0.8) node {\footnotesize{$\frac{6b}{s}$}};
\draw (4,0) -- (4.4,0);
\draw (4.8,0) circle (0.4cm);
\draw (4.8,-0.8) node {\footnotesize{$\frac{3b}{s}$}};
\draw (-0.4,0) -- (-0.8,0);
\draw (-1.2,0) node {\footnotesize{$\dots$}};
\draw (-1.6,0) -- (-2,0);
\draw (0,1.2)[red,fill=red!30] circle (0.4cm);
\draw (0,2) node {\footnotesize{$\frac{a+b}{s}$}};
\draw (0,0.4) -- (0,0.8);
\draw (-0.08,0.4) -- (-0.08,0.8);
\draw (0.08,0.4) -- (0.08,0.8);
\draw (0,0.7) -- (-0.2,0.5);
\draw (0,0.7) -- (0.2,0.5);
\end{tikzpicture} & $\frac{3ab-2(a+b)}{s}$  \\ \hline

	\end{tabular}
	\caption{Exotic minimally unbalanced quivers with $G=SU(n)$, $n=a+b$ and a triple laced edge. $s$ is the greatest common divisor of $a$ and $b$.}
	\label{tab:AseriesExotic2}
\end{table}

\subsection{Exotic Minimally Unbalanced Quivers with $G$ of Type $D_n$}

Let us now classify the exotic minimally unbalanced quivers with $\gf=SO(2n)$. First level of distinction is based on whether the unbalanced node connects via a double or triple laced edge. Second level of distinction is based on the orientation of the non-simply laced edge with respect to the unbalanced node (i.e. inwards or outwards). Finally, we distinguish cases based on whether the unbalanced node attaches to one of the nodes on the main chain or to one of the spinor nodes.\footnote{Attaching the unbalanced node to the co-spinor node instead of the the spinor node yields equivalent cases to those already included.} The results of the classification are divided into two tables. Tables \ref{tab:DseriesExotic1-1} and \ref{tab:DseriesExotic1-2} collect the results for exotic minimally unbalanced quivers of $D$-type with the unbalanced node connected to the rest of the quiver with a double laced edge. Exotic $D$-type minimally unbalanced quivers with the unbalanced node connected by a triple laced edge are reported in tables \ref{tab:DseriesExotic2-1} and \ref{tab:DseriesExotic2-2}.\\

\begin{table}
	\centering
	\begin{tabular}{|c|l|c|}
	\hline
	$a$ &  \multicolumn{1}{c|}{Quiver} & Excess \\ \hline
	$\begin{array}{c}
	a<n-1
	\end{array}$ &\begin{tikzpicture}[scale=0.70]
\draw (0,0) circle (0.4cm);
\draw (0,-0.8) node {\footnotesize{$2a$}};
\draw (1.6,0) -- (2,0);
\draw (1.2,0) node {\footnotesize{\dots}};
\draw (-2.4,0) circle (0.4cm);
\draw (-2.4,-0.8) node {\footnotesize{$6$}};
\draw (-2.8,0) -- (-3.2,0);
\draw (-3.6,0) circle (0.4cm);
\draw (-3.6,-0.8) node {\footnotesize{$4$}};
\draw (-4,0) -- (-4.4,0);
\draw (-4.8,0) circle (0.4cm);
\draw (-4.8,-0.8) node {\footnotesize{$2$}};
\draw (2.4,0) circle (0.4cm);
\draw (2.4,-0.8) node {\footnotesize{$2a$}};
\draw (-0.4,0) -- (-0.8,0);
\draw (0.4,0) -- (0.8,0);
\draw (-1.2,0) node {\footnotesize{$\dots$}};
\draw (-1.6,0) -- (-2,0);
\draw (0,1.2) [red,fill=red!30] circle (0.4cm);
\draw (0,2) node {\footnotesize{$1$}};
\draw (-0.08,0.4) -- (-0.08,0.8);
\draw (0.08,0.4) -- (0.08,0.8);
\draw (0,0.7) -- (-0.2,0.5);
\draw (0,0.7) -- (0.2,0.5);
\draw [decorate,decoration={brace,amplitude=6pt}] (0.3,0.5) to (2.4,0.5);
\draw (1.6,1.1) node {\footnotesize{$n-a-1$}};
\draw (3.4,-1) circle (0.4cm);
\draw (4.26,-1) node {\footnotesize{$a$}};
\draw (2.7,-0.24) -- (3.1,-0.75);
\draw (3.4,1) circle (0.4cm);
\draw (4.26,1) node {\footnotesize{$a$}};
\draw (2.7,0.24) -- (3.1,0.75);

\end{tikzpicture} & $2a-2$\\ \hline	

	$\begin{array}{c}
	a<n-1\\
	a=2m
	\end{array}$ &\begin{tikzpicture}[scale=0.70]
\draw (0,0) circle (0.4cm);
\draw (0,-0.8) node {\footnotesize{$2m$}};
\draw (1.6,0) -- (2,0);
\draw (1.2,0) node {\footnotesize{\dots}};
\draw (-2.4,0) circle (0.4cm);
\draw (-2.4,-0.8) node {\footnotesize{$3$}};
\draw (-2.8,0) -- (-3.2,0);
\draw (-3.6,0) circle (0.4cm);
\draw (-3.6,-0.8) node {\footnotesize{$2$}};
\draw (-4,0) -- (-4.4,0);
\draw (-4.8,0) circle (0.4cm);
\draw (-4.8,-0.8) node {\footnotesize{$1$}};
\draw (2.4,0) circle (0.4cm);
\draw (2.4,-0.8) node {\footnotesize{$2m$}};
\draw (-0.4,0) -- (-0.8,0);
\draw (0.4,0) -- (0.8,0);
\draw (-1.2,0) node {\footnotesize{$\dots$}};
\draw (-1.6,0) -- (-2,0);
\draw (0,1.2) [red,fill=red!30] circle (0.4cm);
\draw (0,2) node {\footnotesize{$1$}};
\draw (-0.08,0.4) -- (-0.08,0.8);
\draw (0.08,0.4) -- (0.08,0.8);
\draw (0,0.5) -- (-0.2,0.7);
\draw (0,0.5) -- (0.2,0.7);
\draw [decorate,decoration={brace,amplitude=6pt}] (0.3,0.5) to (2.4,0.5);
\draw (1.62,1.1) node {\footnotesize{$n-2m-1$}};
\draw (3.4,-1) circle (0.4cm);
\draw (4.26,-1) node {\footnotesize{$m$}};
\draw (2.7,-0.24) -- (3.1,-0.75);
\draw (3.4,1) circle (0.4cm);
\draw (4.26,1) node {\footnotesize{$m$}};
\draw (2.7,0.24) -- (3.1,0.75);
\end{tikzpicture} & $2a-2$\\ \hline	

	$\begin{array}{c}
	a<n-1\\
	a=2m+1
	\end{array}$ &\begin{tikzpicture}[scale=0.70]
\draw (0,0) circle (0.4cm);
\draw (-0.24,-0.8) node {\footnotesize{$2(2m+1)$}};
\draw (1.6,0) -- (2,0);
\draw (1.2,0) node {\footnotesize{\dots}};
\draw (-2.4,0) circle (0.4cm);
\draw (-2.4,-0.8) node {\footnotesize{$6$}};
\draw (-2.8,0) -- (-3.2,0);
\draw (-3.6,0) circle (0.4cm);
\draw (-3.6,-0.8) node {\footnotesize{$4$}};
\draw (-4,0) -- (-4.4,0);
\draw (-4.8,0) circle (0.4cm);
\draw (-4.8,-0.8) node {\footnotesize{$2$}};
\draw (2.4,0) circle (0.4cm);
\draw (2.2,-0.8) node {\footnotesize{$2(2m+1)$}};
\draw (-0.4,0) -- (-0.8,0);
\draw (0.4,0) -- (0.8,0);
\draw (-1.2,0) node {\footnotesize{$\dots$}};
\draw (-1.6,0) -- (-2,0);
\draw (0,1.2) [red,fill=red!30] circle (0.4cm);
\draw (0,2) node {\footnotesize{$2$}};
\draw (-0.08,0.4) -- (-0.08,0.8);
\draw (0.08,0.4) -- (0.08,0.8);
\draw (0,0.5) -- (-0.2,0.7);
\draw (0,0.5) -- (0.2,0.7);
\draw [decorate,decoration={brace,amplitude=6pt}] (0.3,0.5) to (2.4,0.5);
\draw (1.63,1.1) node {\footnotesize{$n-2m-2$}};
\draw (3.8,-1) circle (0.4cm);
\draw (3.8,1) circle (0.4cm);
\draw (5.1,-1) node {\footnotesize{$2m+1$}};
\draw (5.1,1) node {\footnotesize{$2m+1$}};
\draw (2.76,-0.16) -- (3.5,-0.75);
\draw (2.76,0.16) -- (3.5,0.75);

\end{tikzpicture} & $4a-4$\\ \hline
	\end{tabular}
	\caption{Exotic minimally unbalanced quivers with $G=SO(2n)$ and a double laced edge.}
	\label{tab:DseriesExotic1-1}
\end{table}

\begin{table}
	\centering
	\begin{tabular}{|c|l|c|}
	\hline
	$a$ &  \multicolumn{1}{c|}{Quiver} & Excess \\ \hline
$\begin{array}{c}
	a=n\\
	a=2m
	\end{array}$ &
\begin{tikzpicture}[scale=0.70]
\draw (0,0) circle (0.4cm);
\draw (2,1.5) node {\footnotesize{$$}};
\draw (-0.14,0.8) node {\footnotesize{$2m-3$}};
\draw (1.2,0) circle (0.4cm);
\draw (1,-0.8) node {\footnotesize{$2m-2$}};
\draw (-2.4,0) circle (0.4cm);
\draw (-2.4,-0.8) node {\footnotesize{$3$}};
\draw (-2.8,0) -- (-3.2,0);
\draw (-3.6,0) circle (0.4cm);
\draw (-3.6,-0.8) node {\footnotesize{$2$}};
\draw (-4,0) -- (-4.4,0);
\draw (-4.8,0) circle (0.4cm);
\draw (-4.8,-0.8) node {\footnotesize{$1$}};
\draw (-0.4,0) -- (-0.8,0);
\draw (0.4,0) -- (0.8,0);
\draw (-1.2,0) node {\footnotesize{$\dots$}};
\draw (-1.6,0) -- (-2,0);
\draw (3.4,-1) [red,fill=red!30] circle (0.4cm);
\draw (3.4,-1.8) node {\footnotesize{$1$}};
\draw (2.6,-0.92) -- (3,-0.92);
\draw (2.6,-1.08) -- (3,-1.08);
\draw (2.9,-1) -- (2.7,-0.78);
\draw (2.9,-1) -- (2.7,-1.22);
\draw (2.2,-1) circle (0.4cm);
\draw (2.2,-1.8) node {\footnotesize{$m$}};
\draw (1.5,-0.24) -- (1.9,-0.75);
\draw (2.2,1) circle (0.4cm);
\draw (3.3,1.1) node {\footnotesize{$m-1$}};
\draw (1.5,0.24) -- (1.9,0.75);
\end{tikzpicture} & $\frac{a}{2}-2$ \\ \hline
$\begin{array}{c}
	a=n\\
	a=2m+1
	\end{array}$ &
\begin{tikzpicture}[scale=0.70]
\draw (0,0) circle (0.4cm);
\draw (2,1.5) node {\footnotesize{$$}};
\draw (0,0.8) node {\footnotesize{$4m-4$}};
\draw (1.2,0) circle (0.4cm);
\draw (1,-0.8) node {\footnotesize{$4m-2$}};
\draw (-2.4,0) circle (0.4cm);
\draw (-2.4,-0.8) node {\footnotesize{$6$}};
\draw (-2.8,0) -- (-3.2,0);
\draw (-3.6,0) circle (0.4cm);
\draw (-3.6,-0.8) node {\footnotesize{$4$}};
\draw (-4,0) -- (-4.4,0);
\draw (-4.8,0) circle (0.4cm);
\draw (-4.8,-0.8) node {\footnotesize{$2$}};
\draw (-0.4,0) -- (-0.8,0);
\draw (0.4,0) -- (0.8,0);
\draw (-1.2,0) node {\footnotesize{$\dots$}};
\draw (-1.6,0) -- (-2,0);
\draw (3.4,-1) [red,fill=red!30] circle (0.4cm);
\draw (3.4,-1.8) node {\footnotesize{$2$}};
\draw (2.6,-0.92) -- (3,-0.92);
\draw (2.6,-1.08) -- (3,-1.08);
\draw (2.9,-1) -- (2.7,-0.78);
\draw (2.9,-1) -- (2.7,-1.22);
\draw (2.2,-1) circle (0.4cm);
\draw (2.2,-1.8) node {\footnotesize{$2m+1$}};
\draw (1.5,-0.24) -- (1.9,-0.75);
\draw (2.2,1) circle (0.4cm);
\draw (3.48,1.1) node {\footnotesize{$2m-1$}};
\draw (1.5,0.24) -- (1.9,0.75);
\end{tikzpicture} & $a-4$ \\ \hline
	
	$\begin{array}{c}
	a=n\\
	a=2m
	\end{array}$ &
\begin{tikzpicture}[scale=0.70]
\draw (0,0) circle (0.4cm);
\draw (2,1.5) node {\footnotesize{$$}};
\draw (-0.15,0.8) node {\footnotesize{$2m-3$}};
\draw (1.2,0) circle (0.4cm);
\draw (1,-0.8) node {\footnotesize{$2m-2$}};
\draw (-2.4,0) circle (0.4cm);
\draw (-2.4,-0.8) node {\footnotesize{$3$}};
\draw (-2.8,0) -- (-3.2,0);
\draw (-3.6,0) circle (0.4cm);
\draw (-3.6,-0.8) node {\footnotesize{$2$}};
\draw (-4,0) -- (-4.4,0);
\draw (-4.8,0) circle (0.4cm);
\draw (-4.8,-0.8) node {\footnotesize{$1$}};
\draw (-0.4,0) -- (-0.8,0);
\draw (0.4,0) -- (0.8,0);
\draw (-1.2,0) node {\footnotesize{$\dots$}};
\draw (-1.6,0) -- (-2,0);
\draw (3.4,-1) [red,fill=red!30] circle (0.4cm);
\draw (3.4,-1.8) node {\footnotesize{$2$}};
\draw (2.6,-0.92) -- (3,-0.92);
\draw (2.6,-1.08) -- (3,-1.08);
\draw (2.7,-1) -- (2.9,-0.78);
\draw (2.7,-1) -- (2.9,-1.22);
\draw (2.2,-1) circle (0.4cm);
\draw (2.2,-1.8) node {\footnotesize{$m$}};
\draw (1.5,-0.24) -- (1.9,-0.75);
\draw (2.2,1) circle (0.4cm);
\draw (3.3,1.1) node {\footnotesize{$m-1$}};
\draw (1.5,0.24) -- (1.9,0.75);
\end{tikzpicture} & $a-4$ \\ \hline	

	$\begin{array}{c}
	a=n\\
	a=2m+1
	\end{array}$ &
\begin{tikzpicture}[scale=0.70]
\draw (2,1.5) node {\footnotesize{$$}};
\draw (1.2,0) circle (0.4cm);
\draw (1,-0.8) node {\footnotesize{$4m-2$}};
\draw (-2.4,0) circle (0.4cm);
\draw (-2.4,-0.8) node {\footnotesize{$4$}};
\draw (-2.8,0) -- (-3.2,0);
\draw (-3.6,0) circle (0.4cm);
\draw (-3.6,-0.8) node {\footnotesize{$2$}};
\draw (-1.2,0) circle (0.4cm);
\draw (-1.2,-0.8) node {\footnotesize{$6$}};
\draw (-0.4,0) -- (-0.8,0);
\draw (0.4,0) -- (0.8,0);
\draw (0,0) node {\footnotesize{$\dots$}};
\draw (-1.6,0) -- (-2,0);
\draw (3.4,-1) [red,fill=red!30] circle (0.4cm);
\draw (3.4,-1.8) node {\footnotesize{$4$}};
\draw (2.6,-0.92) -- (3,-0.92);
\draw (2.6,-1.08) -- (3,-1.08);
\draw (2.7,-1) -- (2.9,-0.78);
\draw (2.7,-1) -- (2.9,-1.22);
\draw (2.2,-1) circle (0.4cm);
\draw (2.2,-1.8) node {\footnotesize{$2m+1$}};
\draw (1.5,-0.24) -- (1.9,-0.75);
\draw (2.2,1) circle (0.4cm);
\draw (3.5,1.1) node {\footnotesize{$2m-1$}};
\draw (1.5,0.24) -- (1.9,0.75);
\end{tikzpicture} & $2a-8$ \\ \hline	
	\end{tabular}
	\caption{Exotic minimally unbalanced quivers with $G=SO(2n)$ and a double laced edge.}
	\label{tab:DseriesExotic1-2}
\end{table}


\begin{table}
	\centering
	\begin{tabular}{|c|l|c|}
	\hline
	$a$ &  \multicolumn{1}{c|}{Quiver} & Excess \\ \hline

$\begin{array}{c}
	a<n-1\\
	a=2m
	\end{array}$ &
\begin{tikzpicture}[scale=0.70]
\draw (0,0) circle (0.4cm);
\draw (0,-0.8) node {\footnotesize{$6m$}};
\draw (1.6,0) -- (2,0);
\draw (1.2,0) node {\footnotesize{\dots}};
\draw (-2.4,0) circle (0.4cm);
\draw (-2.4,-0.8) node {\footnotesize{$9$}};
\draw (-2.8,0) -- (-3.2,0);
\draw (-3.6,0) circle (0.4cm);
\draw (-3.6,-0.8) node {\footnotesize{$6$}};
\draw (-4,0) -- (-4.4,0);
\draw (-4.8,0) circle (0.4cm);
\draw (-4.8,-0.8) node {\footnotesize{$3$}};
\draw (2.4,0) circle (0.4cm);
\draw (2.4,-0.8) node {\footnotesize{$6m$}};
\draw (-0.4,0) -- (-0.8,0);
\draw (0.4,0) -- (0.8,0);
\draw (-1.2,0) node {\footnotesize{$\dots$}};
\draw (-1.6,0) -- (-2,0);
\draw (0,1.2) [red,fill=red!30] circle (0.4cm);
\draw (0,2) node {\footnotesize{$1$}};
\draw (-0.08,0.4) -- (-0.08,0.8);
\draw (0.08,0.4) -- (0.08,0.8);
\draw (0,0.4) -- (0,0.8);
\draw (0,0.7) -- (-0.2,0.5);
\draw (0,0.7) -- (0.2,0.5);
\draw [decorate,decoration={brace,amplitude=6pt}] (0.3,0.5) to (2.4,0.5);
\draw (1.6,1.1) node {\footnotesize{$n-2m-1$}};
\draw (3.4,-1) circle (0.4cm);
\draw (4.26,-1) node {\footnotesize{$3m$}};
\draw (2.7,-0.24) -- (3.1,-0.75);
\draw (3.4,1) circle (0.4cm);
\draw (4.26,1) node {\footnotesize{$3m$}};
\draw (2.7,0.24) -- (3.1,0.75);
\end{tikzpicture} & $3a-2$ \\ \hline

	$\begin{array}{c}
	a<n-1\\
	a=2m+1
	\end{array}$ &\begin{tikzpicture}[scale=0.70]
\draw (0,0) circle (0.4cm);
\draw (-0.24,-0.8) node {\footnotesize{$6(2m+1)$}};
\draw (1.6,0) -- (2,0);
\draw (1.2,0) node {\footnotesize{\dots}};
\draw (-2.4,0) circle (0.4cm);
\draw (-2.4,-0.8) node {\footnotesize{$18$}};
\draw (-2.8,0) -- (-3.2,0);
\draw (-3.6,0) circle (0.4cm);
\draw (-3.6,-0.8) node {\footnotesize{$12$}};
\draw (-4,0) -- (-4.4,0);
\draw (-4.8,0) circle (0.4cm);
\draw (-4.8,-0.8) node {\footnotesize{$6$}};
\draw (2.4,0) circle (0.4cm);
\draw (2.2,-0.8) node {\footnotesize{$6(2m+1)$}};
\draw (-0.4,0) -- (-0.8,0);
\draw (0.4,0) -- (0.8,0);
\draw (-1.2,0) node {\footnotesize{$\dots$}};
\draw (-1.6,0) -- (-2,0);
\draw (0,1.2) [red,fill=red!30] circle (0.4cm);
\draw (0,2) node {\footnotesize{$2$}};
\draw (-0.08,0.4) -- (-0.08,0.8);
\draw (0.08,0.4) -- (0.08,0.8);
\draw (0,0.4) -- (0,0.8);
\draw (0,0.7) -- (-0.2,0.5);
\draw (0,0.7) -- (0.2,0.5);
\draw [decorate,decoration={brace,amplitude=6pt}] (0.3,0.5) to (2.4,0.5);
\draw (1.63,1.1) node {\footnotesize{$n-2m-2$}};
\draw (3.8,-1) circle (0.4cm);
\draw (3.8,1) circle (0.4cm);
\draw (5.1,-1) node {\footnotesize{$6m+3$}};
\draw (5.1,1) node {\footnotesize{$6m+3$}};
\draw (2.76,-0.16) -- (3.5,-0.75);
\draw (2.76,0.16) -- (3.5,0.75);

\end{tikzpicture} & $6a-4$\\ \hline	

$\begin{array}{c}
	a<n-1\\
	a=2m
	\end{array}$ &
\begin{tikzpicture}[scale=0.70]
\draw (0,0) circle (0.4cm);
\draw (0,-0.8) node {\footnotesize{$2m$}};
\draw (1.6,0) -- (2,0);
\draw (1.2,0) node {\footnotesize{\dots}};
\draw (-2.4,0) circle (0.4cm);
\draw (-2.4,-0.8) node {\footnotesize{$3$}};
\draw (-2.8,0) -- (-3.2,0);
\draw (-3.6,0) circle (0.4cm);
\draw (-3.6,-0.8) node {\footnotesize{$2$}};
\draw (-4,0) -- (-4.4,0);
\draw (-4.8,0) circle (0.4cm);
\draw (-4.8,-0.8) node {\footnotesize{$1$}};
\draw (2.4,0) circle (0.4cm);
\draw (2.4,-0.8) node {\footnotesize{$2m$}};
\draw (-0.4,0) -- (-0.8,0);
\draw (0.4,0) -- (0.8,0);
\draw (-1.2,0) node {\footnotesize{$\dots$}};
\draw (-1.6,0) -- (-2,0);
\draw (0,1.2) [red,fill=red!30] circle (0.4cm);
\draw (0,2) node {\footnotesize{$1$}};
\draw (-0.08,0.4) -- (-0.08,0.8);
\draw (0.08,0.4) -- (0.08,0.8);
\draw (0,0.4) -- (0,0.8);
\draw (0,0.5) -- (-0.2,0.7);
\draw (0,0.5) -- (0.2,0.7);
\draw [decorate,decoration={brace,amplitude=6pt}] (0.3,0.5) to (2.4,0.5);
\draw (1.6,1.1) node {\footnotesize{$n-2m-1$}};
\draw (3.4,-1) circle (0.4cm);
\draw (4.26,-1) node {\footnotesize{$m$}};
\draw (2.7,-0.24) -- (3.1,-0.75);
\draw (3.4,1) circle (0.4cm);
\draw (4.26,1) node {\footnotesize{$m$}};
\draw (2.7,0.24) -- (3.1,0.75);
\end{tikzpicture} & $3a-2$ \\ \hline

	$\begin{array}{c}
	a<n-1\\
	a=2m+1
	\end{array}$ &\begin{tikzpicture}[scale=0.70]
\draw (0,0) circle (0.4cm);
\draw (-0.24,-0.8) node {\footnotesize{$2(2m+1)$}};
\draw (1.6,0) -- (2,0);
\draw (1.2,0) node {\footnotesize{\dots}};
\draw (-2.4,0) circle (0.4cm);
\draw (-2.4,-0.8) node {\footnotesize{$6$}};
\draw (-2.8,0) -- (-3.2,0);
\draw (-3.6,0) circle (0.4cm);
\draw (-3.6,-0.8) node {\footnotesize{$4$}};
\draw (-4,0) -- (-4.4,0);
\draw (-4.8,0) circle (0.4cm);
\draw (-4.8,-0.8) node {\footnotesize{$2$}};
\draw (2.4,0) circle (0.4cm);
\draw (2.2,-0.8) node {\footnotesize{$2(2m+1)$}};
\draw (-0.4,0) -- (-0.8,0);
\draw (0.4,0) -- (0.8,0);
\draw (-1.2,0) node {\footnotesize{$\dots$}};
\draw (-1.6,0) -- (-2,0);
\draw (0,1.2) [red,fill=red!30] circle (0.4cm);
\draw (0,2) node {\footnotesize{$2$}};
\draw (-0.08,0.4) -- (-0.08,0.8);
\draw (0.08,0.4) -- (0.08,0.8);
\draw (0,0.4) -- (0,0.8);
\draw (0,0.5) -- (-0.2,0.7);
\draw (0,0.5) -- (0.2,0.7);
\draw [decorate,decoration={brace,amplitude=6pt}] (0.3,0.5) to (2.4,0.5);
\draw (1.63,1.1) node {\footnotesize{$n-2m-2$}};
\draw (3.8,-1) circle (0.4cm);
\draw (3.8,1) circle (0.4cm);
\draw (5.1,-1) node {\footnotesize{$2m+1$}};
\draw (5.1,1) node {\footnotesize{$2m+1$}};
\draw (2.76,-0.16) -- (3.5,-0.75);
\draw (2.76,0.16) -- (3.5,0.75);

\end{tikzpicture} & $6a-4$\\ \hline

	\end{tabular}
	\caption{Exotic minimally unbalanced quivers with $G=SO(2n)$ and a triple laced edge.}
	\label{tab:DseriesExotic2-1}
\end{table}

\begin{table}
	\centering
	\begin{tabular}{|c|l|c|}
	\hline
	$a$ &  \multicolumn{1}{c|}{Quiver} & Excess \\ \hline
	$\begin{array}{c}
	a=n\\
	a=2m
	\end{array}$ &
\begin{tikzpicture}[scale=0.70]
\draw (0,0) circle (0.4cm);
\draw (2,1.5) node {\footnotesize{$$}};
\draw (-0.2,0.8) node {\footnotesize{$6m-9$}};
\draw (1.2,0) circle (0.4cm);
\draw (1,-0.8) node {\footnotesize{$6m-6$}};
\draw (-2.4,0) circle (0.4cm);
\draw (-2.4,-0.8) node {\footnotesize{$9$}};
\draw (-2.8,0) -- (-3.2,0);
\draw (-3.6,0) circle (0.4cm);
\draw (-3.6,-0.8) node {\footnotesize{$6$}};
\draw (-4,0) -- (-4.4,0);
\draw (-4.8,0) circle (0.4cm);
\draw (-4.8,-0.8) node {\footnotesize{$3$}};
\draw (-0.4,0) -- (-0.8,0);
\draw (0.4,0) -- (0.8,0);
\draw (-1.2,0) node {\footnotesize{$\dots$}};
\draw (-1.6,0) -- (-2,0);
\draw (3.4,-1) [red,fill=red!30] circle (0.4cm);
\draw (3.4,-1.8) node {\footnotesize{$2$}};
\draw (2.6,-0.92) -- (3,-0.92);
\draw (2.6,-1.08) -- (3,-1.08);
\draw (2.6,-1) -- (3,-1);
\draw (2.9,-1) -- (2.7,-0.78);
\draw (2.9,-1) -- (2.7,-1.22);
\draw (2.2,-1) circle (0.4cm);
\draw (2.2,-1.8) node {\footnotesize{$3m$}};
\draw (1.5,-0.24) -- (1.9,-0.75);
\draw (2.2,1) circle (0.4cm);
\draw (3.5,1.1) node {\footnotesize{$3m-3$}};
\draw (1.5,0.24) -- (1.9,0.75);
\end{tikzpicture} & $\frac{3a}{2}-4$ \\ \hline	

	$\begin{array}{c}
	a=n\\
	a=2m+1
	\end{array}$ &
\begin{tikzpicture}[scale=0.70]
\draw (2,1.5) node {\footnotesize{$$}};
\draw (1.2,0) circle (0.4cm);
\draw (0.8,-0.8) node {\footnotesize{$12m-6$}};
\draw (-2.4,0) circle (0.4cm);
\draw (-2.4,-0.8) node {\footnotesize{$12$}};
\draw (-2.8,0) -- (-3.2,0);
\draw (-3.6,0) circle (0.4cm);
\draw (-3.6,-0.8) node {\footnotesize{$6$}};
\draw (-1.2,0) circle (0.4cm);
\draw (-1.2,-0.8) node {\footnotesize{$18$}};
\draw (-0.4,0) -- (-0.8,0);
\draw (0.4,0) -- (0.8,0);
\draw (0,0) node {\footnotesize{$\dots$}};
\draw (-1.6,0) -- (-2,0);
\draw (3.4,-1) [red,fill=red!30] circle (0.4cm);
\draw (3.4,-1.8) node {\footnotesize{$4$}};
\draw (2.6,-0.92) -- (3,-0.92);
\draw (2.6,-1.08) -- (3,-1.08);
\draw (2.6,-1) -- (3,-1);
\draw (2.9,-1) -- (2.7,-0.78);
\draw (2.9,-1) -- (2.7,-1.22);
\draw (2.2,-1) circle (0.4cm);
\draw (2.2,-1.8) node {\footnotesize{$6m+3$}};
\draw (1.5,-0.24) -- (1.9,-0.75);
\draw (2.2,1) circle (0.4cm);
\draw (3.5,1.1) node {\footnotesize{$6m-3$}};
\draw (1.5,0.24) -- (1.9,0.75);
\end{tikzpicture} & $3a-12$ \\ \hline

	$\begin{array}{c}
	a=n\\
	a=2m
	\end{array}$ &
\begin{tikzpicture}[scale=0.70]
\draw (0,0) circle (0.4cm);
\draw (2,1.5) node {\footnotesize{$$}};
\draw (-0.15,0.8) node {\footnotesize{$2m-3$}};
\draw (1.2,0) circle (0.4cm);
\draw (1,-0.8) node {\footnotesize{$2m-2$}};
\draw (-2.4,0) circle (0.4cm);
\draw (-2.4,-0.8) node {\footnotesize{$3$}};
\draw (-2.8,0) -- (-3.2,0);
\draw (-3.6,0) circle (0.4cm);
\draw (-3.6,-0.8) node {\footnotesize{$2$}};
\draw (-4,0) -- (-4.4,0);
\draw (-4.8,0) circle (0.4cm);
\draw (-4.8,-0.8) node {\footnotesize{$1$}};
\draw (-0.4,0) -- (-0.8,0);
\draw (0.4,0) -- (0.8,0);
\draw (-1.2,0) node {\footnotesize{$\dots$}};
\draw (-1.6,0) -- (-2,0);
\draw (3.4,-1) [red,fill=red!30] circle (0.4cm);
\draw (3.4,-1.8) node {\footnotesize{$2$}};
\draw (2.6,-0.92) -- (3,-0.92);
\draw (2.6,-1.08) -- (3,-1.08);
\draw (2.6,-1) -- (3,-1);
\draw (2.7,-1) -- (2.9,-0.78);
\draw (2.7,-1) -- (2.9,-1.22);
\draw (2.2,-1) circle (0.4cm);
\draw (2.2,-1.8) node {\footnotesize{$m$}};
\draw (1.5,-0.24) -- (1.9,-0.75);
\draw (2.2,1) circle (0.4cm);
\draw (3.4,1.1) node {\footnotesize{$m-1$}};
\draw (1.5,0.24) -- (1.9,0.75);
\end{tikzpicture} & $\frac{3a}{2}-4$ \\ \hline

	$\begin{array}{c}
	a=n\\
	a=2m+1
	\end{array}$ &
\begin{tikzpicture}[scale=0.70]
\draw (2,1.5) node {\footnotesize{$$}};
\draw (1.2,0) circle (0.4cm);
\draw (0.9,-0.8) node {\footnotesize{$4m-2$}};
\draw (-2.4,0) circle (0.4cm);
\draw (-2.4,-0.8) node {\footnotesize{$4$}};
\draw (-2.8,0) -- (-3.2,0);
\draw (-3.6,0) circle (0.4cm);
\draw (-3.6,-0.8) node {\footnotesize{$2$}};
\draw (-1.2,0) circle (0.4cm);
\draw (-1.2,-0.8) node {\footnotesize{$6$}};
\draw (-0.4,0) -- (-0.8,0);
\draw (0.4,0) -- (0.8,0);
\draw (0,0) node {\footnotesize{$\dots$}};
\draw (-1.6,0) -- (-2,0);
\draw (3.4,-1) [red,fill=red!30] circle (0.4cm);
\draw (3.4,-1.8) node {\footnotesize{$4$}};
\draw (2.6,-0.92) -- (3,-0.92);
\draw (2.6,-1.08) -- (3,-1.08);
\draw (2.6,-1) -- (3,-1);
\draw (2.7,-1) -- (2.9,-0.78);
\draw (2.7,-1) -- (2.9,-1.22);
\draw (2.2,-1) circle (0.4cm);
\draw (2.2,-1.8) node {\footnotesize{$2m+1$}};
\draw (1.5,-0.24) -- (1.9,-0.75);
\draw (2.2,1) circle (0.4cm);
\draw (3.5,1.1) node {\footnotesize{$2m-1$}};
\draw (1.5,0.24) -- (1.9,0.75);
\end{tikzpicture} & $3a-8$ \\ \hline

	\end{tabular}
	\caption{Exotic minimally unbalanced quivers with $G=SO(2n)$ and a triple laced edge.}
	\label{tab:DseriesExotic2-2}
\end{table}

\clearpage
\subsection{Exotic Minimally Unbalanced Quivers with $G$ of Type $E_n$}

In this subsection, we classify all exotic minimally unbalanced quivers of $E_n$-type with the unbalanced node connected to the balanced part of the quiver by a non-simply laced edge. 

\subsubsection{Exotic Minimally Unbalanced Quivers with $G$ of Type $E_6$}

We start by showing the results for all exotic minimally unbalanced quivers with $G=E_6$ and with the unbalanced node connected by a double and triple laced edge. The former quivers are collected in table \ref{tab:E6seriesExoticDouble} and the latter in table \ref{tab:E6seriesExoticTriple}, respectively. Note that, within the tables, a further distinction involves the orientation of the non-simply laced edge and the position of the unbalanced node. The orientation is always considered with respect to the unbalanced node.

\begin{table}
	\centering
	\begin{tabular}{|c|c|}
	\hline
Quiver & Excess \\ \hline
\begin{tikzpicture}[scale=0.70]
\draw (0.4,0) -- (0.8,0);
\draw (0,0) circle (0.4cm);
\draw (0,-0.8) node {\footnotesize{$6$}};
\draw (1.6,0) -- (2,0);
\draw (2.4,0) circle (0.4cm);
\draw (1.2,-0.8) node {\footnotesize{$4$}};
\draw(1.2,0) circle (0.4cm);
\draw (2.4,-0.8) node {\footnotesize{$2$}};
\draw (-2.4,0) circle (0.4cm);
\draw (-2.4,-0.8) node {\footnotesize{$4$}};
\draw (-0.4,0) -- (-0.8,0);
\draw (-1.2,0) circle (0.4cm);
\draw (-1.2,-0.8) node {\footnotesize{$5$}};
\draw (-1.6,0) -- (-2,0);
\draw (0,0.4) -- (0,0.8);
\draw (0,1.2) circle (0.4cm);
\draw (0,2) node {\footnotesize{$3$}};
\draw (-2.32,0.4) -- (-2.32,0.8);
\draw (-2.48,0.4) -- (-2.48,0.8);
\draw (-2.2,0.7) -- (-2.4,0.5);
\draw (-2.6,0.7) -- (-2.4,0.5);
\draw (-2.4,1.2)[red, fill=red!30] circle (0.4cm);
\draw (-2.4,2) node {\footnotesize{$3$}};

\end{tikzpicture} & $2$\\ \hline	
\begin{tikzpicture}[scale=0.70]
\draw (0.4,0) -- (0.8,0);
\draw (0,0) circle (0.4cm);
\draw (0,-0.8) node {\footnotesize{$12$}};
\draw (1.6,0) -- (2,0);
\draw (2.4,0) circle (0.4cm);
\draw (1.2,-0.8) node {\footnotesize{$8$}};
\draw(1.2,0) circle (0.4cm);
\draw (2.4,-0.8) node {\footnotesize{$4$}};
\draw (-2.4,0) circle (0.4cm);
\draw (-2.4,-0.8) node {\footnotesize{$5$}};
\draw (-0.4,0) -- (-0.8,0);
\draw (-1.2,0) circle (0.4cm);
\draw (-1.2,-0.8) node {\footnotesize{$10$}};
\draw (-1.6,0) -- (-2,0);
\draw (0,0.4) -- (0,0.8);
\draw (0,1.2) circle (0.4cm);
\draw (0,2) node {\footnotesize{$6$}};
\draw (-1.12,0.4) -- (-1.12,0.8);
\draw (-1.28,0.4) -- (-1.28,0.8);
\draw (-1,0.7) -- (-1.2,0.5);
\draw (-1.4,0.7) -- (-1.2,0.5);
\draw (-1.2,1.2) [red, fill=red!30] circle (0.4cm);
\draw (-1.2,2) node {\footnotesize{$3$}};

\end{tikzpicture} & $14$\\ \hline	

\begin{tikzpicture}[scale=0.70]
\draw (0.4,0) -- (0.8,0);
\draw (0,0) circle (0.4cm);
\draw (0,-0.8) node {\footnotesize{$6$}};
\draw (1.2,-0.8) node {\footnotesize{$4$}};
\draw (2.4,0) circle (0.4cm);
\draw (2.4,-0.8) node {\footnotesize{$2$}};
\draw(1.2,0) circle (0.4cm);
\draw (-2.4,0) circle (0.4cm);
\draw (-2.4,-0.8) node {\footnotesize{$2$}};
\draw (-0.4,0) -- (-0.8,0);
\draw (1.6,0) -- (2,0);
\draw (-1.2,0) circle (0.4cm);
\draw (-1.2,-0.8) node {\footnotesize{$4$}};
\draw (-1.6,0) -- (-2,0);
\draw (-0.36,0.56) -- (-0.64,0.56);
\draw (-0.36,0.56) -- (-0.288,0.8);
\draw (-0.14,0.38) -- (-0.52,0.92);
\draw (-0.28,0.28) -- (-0.68,0.83);
\draw (-0.8,1.2) [red, fill=red!30] circle (0.4cm);
\draw (-0.8,2) node {\footnotesize{$1$}};
\draw (0.26,0.3) -- (0.65,0.83);
\draw (0.8,1.2) circle (0.4cm);
\draw (0.8,2) node {\footnotesize{$3$}};

\end{tikzpicture} & $10$\\ \hline	

\begin{tikzpicture}[scale=0.70]
\draw (0.4,0) -- (0.8,0);
\draw (0,0) circle (0.4cm);
\draw (0,-0.8) node {\footnotesize{$3$}};
\draw (1.6,0) -- (2,0);
\draw (2.4,0) circle (0.4cm);
\draw (1.2,-0.8) node {\footnotesize{$2$}};
\draw(1.2,0) circle (0.4cm);
\draw (2.4,-0.8) node {\footnotesize{$1$}};
\draw (-2.4,0) circle (0.4cm);
\draw (-2.4,-0.8) node {\footnotesize{$1$}};
\draw (-0.4,0) -- (-0.8,0);
\draw (-1.2,0) circle (0.4cm);
\draw (-1.2,-0.8) node {\footnotesize{$2$}};
\draw (-1.6,0) -- (-2,0);
\draw (0,0.4) -- (0,0.8);
\draw (0,1.2) circle (0.4cm);
\draw (0.8,1.2) node {\footnotesize{$2$}};
\draw (-0.08,1.6) -- (-0.08,2);
\draw (0.08,1.6) -- (0.08,2);
\draw (-0.2,1.9) -- (0,1.7);
\draw (0.2,1.9) -- (0,1.7);
\draw (0,2.4) [red, fill=red!30] circle (0.4cm);
\draw (0.8,2.4) node {\footnotesize{$1$}};
\draw (0,3) node {\footnotesize{$$}};

\end{tikzpicture} & $2$ \\ \hline	\hline	

\begin{tikzpicture}[scale=0.70]
\draw (0.4,0) -- (0.8,0);
\draw (0,0) circle (0.4cm);
\draw (0,-0.8) node {\footnotesize{$12$}};
\draw (1.6,0) -- (2,0);
\draw (2.4,0) circle (0.4cm);
\draw (1.2,-0.8) node {\footnotesize{$8$}};
\draw(1.2,0) circle (0.4cm);
\draw (2.4,-0.8) node {\footnotesize{$4$}};
\draw (-2.4,0) circle (0.4cm);
\draw (-2.4,-0.8) node {\footnotesize{$8$}};
\draw (-0.4,0) -- (-0.8,0);
\draw (-1.2,0) circle (0.4cm);
\draw (-1.2,-0.8) node {\footnotesize{$10$}};
\draw (-1.6,0) -- (-2,0);
\draw (0,0.4) -- (0,0.8);
\draw (0,1.2) circle (0.4cm);
\draw (0,2) node {\footnotesize{$6$}};
\draw (-2.32,0.4) -- (-2.32,0.8);
\draw (-2.48,0.4) -- (-2.48,0.8);
\draw (-2.2,0.5) -- (-2.4,0.7);
\draw (-2.6,0.5) -- (-2.4,0.7);
\draw (-2.4,1.2)[red, fill=red!30] circle (0.4cm);
\draw (-2.4,2) node {\footnotesize{$3$}};

\end{tikzpicture} & $-2$ \\ \hline

\begin{tikzpicture}[scale=0.70]
\draw (0.4,0) -- (0.8,0);
\draw (0,0) circle (0.4cm);
\draw (0,-0.8) node {\footnotesize{$24$}};
\draw (1.6,0) -- (2,0);
\draw (2.4,0) circle (0.4cm);
\draw (1.2,-0.8) node {\footnotesize{$16$}};
\draw(1.2,0) circle (0.4cm);
\draw (2.4,-0.8) node {\footnotesize{$8$}};
\draw (-2.4,0) circle (0.4cm);
\draw (-2.4,-0.8) node {\footnotesize{$10$}};
\draw (-0.4,0) -- (-0.8,0);
\draw (-1.2,0) circle (0.4cm);
\draw (-1.2,-0.8) node {\footnotesize{$20$}};
\draw (-1.6,0) -- (-2,0);
\draw (0,0.4) -- (0,0.8);
\draw (0,1.2) circle (0.4cm);
\draw (0,2) node {\footnotesize{$12$}};
\draw (-1.12,0.4) -- (-1.12,0.8);
\draw (-1.28,0.4) -- (-1.28,0.8);
\draw (-1,0.5) -- (-1.2,0.7);
\draw (-1.4,0.5) -- (-1.2,0.7);
\draw (-1.2,1.2) [red, fill=red!30] circle (0.4cm);
\draw (-1.2,2) node {\footnotesize{$3$}};

\end{tikzpicture} & $14$ \\ \hline

\begin{tikzpicture}[scale=0.70]	
\draw (0.4,0) -- (0.8,0);
\draw (0,0) circle (0.4cm);
\draw (0,-0.8) node {\footnotesize{$12$}};
\draw (1.2,-0.8) node {\footnotesize{$8$}};
\draw (2.4,0) circle (0.4cm);
\draw (2.4,-0.8) node {\footnotesize{$4$}};
\draw(1.2,0) circle (0.4cm);
\draw (-2.4,0) circle (0.4cm);
\draw (-2.4,-0.8) node {\footnotesize{$4$}};
\draw (-0.4,0) -- (-0.8,0);
\draw (1.6,0) -- (2,0);
\draw (-1.2,0) circle (0.4cm);
\draw (-1.2,-0.8) node {\footnotesize{$8$}};
\draw (-1.6,0) -- (-2,0);
\draw (-0.432,0.648) -- (-0.14,0.62);
\draw (-0.432,0.648) -- (-0.5,0.35);
\draw (-0.14,0.38) -- (-0.52,0.92);
\draw (-0.28,0.28) -- (-0.68,0.83);
\draw (-0.8,1.2) [red, fill=red!30] circle (0.4cm);
\draw (-0.8,2) node {\footnotesize{$1$}};
\draw (0.26,0.3) -- (0.65,0.83);
\draw (0.8,1.2) circle (0.4cm);
\draw (0.8,2) node {\footnotesize{$6$}};

\end{tikzpicture} & $10$ \\ \hline

\begin{tikzpicture}[scale=0.70]
\draw (0.4,0) -- (0.8,0);
\draw (0,0) circle (0.4cm);
\draw (0,-0.8) node {\footnotesize{$6$}};
\draw (1.6,0) -- (2,0);
\draw (2.4,0) circle (0.4cm);
\draw (1.2,-0.8) node {\footnotesize{$4$}};
\draw(1.2,0) circle (0.4cm);
\draw (2.4,-0.8) node {\footnotesize{$2$}};
\draw (-2.4,0) circle (0.4cm);
\draw (-2.4,-0.8) node {\footnotesize{$2$}};
\draw (-0.4,0) -- (-0.8,0);
\draw (-1.2,0) circle (0.4cm);
\draw (-1.2,-0.8) node {\footnotesize{$4$}};
\draw (-1.6,0) -- (-2,0);
\draw (0,0.4) -- (0,0.8);
\draw (0,1.2) circle (0.4cm);
\draw (0.8,1.2) node {\footnotesize{$4$}};
\draw (-0.08,1.6) -- (-0.08,2);
\draw (0.08,1.6) -- (0.08,2);
\draw (-0.2,1.7) -- (0,1.9);
\draw (0.2,1.7) -- (0,1.9);
\draw (0,2.4) [red, fill=red!30] circle (0.4cm);
\draw (0.8,2.4) node {\footnotesize{$1$}};
\draw (0,3) node {\footnotesize{$$}};
\end{tikzpicture} & $2$ \\ \hline
	\end{tabular}
	\caption{Exotic minimally unbalanced quivers of $E_6$-type with a double laced edge.}
	\label{tab:E6seriesExoticDouble}
\end{table}


\begin{table}
	\centering
	\begin{tabular}{|c|c|}
	\hline
	Quiver & Excess \\ \hline
\begin{tikzpicture}[scale=0.70]
\draw (0.4,0) -- (0.8,0);
\draw (0,0) circle (0.4cm);
\draw (0,-0.8) node {\footnotesize{$6$}};
\draw (1.6,0) -- (2,0);
\draw (2.4,0) circle (0.4cm);
\draw (1.2,-0.8) node {\footnotesize{$4$}};
\draw(1.2,0) circle (0.4cm);
\draw (2.4,-0.8) node {\footnotesize{$2$}};
\draw (-2.4,0) circle (0.4cm);
\draw (-2.4,-0.8) node {\footnotesize{$4$}};
\draw (-0.4,0) -- (-0.8,0);
\draw (-1.2,0) circle (0.4cm);
\draw (-1.2,-0.8) node {\footnotesize{$5$}};
\draw (-1.6,0) -- (-2,0);
\draw (0,0.4) -- (0,0.8);
\draw (0,1.2) circle (0.4cm);
\draw (0,2) node {\footnotesize{$3$}};
\draw (-2.32,0.4) -- (-2.32,0.8);
\draw (-2.48,0.4) -- (-2.48,0.8);
\draw (-2.4,0.4) -- (-2.4,0.8);
\draw (-2.2,0.7) -- (-2.4,0.5);
\draw (-2.6,0.7) -- (-2.4,0.5);
\draw (-2.4,1.2)[red, fill=red!30] circle (0.4cm);
\draw (-2.4,2) node {\footnotesize{$3$}};

\end{tikzpicture} & $6$\\ \hline	
	
\begin{tikzpicture}[scale=0.70]
\draw (0.4,0) -- (0.8,0);
\draw (0,0) circle (0.4cm);
\draw (0,-0.8) node {\footnotesize{$12$}};
\draw (1.6,0) -- (2,0);
\draw (2.4,0) circle (0.4cm);
\draw (1.2,-0.8) node {\footnotesize{$8$}};
\draw(1.2,0) circle (0.4cm);
\draw (2.4,-0.8) node {\footnotesize{$4$}};
\draw (-2.4,0) circle (0.4cm);
\draw (-2.4,-0.8) node {\footnotesize{$5$}};
\draw (-0.4,0) -- (-0.8,0);
\draw (-1.2,0) circle (0.4cm);
\draw (-1.2,-0.8) node {\footnotesize{$10$}};
\draw (-1.6,0) -- (-2,0);
\draw (0,0.4) -- (0,0.8);
\draw (0,1.2) circle (0.4cm);
\draw (0,2) node {\footnotesize{$6$}};
\draw (-1.12,0.4) -- (-1.12,0.8);
\draw (-1.28,0.4) -- (-1.28,0.8);
\draw (-1.2,0.4) -- (-1.2,0.8);
\draw (-1,0.7) -- (-1.2,0.5);
\draw (-1.4,0.7) -- (-1.2,0.5);
\draw (-1.2,1.2) [red, fill=red!30] circle (0.4cm);
\draw (-1.2,2) node {\footnotesize{$3$}};

\end{tikzpicture} & $24$\\ \hline	

\begin{tikzpicture}[scale=0.70]
\draw (0.4,0) -- (0.8,0);
\draw (0,0) circle (0.4cm);
\draw (0,-0.8) node {\footnotesize{$6$}};
\draw (1.2,-0.8) node {\footnotesize{$4$}};
\draw (2.4,0) circle (0.4cm);
\draw (2.4,-0.8) node {\footnotesize{$2$}};
\draw(1.2,0) circle (0.4cm);
\draw (-2.4,0) circle (0.4cm);
\draw (-2.4,-0.8) node {\footnotesize{$2$}};
\draw (-0.4,0) -- (-0.8,0);
\draw (1.6,0) -- (2,0);
\draw (-1.2,0) circle (0.4cm);
\draw (-1.2,-0.8) node {\footnotesize{$4$}};
\draw (-1.6,0) -- (-2,0);
\draw (-0.362,0.557) -- (-0.64,0.56);
\draw (-0.362,0.557) -- (-0.288,0.8);
\draw (-0.14,0.38) -- (-0.52,0.92);
\draw (-0.28,0.28) -- (-0.68,0.83);
\draw (-0.21,0.33) -- (-0.6,0.875);
\draw (-0.8,1.2) [red, fill=red!30] circle (0.4cm);
\draw (-0.8,2) node {\footnotesize{$1$}};
\draw (0.26,0.3) -- (0.65,0.83);
\draw (0.8,1.2) circle (0.4cm);
\draw (0.8,2) node {\footnotesize{$3$}};

\end{tikzpicture} & $16$\\ \hline	

\begin{tikzpicture}[scale=0.70]
\draw (0.4,0) -- (0.8,0);
\draw (0,0) circle (0.4cm);
\draw (0,-0.8) node {\footnotesize{$3$}};
\draw (1.6,0) -- (2,0);
\draw (2.4,0) circle (0.4cm);
\draw (1.2,-0.8) node {\footnotesize{$2$}};
\draw(1.2,0) circle (0.4cm);
\draw (2.4,-0.8) node {\footnotesize{$1$}};
\draw (-2.4,0) circle (0.4cm);
\draw (-2.4,-0.8) node {\footnotesize{$1$}};
\draw (-0.4,0) -- (-0.8,0);
\draw (-1.2,0) circle (0.4cm);
\draw (-1.2,-0.8) node {\footnotesize{$2$}};
\draw (-1.6,0) -- (-2,0);
\draw (0,0.4) -- (0,0.8);
\draw (0,1.2) circle (0.4cm);
\draw (0.8,1.2) node {\footnotesize{$2$}};
\draw (-0.08,1.6) -- (-0.08,2);
\draw (0.08,1.6) -- (0.08,2);
\draw (0,1.6) -- (0,2);
\draw (-0.2,1.9) -- (0,1.7);
\draw (0.2,1.9) -- (0,1.7);
\draw (0,2.4) [red, fill=red!30] circle (0.4cm);
\draw (0.8,2.4) node {\footnotesize{$1$}};
\draw (0,3) node {\footnotesize{$$}};

\end{tikzpicture} & $4$ \\ \hline	\hline

\begin{tikzpicture}[scale=0.70]
\draw (0.4,0) -- (0.8,0);
\draw (0,0) circle (0.4cm);
\draw (0,-0.8) node {\footnotesize{$6$}};
\draw (1.6,0) -- (2,0);
\draw (2.4,0) circle (0.4cm);
\draw (1.2,-0.8) node {\footnotesize{$4$}};
\draw(1.2,0) circle (0.4cm);
\draw (2.4,-0.8) node {\footnotesize{$2$}};
\draw (-2.4,0) circle (0.4cm);
\draw (-2.4,-0.8) node {\footnotesize{$4$}};
\draw (-0.4,0) -- (-0.8,0);
\draw (-1.2,0) circle (0.4cm);
\draw (-1.2,-0.8) node {\footnotesize{$5$}};
\draw (-1.6,0) -- (-2,0);
\draw (0,0.4) -- (0,0.8);
\draw (0,1.2) circle (0.4cm);
\draw (0,2) node {\footnotesize{$3$}};
\draw (-2.32,0.4) -- (-2.32,0.8);
\draw (-2.48,0.4) -- (-2.48,0.8);
\draw (-2.4,0.4) -- (-2.4,0.8);
\draw (-2.2,0.5) -- (-2.4,0.7);
\draw (-2.6,0.5) -- (-2.4,0.7);
\draw (-2.4,1.2)[red, fill=red!30] circle (0.4cm);
\draw (-2.4,2) node {\footnotesize{$1$}};

\end{tikzpicture} & $2$ \\ \hline

\begin{tikzpicture}[scale=0.70]
\draw (0.4,0) -- (0.8,0);
\draw (0,0) circle (0.4cm);
\draw (0,-0.8) node {\footnotesize{$12$}};
\draw (1.6,0) -- (2,0);
\draw (2.4,0) circle (0.4cm);
\draw (1.2,-0.8) node {\footnotesize{$8$}};
\draw(1.2,0) circle (0.4cm);
\draw (2.4,-0.8) node {\footnotesize{$4$}};
\draw (-2.4,0) circle (0.4cm);
\draw (-2.4,-0.8) node {\footnotesize{$5$}};
\draw (-0.4,0) -- (-0.8,0);
\draw (-1.2,0) circle (0.4cm);
\draw (-1.2,-0.8) node {\footnotesize{$10$}};
\draw (-1.6,0) -- (-2,0);
\draw (0,0.4) -- (0,0.8);
\draw (0,1.2) circle (0.4cm);
\draw (0,2) node {\footnotesize{$6$}};
\draw (-1.12,0.4) -- (-1.12,0.8);
\draw (-1.28,0.4) -- (-1.28,0.8);
\draw (-1.2,0.4) -- (-1.2,0.8);
\draw (-1,0.5) -- (-1.2,0.7);
\draw (-1.4,0.5) -- (-1.2,0.7);
\draw (-1.2,1.2) [red, fill=red!30] circle (0.4cm);
\draw (-1.2,2) node {\footnotesize{$1$}};

\end{tikzpicture} & $8$ \\ \hline

\begin{tikzpicture}[scale=0.70]	
\draw (0.4,0) -- (0.8,0);
\draw (0,0) circle (0.4cm);
\draw (0,-0.8) node {\footnotesize{$18$}};
\draw (1.2,-0.8) node {\footnotesize{$12$}};
\draw (2.4,0) circle (0.4cm);
\draw (2.4,-0.8) node {\footnotesize{$6$}};
\draw(1.2,0) circle (0.4cm);
\draw (-2.4,0) circle (0.4cm);
\draw (-2.4,-0.8) node {\footnotesize{$6$}};
\draw (-0.4,0) -- (-0.8,0);
\draw (1.6,0) -- (2,0);
\draw (-1.2,0) circle (0.4cm);
\draw (-1.2,-0.8) node {\footnotesize{$12$}};
\draw (-1.6,0) -- (-2,0);
\draw (-0.432,0.648) -- (-0.14,0.62);
\draw (-0.432,0.648) -- (-0.5,0.35);
\draw (-0.14,0.38) -- (-0.52,0.92);
\draw (-0.28,0.28) -- (-0.68,0.83);
\draw (-0.21,0.33) -- (-0.6,0.875);
\draw (-0.8,1.2) [red, fill=red!30] circle (0.4cm);
\draw (-0.8,2) node {\footnotesize{$1$}};
\draw (0.26,0.3) -- (0.65,0.83);
\draw (0.8,1.2) circle (0.4cm);
\draw (0.8,2) node {\footnotesize{$9$}};

\end{tikzpicture} & $16$ \\ \hline

\begin{tikzpicture}[scale=0.70]
\draw (0.4,0) -- (0.8,0);
\draw (0,0) circle (0.4cm);
\draw (0,-0.8) node {\footnotesize{$9$}};
\draw (1.6,0) -- (2,0);
\draw (2.4,0) circle (0.4cm);
\draw (1.2,-0.8) node {\footnotesize{$6$}};
\draw(1.2,0) circle (0.4cm);
\draw (2.4,-0.8) node {\footnotesize{$3$}};
\draw (-2.4,0) circle (0.4cm);
\draw (-2.4,-0.8) node {\footnotesize{$3$}};
\draw (-0.4,0) -- (-0.8,0);
\draw (-1.2,0) circle (0.4cm);
\draw (-1.2,-0.8) node {\footnotesize{$6$}};
\draw (-1.6,0) -- (-2,0);
\draw (0,0.4) -- (0,0.8);
\draw (0,1.2) circle (0.4cm);
\draw (0.8,1.2) node {\footnotesize{$6$}};
\draw (-0.08,1.6) -- (-0.08,2);
\draw (0.08,1.6) -- (0.08,2);
\draw (0,1.6) -- (0,2);
\draw (-0.2,1.7) -- (0,1.9);
\draw (0.2,1.7) -- (0,1.9);
\draw (0,2.4) [red, fill=red!30] circle (0.4cm);
\draw (0.8,2.4) node {\footnotesize{$1$}};
\draw (0,3) node {\footnotesize{$$}};

\end{tikzpicture} & $4$ \\ \hline
	\end{tabular}
	\caption{Exotic minimally unbalanced quivers of $E_6$-type with a triple laced edge.}
	\label{tab:E6seriesExoticTriple}
\end{table}

\clearpage
\newpage

\subsubsection{Exotic Minimally Unbalanced Quivers with $G$ of Type $E_7$}

We repeat the same program for exotic minimally unbalanced quivers with $G=E_7$. First, we distinguish whether the unbalanced node is connected via a double laced edge directed; outwards from the unbalanced node, table \ref{tab:E7seriesExoticDoubleOut}, or inwards with respect to the unbalanced node, table \ref{tab:E7seriesExoticDoubleIn}. Next, we turn to the quivers with the unbalanced node connected via a triple laced edge. The resulting quivers with a triple laced edge pointing outwards and inwards with respect to the unbalanced node are reported in table \ref{tab:E7seriesExoticTripleOut} and \ref{tab:E7seriesExoticTripleIn}, respectively.


\begin{table}
	\centering
	\begin{tabular}{|c|c|}
	\hline
	Quiver & Excess \\ \hline

\begin{tikzpicture}[scale=0.70]
\draw (0.4,0) -- (0.8,0);
\draw (2.8,0) -- (3.2,0);
\draw (0,0) circle (0.4cm);
\draw (0,-0.8) node {\footnotesize{$4$}};
\draw (1.6,0) -- (2,0);
\draw (2.4,0) circle (0.4cm);
\draw (1.2,-0.8) node {\footnotesize{$3$}};
\draw (3.6,0) circle (0.4cm);
\draw (3.6,-0.8) node {\footnotesize{$1$}};
\draw (-2.4,1.2) [red,fill=red!30] circle (0.4cm);
\draw (-2.4,2) node {\footnotesize{$1$}};
\draw (-2.32,0.4) -- (-2.32,0.8);
\draw (-2.48,0.4) -- (-2.48,0.8);
\draw (-2.4,0.5) -- (-2.6,0.7);
\draw (-2.4,0.5) -- (-2.2,0.7);
\draw(1.2,0) circle (0.4cm);
\draw (2.4,-0.8) node {\footnotesize{$2$}};
\draw (-2.4,0) circle (0.4cm);
\draw (-2.4,-0.8) node {\footnotesize{$2$}};
\draw (-0.4,0) -- (-0.8,0);
\draw (-1.2,0) circle (0.4cm);
\draw (-1.2,-0.8) node {\footnotesize{$3$}};
\draw (-1.6,0) -- (-2,0);
\draw (0,0.4) -- (0,0.8);
\draw (0,1.2) circle (0.4cm);
\draw (0,2) node {\footnotesize{$2$}};

\end{tikzpicture} & $2$\\ \hline	
	
\begin{tikzpicture}[scale=0.70]
\draw (0.4,0) -- (0.8,0);
\draw (2.8,0) -- (3.2,0);
\draw (0,0) circle (0.4cm);
\draw (0,-0.8) node {\footnotesize{$8$}};
\draw (1.6,0) -- (2,0);
\draw (2.4,0) circle (0.4cm);
\draw (1.2,-0.8) node {\footnotesize{$6$}};
\draw (3.6,0) circle (0.4cm);
\draw (3.6,-0.8) node {\footnotesize{$2$}};
\draw (-1.2,1.2) [red,fill=red!30] circle (0.4cm);
\draw (-1.2,2) node {\footnotesize{$1$}};
\draw (-1.12,0.4) -- (-1.12,0.8);
\draw (-1.28,0.4) -- (-1.28,0.8);
\draw (-1,0.7) -- (-1.2,0.5);
\draw (-1.4,0.7) -- (-1.2,0.5);
\draw(1.2,0) circle (0.4cm);
\draw (2.4,-0.8) node {\footnotesize{$4$}};
\draw (-2.4,0) circle (0.4cm);
\draw (-2.4,-0.8) node {\footnotesize{$3$}};
\draw (-0.4,0) -- (-0.8,0);
\draw (-1.2,0) circle (0.4cm);
\draw (-1.2,-0.8) node {\footnotesize{$6$}};
\draw (-1.6,0) -- (-2,0);
\draw (0,0.4) -- (0,0.8);
\draw (0,1.2) circle (0.4cm);
\draw (0,2) node {\footnotesize{$4$}};

\end{tikzpicture} & $10$\\ \hline	
	
\begin{tikzpicture}[scale=0.70]
\draw (0.4,0) -- (0.8,0);
\draw (2.8,0) -- (3.2,0);
\draw (0,0) circle (0.4cm);
\draw (0,-0.8) node {\footnotesize{$12$}};
\draw (1.6,0) -- (2,0);
\draw (2.4,0) circle (0.4cm);
\draw (1.2,-0.8) node {\footnotesize{$9$}};
\draw (3.6,0) circle (0.4cm);
\draw (3.6,-0.8) node {\footnotesize{$3$}};
\draw (0,2.4) [red,fill=red!30] circle (0.4cm);
\draw (0.8,2.4) node {\footnotesize{$2$}};
\draw (-0.08,1.6) -- (-0.08,2);
\draw (0.08,1.6) -- (0.08,2);
\draw (-0.2,1.9) -- (0,1.7);
\draw (0.2,1.9) -- (0,1.7);
\draw(1.2,0) circle (0.4cm);
\draw (2.4,-0.8) node {\footnotesize{$6$}};
\draw (-2.4,0) circle (0.4cm);
\draw (-2.4,-0.8) node {\footnotesize{$4$}};
\draw (-0.4,0) -- (-0.8,0);
\draw (-1.2,0) circle (0.4cm);
\draw (-1.2,-0.8) node {\footnotesize{$8$}};
\draw (-1.6,0) -- (-2,0);
\draw (0,0.4) -- (0,0.8);
\draw (0,1.2) circle (0.4cm);
\draw (0.8,1.2) node {\footnotesize{$7$}};
\draw (0,3) node {\footnotesize{$$}};

\end{tikzpicture} & $10$\\ \hline	

\begin{tikzpicture}[scale=0.70]
\draw (0.4,0) -- (0.8,0);
\draw (0,0) circle (0.4cm);
\draw (0,-0.8) node {\footnotesize{$12$}};
\draw (1.2,-0.8) node {\footnotesize{$9$}};
\draw (2.4,0) circle (0.4cm);
\draw (2.4,-0.8) node {\footnotesize{$6$}};
\draw(1.2,0) circle (0.4cm);
\draw (-2.4,0) circle (0.4cm);
\draw (-2.4,-0.8) node {\footnotesize{$4$}};
\draw (-0.4,0) -- (-0.8,0);
\draw (1.6,0) -- (2,0);
\draw (-1.2,0) circle (0.4cm);
\draw (-1.2,-0.8) node {\footnotesize{$8$}};
\draw (2.8,0) -- (3.2,0);
\draw (3.6,0) circle (0.4cm);
\draw (3.6,-0.8) node {\footnotesize{$3$}};
\draw (-1.6,0) -- (-2,0);
\draw (-0.36,0.56) -- (-0.64,0.56);
\draw (-0.36,0.56) -- (-0.288,0.8);
\draw (-0.14,0.38) -- (-0.52,0.92);
\draw (-0.26,0.3) -- (-0.65,0.82);
\draw (-0.8,1.2) [red, fill=red!30] circle (0.4cm);
\draw (-0.8,2) node {\footnotesize{$1$}};
\draw (0.26,0.3) -- (0.65,0.82);
\draw (0.8,1.2) circle (0.4cm);
\draw (0.8,2) node {\footnotesize{$6$}};

\end{tikzpicture} & $22$ \\ \hline	

\begin{tikzpicture}[scale=0.70]
\draw (0.4,0) -- (0.8,0);
\draw (2.8,0) -- (3.2,0);
\draw (0,0) circle (0.4cm);
\draw (0,-0.8) node {\footnotesize{$18$}};
\draw (1.6,0) -- (2,0);
\draw (2.4,0) circle (0.4cm);
\draw (1.2,-0.8) node {\footnotesize{$15$}};
\draw (3.6,0) circle (0.4cm);
\draw (3.6,-0.8) node {\footnotesize{$5$}};
\draw (1.2,1.2) [red,fill=red!30] circle (0.4cm);
\draw (1.2,2) node {\footnotesize{$1$}};
\draw(1.2,0) circle (0.4cm);
\draw (2.4,-0.8) node {\footnotesize{$10$}};
\draw (-2.4,0) circle (0.4cm);
\draw (-2.4,-0.8) node {\footnotesize{$6$}};
\draw (-0.4,0) -- (-0.8,0);
\draw (-1.2,0) circle (0.4cm);
\draw (-1.2,-0.8) node {\footnotesize{$12$}};
\draw (-1.6,0) -- (-2,0);
\draw (0,0.4) -- (0,0.8);
\draw (0,1.2) circle (0.4cm);
\draw (0,2) node {\footnotesize{$9$}};
\draw (1.12,0.4) -- (1.12,0.8);
\draw (1.28,0.4) -- (1.28,0.8);
\draw (1,0.7) -- (1.2,0.5);
\draw (1.4,0.7) -- (1.2,0.5);

\end{tikzpicture} & $28$ \\ \hline

\begin{tikzpicture}[scale=0.70]
\draw (0.4,0) -- (0.8,0);
\draw (2.8,0) -- (3.2,0);
\draw (0,0) circle (0.4cm);
\draw (0,-0.8) node {\footnotesize{$6$}};
\draw (1.6,0) -- (2,0);
\draw (2.4,0) circle (0.4cm);
\draw (1.2,-0.8) node {\footnotesize{$5$}};
\draw (3.6,0) circle (0.4cm);
\draw (3.6,-0.8) node {\footnotesize{$2$}};
\draw (2.4,1.2) [red,fill=red!30] circle (0.4cm);
\draw (2.4,2) node {\footnotesize{$1$}};
\draw(1.2,0) circle (0.4cm);
\draw (2.4,-0.8) node {\footnotesize{$4$}};
\draw (-2.4,0) circle (0.4cm);
\draw (-2.4,-0.8) node {\footnotesize{$2$}};
\draw (-0.4,0) -- (-0.8,0);
\draw (-1.2,0) circle (0.4cm);
\draw (-1.2,-0.8) node {\footnotesize{$4$}};
\draw (-1.6,0) -- (-2,0);
\draw (0,0.4) -- (0,0.8);
\draw (0,1.2) circle (0.4cm);
\draw (0,2) node {\footnotesize{$3$}};
\draw (2.32,0.4) -- (2.32,0.8);
\draw (2.48,0.4) -- (2.48,0.8);
\draw (2.4,0.5) -- (2.6,0.7);
\draw (2.4,0.5) -- (2.2,0.7);

\end{tikzpicture} & $6$ \\ \hline

\begin{tikzpicture}[scale=0.70]	
\draw (0.4,0) -- (0.8,0);
\draw (2.8,0) -- (3.2,0);
\draw (0,0) circle (0.4cm);
\draw (0,-0.8) node {\footnotesize{$6$}};
\draw (1.6,0) -- (2,0);
\draw (2.4,0) circle (0.4cm);
\draw (1.2,-0.8) node {\footnotesize{$5$}};
\draw (3.6,0) circle (0.4cm);
\draw (3.6,-0.8) node {\footnotesize{$3$}};
\draw (3.6,1.2) [red,fill=red!30] circle (0.4cm);
\draw (3.6,2) node {\footnotesize{$2$}};
\draw(1.2,0) circle (0.4cm);
\draw (2.4,-0.8) node {\footnotesize{$4$}};
\draw (-2.4,0) circle (0.4cm);
\draw (-2.4,-0.8) node {\footnotesize{$2$}};
\draw (-0.4,0) -- (-0.8,0);
\draw (-1.2,0) circle (0.4cm);
\draw (-1.2,-0.8) node {\footnotesize{$4$}};
\draw (-1.6,0) -- (-2,0);
\draw (0,0.4) -- (0,0.8);
\draw (0,1.2) circle (0.4cm);
\draw (0,2) node {\footnotesize{$3$}};
\draw (3.52,0.4) -- (3.52,0.8);
\draw (3.68,0.4) -- (3.68,0.8);
\draw (3.6,0.5) -- (3.4,0.7);
\draw (3.6,0.5) -- (3.8,0.7);

\end{tikzpicture} & $2$ \\ \hline
	\end{tabular}
	\caption{Exotic minimally unbalanced quivers of $E_7$-type with outward double laced edge.}
	\label{tab:E7seriesExoticDoubleOut}
\end{table}


\begin{table}
	\centering
	\begin{tabular}{|c|c|}
	\hline
	Quiver & Excess \\ \hline
	
\begin{tikzpicture}[scale=0.70]
\draw (0.4,0) -- (0.8,0);
\draw (2.8,0) -- (3.2,0);
\draw (0,0) circle (0.4cm);
\draw (0,-0.8) node {\footnotesize{$8$}};
\draw (1.6,0) -- (2,0);
\draw (2.4,0) circle (0.4cm);
\draw (1.2,-0.8) node {\footnotesize{$6$}};
\draw (3.6,0) circle (0.4cm);
\draw (3.6,-0.8) node {\footnotesize{$2$}};
\draw (-2.4,1.2) [red,fill=red!30] circle (0.4cm);
\draw (-2.4,2) node {\footnotesize{$1$}};
\draw (-2.32,0.4) -- (-2.32,0.8);
\draw (-2.48,0.4) -- (-2.48,0.8);
\draw (-2.4,0.7) -- (-2.6,0.5);
\draw (-2.4,0.7) -- (-2.2,0.5);
\draw(1.2,0) circle (0.4cm);
\draw (2.4,-0.8) node {\footnotesize{$4$}};
\draw (-2.4,0) circle (0.4cm);
\draw (-2.4,-0.8) node {\footnotesize{$4$}};
\draw (-0.4,0) -- (-0.8,0);
\draw (-1.2,0) circle (0.4cm);
\draw (-1.2,-0.8) node {\footnotesize{$6$}};
\draw (-1.6,0) -- (-2,0);
\draw (0,0.4) -- (0,0.8);
\draw (0,1.2) circle (0.4cm);
\draw (0,2) node {\footnotesize{$4$}};

\end{tikzpicture} & $2$\\ \hline	

\begin{tikzpicture}[scale=0.70]
\draw (0.4,0) -- (0.8,0);
\draw (2.8,0) -- (3.2,0);
\draw (0,0) circle (0.4cm);
\draw (0,-0.8) node {\footnotesize{$16$}};
\draw (1.6,0) -- (2,0);
\draw (2.4,0) circle (0.4cm);
\draw (1.2,-0.8) node {\footnotesize{$12$}};
\draw (3.6,0) circle (0.4cm);
\draw (3.6,-0.8) node {\footnotesize{$4$}};
\draw (-1.2,1.2) [red,fill=red!30] circle (0.4cm);
\draw (-1.2,2) node {\footnotesize{$1$}};
\draw (-1.12,0.4) -- (-1.12,0.8);
\draw (-1.28,0.4) -- (-1.28,0.8);
\draw (-1,0.5) -- (-1.2,0.7);
\draw (-1.4,0.5) -- (-1.2,0.7);
\draw(1.2,0) circle (0.4cm);
\draw (2.4,-0.8) node {\footnotesize{$8$}};
\draw (-2.4,0) circle (0.4cm);
\draw (-2.4,-0.8) node {\footnotesize{$6$}};
\draw (-0.4,0) -- (-0.8,0);
\draw (-1.2,0) circle (0.4cm);
\draw (-1.2,-0.8) node {\footnotesize{$12$}};
\draw (-1.6,0) -- (-2,0);
\draw (0,0.4) -- (0,0.8);
\draw (0,1.2) circle (0.4cm);
\draw (0,2) node {\footnotesize{$8$}};

\end{tikzpicture} & $10$\\ \hline	
	
\begin{tikzpicture}[scale=0.70]
\draw (0.4,0) -- (0.8,0);
\draw (2.8,0) -- (3.2,0);
\draw (0,0) circle (0.4cm);
\draw (0,-0.8) node {\footnotesize{$12$}};
\draw (1.6,0) -- (2,0);
\draw (2.4,0) circle (0.4cm);
\draw (1.2,-0.8) node {\footnotesize{$9$}};
\draw (3.6,0) circle (0.4cm);
\draw (3.6,-0.8) node {\footnotesize{$3$}};
\draw (0,2.4) [red,fill=red!30] circle (0.4cm);
\draw (0.8,2.4) node {\footnotesize{$1$}};
\draw (-0.08,1.6) -- (-0.08,2);
\draw (0.08,1.6) -- (0.08,2);
\draw (-0.2,1.7) -- (0,1.9);
\draw (0.2,1.7) -- (0,1.9);
\draw(1.2,0) circle (0.4cm);
\draw (2.4,-0.8) node {\footnotesize{$6$}};
\draw (-2.4,0) circle (0.4cm);
\draw (-2.4,-0.8) node {\footnotesize{$4$}};
\draw (-0.4,0) -- (-0.8,0);
\draw (-1.2,0) circle (0.4cm);
\draw (-1.2,-0.8) node {\footnotesize{$8$}};
\draw (-1.6,0) -- (-2,0);
\draw (0,0.4) -- (0,0.8);
\draw (0,1.2) circle (0.4cm);
\draw (0.8,1.2) node {\footnotesize{$7$}};
\draw (0,3) node {\footnotesize{$$}};

\end{tikzpicture} & $5$\\ \hline	

\begin{tikzpicture}[scale=0.70]
\draw (0.4,0) -- (0.8,0);
\draw (0,0) circle (0.4cm);
\draw (0,-0.8) node {\footnotesize{$24$}};
\draw (1.2,-0.8) node {\footnotesize{$18$}};
\draw (2.4,0) circle (0.4cm);
\draw (2.4,-0.8) node {\footnotesize{$12$}};
\draw(1.2,0) circle (0.4cm);
\draw (-2.4,0) circle (0.4cm);
\draw (-2.4,-0.8) node {\footnotesize{$8$}};
\draw (-0.4,0) -- (-0.8,0);
\draw (1.6,0) -- (2,0);
\draw (-1.2,0) circle (0.4cm);
\draw (-1.2,-0.8) node {\footnotesize{$16$}};
\draw (2.8,0) -- (3.2,0);
\draw (3.6,0) circle (0.4cm);
\draw (3.6,-0.8) node {\footnotesize{$6$}};
\draw (-1.6,0) -- (-2,0);
\draw (-0.14,0.38) -- (-0.52,0.92);
\draw (-0.432,0.648) -- (-0.14,0.62);
\draw (-0.432,0.648) -- (-0.5,0.35);
\draw (-0.26,0.3) -- (-0.65,0.82);
\draw (-0.8,1.2) [red, fill=red!30] circle (0.4cm);
\draw (-0.8,2) node {\footnotesize{$1$}};
\draw (0.26,0.3) -- (0.65,0.82);
\draw (0.8,1.2) circle (0.4cm);
\draw (0.8,2) node {\footnotesize{$12$}};

\end{tikzpicture} & $22$ \\ \hline	

\begin{tikzpicture}[scale=0.70]
\draw (0.4,0) -- (0.8,0);
\draw (2.8,0) -- (3.2,0);
\draw (0,0) circle (0.4cm);
\draw (0,-0.8) node {\footnotesize{$18$}};
\draw (1.6,0) -- (2,0);
\draw (2.4,0) circle (0.4cm);
\draw (1.2,-0.8) node {\footnotesize{$15$}};
\draw (3.6,0) circle (0.4cm);
\draw (3.6,-0.8) node {\footnotesize{$5$}};
\draw (1.2,1.2) [red,fill=red!30] circle (0.4cm);
\draw (1.2,2) node {\footnotesize{$1$}};
\draw(1.2,0) circle (0.4cm);
\draw (2.4,-0.8) node {\footnotesize{$10$}};
\draw (-2.4,0) circle (0.4cm);
\draw (-2.4,-0.8) node {\footnotesize{$6$}};
\draw (-0.4,0) -- (-0.8,0);
\draw (-1.2,0) circle (0.4cm);
\draw (-1.2,-0.8) node {\footnotesize{$12$}};
\draw (-1.6,0) -- (-2,0);
\draw (0,0.4) -- (0,0.8);
\draw (0,1.2) circle (0.4cm);
\draw (0,2) node {\footnotesize{$9$}};
\draw (1.12,0.4) -- (1.12,0.8);
\draw (1.28,0.4) -- (1.28,0.8);
\draw (1,0.5) -- (1.2,0.7);
\draw (1.4,0.5) -- (1.2,0.7);

\end{tikzpicture} & $13$ \\ \hline

\begin{tikzpicture}[scale=0.70]
\draw (0.4,0) -- (0.8,0);
\draw (2.8,0) -- (3.2,0);
\draw (0,0) circle (0.4cm);
\draw (0,-0.8) node {\footnotesize{$12$}};
\draw (1.6,0) -- (2,0);
\draw (2.4,0) circle (0.4cm);
\draw (1.2,-0.8) node {\footnotesize{$10$}};
\draw (3.6,0) circle (0.4cm);
\draw (3.6,-0.8) node {\footnotesize{$4$}};
\draw (2.4,1.2) [red,fill=red!30] circle (0.4cm);
\draw (2.4,2) node {\footnotesize{$1$}};
\draw(1.2,0) circle (0.4cm);
\draw (2.4,-0.8) node {\footnotesize{$8$}};
\draw (-2.4,0) circle (0.4cm);
\draw (-2.4,-0.8) node {\footnotesize{$4$}};
\draw (-0.4,0) -- (-0.8,0);
\draw (-1.2,0) circle (0.4cm);
\draw (-1.2,-0.8) node {\footnotesize{$8$}};
\draw (-1.6,0) -- (-2,0);
\draw (0,0.4) -- (0,0.8);
\draw (0,1.2) circle (0.4cm);
\draw (0,2) node {\footnotesize{$6$}};
\draw (2.32,0.4) -- (2.32,0.8);
\draw (2.48,0.4) -- (2.48,0.8);
\draw (2.4,0.7) -- (2.6,0.5);
\draw (2.4,0.7) -- (2.2,0.5);

\end{tikzpicture} & $6$ \\ \hline

\begin{tikzpicture}[scale=0.70]	
\draw (0.4,0) -- (0.8,0);
\draw (2.8,0) -- (3.2,0);
\draw (0,0) circle (0.4cm);
\draw (0,-0.8) node {\footnotesize{$6$}};
\draw (1.6,0) -- (2,0);
\draw (2.4,0) circle (0.4cm);
\draw (1.2,-0.8) node {\footnotesize{$5$}};
\draw (3.6,0) circle (0.4cm);
\draw (3.6,-0.8) node {\footnotesize{$3$}};
\draw (3.6,1.2) [red,fill=red!30] circle (0.4cm);
\draw (3.6,2) node {\footnotesize{$1$}};
\draw(1.2,0) circle (0.4cm);
\draw (2.4,-0.8) node {\footnotesize{$4$}};
\draw (-2.4,0) circle (0.4cm);
\draw (-2.4,-0.8) node {\footnotesize{$2$}};
\draw (-0.4,0) -- (-0.8,0);
\draw (-1.2,0) circle (0.4cm);
\draw (-1.2,-0.8) node {\footnotesize{$4$}};
\draw (-1.6,0) -- (-2,0);
\draw (0,0.4) -- (0,0.8);
\draw (0,1.2) circle (0.4cm);
\draw (0,2) node {\footnotesize{$3$}};
\draw (3.52,0.4) -- (3.52,0.8);
\draw (3.68,0.4) -- (3.68,0.8);
\draw (3.6,0.7) -- (3.4,0.5);
\draw (3.6,0.7) -- (3.8,0.5);

\end{tikzpicture} & $1$ \\ \hline
	\end{tabular}
	\caption{Exotic minimally unbalanced quivers of $E_7$-type with inward double laced edge.}
	\label{tab:E7seriesExoticDoubleIn}
\end{table}

\clearpage


\begin{table}
	\centering
	\begin{tabular}{|c|c|}
	\hline
	Quiver & Excess \\ \hline
	
\begin{tikzpicture}[scale=0.70]
\draw (0.4,0) -- (0.8,0);
\draw (2.8,0) -- (3.2,0);
\draw (0,0) circle (0.4cm);
\draw (0,-0.8) node {\footnotesize{$4$}};
\draw (1.6,0) -- (2,0);
\draw (2.4,0) circle (0.4cm);
\draw (1.2,-0.8) node {\footnotesize{$3$}};
\draw (3.6,0) circle (0.4cm);
\draw (3.6,-0.8) node {\footnotesize{$1$}};
\draw (-2.4,1.2) [red,fill=red!30] circle (0.4cm);
\draw (-2.4,2) node {\footnotesize{$1$}};
\draw (-2.32,0.4) -- (-2.32,0.8);
\draw (-2.48,0.4) -- (-2.48,0.8);
\draw (-2.4,0.4) -- (-2.4,0.8);
\draw (-2.4,0.5) -- (-2.6,0.7);
\draw (-2.4,0.5) -- (-2.2,0.7);
\draw(1.2,0) circle (0.4cm);
\draw (2.4,-0.8) node {\footnotesize{$2$}};
\draw (-2.4,0) circle (0.4cm);
\draw (-2.4,-0.8) node {\footnotesize{$2$}};
\draw (-0.4,0) -- (-0.8,0);
\draw (-1.2,0) circle (0.4cm);
\draw (-1.2,-0.8) node {\footnotesize{$3$}};
\draw (-1.6,0) -- (-2,0);
\draw (0,0.4) -- (0,0.8);
\draw (0,1.2) circle (0.4cm);
\draw (0,2) node {\footnotesize{$2$}};

\end{tikzpicture} & $4$\\ \hline	

\begin{tikzpicture}[scale=0.70]
\draw (0.4,0) -- (0.8,0);
\draw (2.8,0) -- (3.2,0);
\draw (0,0) circle (0.4cm);
\draw (0,-0.8) node {\footnotesize{$8$}};
\draw (1.6,0) -- (2,0);
\draw (2.4,0) circle (0.4cm);
\draw (1.2,-0.8) node {\footnotesize{$6$}};
\draw (3.6,0) circle (0.4cm);
\draw (3.6,-0.8) node {\footnotesize{$2$}};
\draw (-1.2,1.2) [red,fill=red!30] circle (0.4cm);
\draw (-1.2,2) node {\footnotesize{$1$}};
\draw (-1.12,0.4) -- (-1.12,0.8);
\draw (-1.28,0.4) -- (-1.28,0.8);
\draw (-1.2,0.4) -- (-1.2,0.8);
\draw (-1,0.7) -- (-1.2,0.5);
\draw (-1.4,0.7) -- (-1.2,0.5);
\draw(1.2,0) circle (0.4cm);
\draw (2.4,-0.8) node {\footnotesize{$4$}};
\draw (-2.4,0) circle (0.4cm);
\draw (-2.4,-0.8) node {\footnotesize{$3$}};
\draw (-0.4,0) -- (-0.8,0);
\draw (-1.2,0) circle (0.4cm);
\draw (-1.2,-0.8) node {\footnotesize{$6$}};
\draw (-1.6,0) -- (-2,0);
\draw (0,0.4) -- (0,0.8);
\draw (0,1.2) circle (0.4cm);
\draw (0,2) node {\footnotesize{$4$}};

\end{tikzpicture} & $16$\\ \hline	
	
\begin{tikzpicture}[scale=0.70]
\draw (0.4,0) -- (0.8,0);
\draw (2.8,0) -- (3.2,0);
\draw (0,0) circle (0.4cm);
\draw (0,-0.8) node {\footnotesize{$12$}};
\draw (1.6,0) -- (2,0);
\draw (2.4,0) circle (0.4cm);
\draw (1.2,-0.8) node {\footnotesize{$9$}};
\draw (3.6,0) circle (0.4cm);
\draw (3.6,-0.8) node {\footnotesize{$3$}};
\draw (0,2.4) [red,fill=red!30] circle (0.4cm);
\draw (0.8,2.4) node {\footnotesize{$2$}};
\draw (-0.08,1.6) -- (-0.08,2);
\draw (0.08,1.6) -- (0.08,2);
\draw (0,1.6) -- (0,2);
\draw (-0.2,1.9) -- (0,1.7);
\draw (0.2,1.9) -- (0,1.7);
\draw(1.2,0) circle (0.4cm);
\draw (2.4,-0.8) node {\footnotesize{$6$}};
\draw (-2.4,0) circle (0.4cm);
\draw (-2.4,-0.8) node {\footnotesize{$4$}};
\draw (-0.4,0) -- (-0.8,0);
\draw (-1.2,0) circle (0.4cm);
\draw (-1.2,-0.8) node {\footnotesize{$8$}};
\draw (-1.6,0) -- (-2,0);
\draw (0,0.4) -- (0,0.8);
\draw (0,1.2) circle (0.4cm);
\draw (0.8,1.2) node {\footnotesize{$7$}};
\draw (0,3) node {\footnotesize{$$}};

\end{tikzpicture} & $17$\\ \hline	
	
\begin{tikzpicture}[scale=0.70]
\draw (0.4,0) -- (0.8,0);
\draw (0,0) circle (0.4cm);
\draw (0,-0.8) node {\footnotesize{$12$}};
\draw (1.2,-0.8) node {\footnotesize{$9$}};
\draw (2.4,0) circle (0.4cm);
\draw (2.4,-0.8) node {\footnotesize{$6$}};
\draw(1.2,0) circle (0.4cm);
\draw (-2.4,0) circle (0.4cm);
\draw (-2.4,-0.8) node {\footnotesize{$4$}};
\draw (-0.4,0) -- (-0.8,0);
\draw (1.6,0) -- (2,0);
\draw (-1.2,0) circle (0.4cm);
\draw (-1.2,-0.8) node {\footnotesize{$8$}};
\draw (2.8,0) -- (3.2,0);
\draw (3.6,0) circle (0.4cm);
\draw (3.6,-0.8) node {\footnotesize{$3$}};
\draw (-1.6,0) -- (-2,0);
\draw (-0.36,0.56) -- (-0.64,0.56);
\draw (-0.36,0.56) -- (-0.288,0.8);
\draw (-0.14,0.38) -- (-0.52,0.92);
\draw (-0.27,0.29) -- (-0.65,0.82);
\draw (-0.206,0.332) -- (-0.595,0.88);
\draw (-0.8,1.2) [red, fill=red!30] circle (0.4cm);
\draw (-0.8,2) node {\footnotesize{$1$}};
\draw (0.26,0.3) -- (0.65,0.82);
\draw (0.8,1.2) circle (0.4cm);
\draw (0.8,2) node {\footnotesize{$6$}};

\end{tikzpicture} & $34$ \\ \hline	

\begin{tikzpicture}[scale=0.70]
\draw (0.4,0) -- (0.8,0);
\draw (2.8,0) -- (3.2,0);
\draw (0,0) circle (0.4cm);
\draw (0,-0.8) node {\footnotesize{$18$}};
\draw (1.6,0) -- (2,0);
\draw (2.4,0) circle (0.4cm);
\draw (1.2,-0.8) node {\footnotesize{$15$}};
\draw (3.6,0) circle (0.4cm);
\draw (3.6,-0.8) node {\footnotesize{$5$}};
\draw (1.2,1.2) [red,fill=red!30] circle (0.4cm);
\draw (1.2,2) node {\footnotesize{$1$}};
\draw(1.2,0) circle (0.4cm);
\draw (2.4,-0.8) node {\footnotesize{$10$}};
\draw (-2.4,0) circle (0.4cm);
\draw (-2.4,-0.8) node {\footnotesize{$6$}};
\draw (-0.4,0) -- (-0.8,0);
\draw (-1.2,0) circle (0.4cm);
\draw (-1.2,-0.8) node {\footnotesize{$12$}};
\draw (-1.6,0) -- (-2,0);
\draw (0,0.4) -- (0,0.8);
\draw (0,1.2) circle (0.4cm);
\draw (0,2) node {\footnotesize{$9$}};
\draw (1.12,0.4) -- (1.12,0.8);
\draw (1.28,0.4) -- (1.28,0.8);
\draw (1.2,0.4) -- (1.2,0.8);
\draw (1,0.7) -- (1.2,0.5);
\draw (1.4,0.7) -- (1.2,0.5);

\end{tikzpicture} & $43$ \\ \hline

\begin{tikzpicture}[scale=0.70]
\draw (0.4,0) -- (0.8,0);
\draw (2.8,0) -- (3.2,0);
\draw (0,0) circle (0.4cm);
\draw (0,-0.8) node {\footnotesize{$6$}};
\draw (1.6,0) -- (2,0);
\draw (2.4,0) circle (0.4cm);
\draw (1.2,-0.8) node {\footnotesize{$5$}};
\draw (3.6,0) circle (0.4cm);
\draw (3.6,-0.8) node {\footnotesize{$2$}};
\draw (2.4,1.2) [red,fill=red!30] circle (0.4cm);
\draw (2.4,2) node {\footnotesize{$1$}};
\draw(1.2,0) circle (0.4cm);
\draw (2.4,-0.8) node {\footnotesize{$4$}};
\draw (-2.4,0) circle (0.4cm);
\draw (-2.4,-0.8) node {\footnotesize{$2$}};
\draw (-0.4,0) -- (-0.8,0);
\draw (-1.2,0) circle (0.4cm);
\draw (-1.2,-0.8) node {\footnotesize{$4$}};
\draw (-1.6,0) -- (-2,0);
\draw (0,0.4) -- (0,0.8);
\draw (0,1.2) circle (0.4cm);
\draw (0,2) node {\footnotesize{$3$}};
\draw (2.32,0.4) -- (2.32,0.8);
\draw (2.48,0.4) -- (2.48,0.8);
\draw (2.4,0.4) -- (2.4,0.8);
\draw (2.4,0.5) -- (2.6,0.7);
\draw (2.4,0.5) -- (2.2,0.7);

\end{tikzpicture} & $10$ \\ \hline

\begin{tikzpicture}[scale=0.70]	
\draw (0.4,0) -- (0.8,0);
\draw (2.8,0) -- (3.2,0);
\draw (0,0) circle (0.4cm);
\draw (0,-0.8) node {\footnotesize{$6$}};
\draw (1.6,0) -- (2,0);
\draw (2.4,0) circle (0.4cm);
\draw (1.2,-0.8) node {\footnotesize{$5$}};
\draw (3.6,0) circle (0.4cm);
\draw (3.6,-0.8) node {\footnotesize{$3$}};
\draw (3.6,1.2) [red,fill=red!30] circle (0.4cm);
\draw (3.6,2) node {\footnotesize{$2$}};
\draw(1.2,0) circle (0.4cm);
\draw (2.4,-0.8) node {\footnotesize{$4$}};
\draw (-2.4,0) circle (0.4cm);
\draw (-2.4,-0.8) node {\footnotesize{$2$}};
\draw (-0.4,0) -- (-0.8,0);
\draw (-1.2,0) circle (0.4cm);
\draw (-1.2,-0.8) node {\footnotesize{$4$}};
\draw (-1.6,0) -- (-2,0);
\draw (0,0.4) -- (0,0.8);
\draw (0,1.2) circle (0.4cm);
\draw (0,2) node {\footnotesize{$3$}};
\draw (3.52,0.4) -- (3.52,0.8);
\draw (3.68,0.4) -- (3.68,0.8);
\draw (3.6,0.4) -- (3.6,0.8);
\draw (3.6,0.5) -- (3.4,0.7);
\draw (3.6,0.5) -- (3.8,0.7);

\end{tikzpicture} & $5$ \\ \hline
	\end{tabular}
	\caption{Exotic minimally unbalanced quivers of $E_7$-type with outward triple laced edge.}
	\label{tab:E7seriesExoticTripleOut}
\end{table}


\begin{table}
	\centering
	\begin{tabular}{|c|c|}
	\hline
Quiver & Excess \\ \hline
	
\begin{tikzpicture}[scale=0.70]
\draw (0.4,0) -- (0.8,0);
\draw (2.8,0) -- (3.2,0);
\draw (0,0) circle (0.4cm);
\draw (0,-0.8) node {\footnotesize{$12$}};
\draw (1.6,0) -- (2,0);
\draw (2.4,0) circle (0.4cm);
\draw (1.2,-0.8) node {\footnotesize{$9$}};
\draw (3.6,0) circle (0.4cm);
\draw (3.6,-0.8) node {\footnotesize{$3$}};
\draw (-2.4,1.2) [red,fill=red!30] circle (0.4cm);
\draw (-2.4,2) node {\footnotesize{$1$}};
\draw (-2.32,0.4) -- (-2.32,0.8);
\draw (-2.48,0.4) -- (-2.48,0.8);
\draw (-2.4,0.4) -- (-2.4,0.8);
\draw (-2.4,0.7) -- (-2.6,0.5);
\draw (-2.4,0.7) -- (-2.2,0.5);
\draw(1.2,0) circle (0.4cm);
\draw (2.4,-0.8) node {\footnotesize{$6$}};
\draw (-2.4,0) circle (0.4cm);
\draw (-2.4,-0.8) node {\footnotesize{$6$}};
\draw (-0.4,0) -- (-0.8,0);
\draw (-1.2,0) circle (0.4cm);
\draw (-1.2,-0.8) node {\footnotesize{$9$}};
\draw (-1.6,0) -- (-2,0);
\draw (0,0.4) -- (0,0.8);
\draw (0,1.2) circle (0.4cm);
\draw (0,2) node {\footnotesize{$6$}};

\end{tikzpicture} & $4$\\ \hline	
	
\begin{tikzpicture}[scale=0.70]
\draw (0.4,0) -- (0.8,0);
\draw (2.8,0) -- (3.2,0);
\draw (0,0) circle (0.4cm);
\draw (0,-0.8) node {\footnotesize{$24$}};
\draw (1.6,0) -- (2,0);
\draw (2.4,0) circle (0.4cm);
\draw (1.2,-0.8) node {\footnotesize{$18$}};
\draw (3.6,0) circle (0.4cm);
\draw (3.6,-0.8) node {\footnotesize{$6$}};
\draw (-1.2,1.2) [red,fill=red!30] circle (0.4cm);
\draw (-1.2,2) node {\footnotesize{$1$}};
\draw (-1.12,0.4) -- (-1.12,0.8);
\draw (-1.28,0.4) -- (-1.28,0.8);
\draw (-1.2,0.4) -- (-1.2,0.8);
\draw (-1,0.5) -- (-1.2,0.7);
\draw (-1.4,0.5) -- (-1.2,0.7);
\draw(1.2,0) circle (0.4cm);
\draw (2.4,-0.8) node {\footnotesize{$12$}};
\draw (-2.4,0) circle (0.4cm);
\draw (-2.4,-0.8) node {\footnotesize{$9$}};
\draw (-0.4,0) -- (-0.8,0);
\draw (-1.2,0) circle (0.4cm);
\draw (-1.2,-0.8) node {\footnotesize{$18$}};
\draw (-1.6,0) -- (-2,0);
\draw (0,0.4) -- (0,0.8);
\draw (0,1.2) circle (0.4cm);
\draw (0,2) node {\footnotesize{$12$}};

\end{tikzpicture} & $16$\\ \hline	
	
\begin{tikzpicture}[scale=0.70]
\draw (0.4,0) -- (0.8,0);
\draw (2.8,0) -- (3.2,0);
\draw (0,0) circle (0.4cm);
\draw (0,-0.8) node {\footnotesize{$36$}};
\draw (1.6,0) -- (2,0);
\draw (2.4,0) circle (0.4cm);
\draw (1.2,-0.8) node {\footnotesize{$27$}};
\draw (3.6,0) circle (0.4cm);
\draw (3.6,-0.8) node {\footnotesize{$9$}};
\draw (0,2.4) [red,fill=red!30] circle (0.4cm);
\draw (0.8,2.4) node {\footnotesize{$2$}};
\draw (-0.08,1.6) -- (-0.08,2);
\draw (0.08,1.6) -- (0.08,2);
\draw (0,1.6) -- (0,2);
\draw (-0.2,1.7) -- (0,1.9);
\draw (0.2,1.7) -- (0,1.9);
\draw(1.2,0) circle (0.4cm);
\draw (2.4,-0.8) node {\footnotesize{$18$}};
\draw (-2.4,0) circle (0.4cm);
\draw (-2.4,-0.8) node {\footnotesize{$12$}};
\draw (-0.4,0) -- (-0.8,0);
\draw (-1.2,0) circle (0.4cm);
\draw (-1.2,-0.8) node {\footnotesize{$24$}};
\draw (-1.6,0) -- (-2,0);
\draw (0,0.4) -- (0,0.8);
\draw (0,1.2) circle (0.4cm);
\draw (0.8,1.2) node {\footnotesize{$21$}};
\draw (0,3) node {\footnotesize{$$}};

\end{tikzpicture} & $17$\\ \hline	
	
\begin{tikzpicture}[scale=0.70]
\draw (0.4,0) -- (0.8,0);
\draw (0,0) circle (0.4cm);
\draw (0,-0.8) node {\footnotesize{$36$}};
\draw (1.2,-0.8) node {\footnotesize{$27$}};
\draw (2.4,0) circle (0.4cm);
\draw (2.4,-0.8) node {\footnotesize{$18$}};
\draw(1.2,0) circle (0.4cm);
\draw (-2.4,0) circle (0.4cm);
\draw (-2.4,-0.8) node {\footnotesize{$12$}};
\draw (-0.4,0) -- (-0.8,0);
\draw (1.6,0) -- (2,0);
\draw (-1.2,0) circle (0.4cm);
\draw (-1.2,-0.8) node {\footnotesize{$24$}};
\draw (2.8,0) -- (3.2,0);
\draw (3.6,0) circle (0.4cm);
\draw (3.6,-0.8) node {\footnotesize{$9$}};
\draw (-1.6,0) -- (-2,0);
\draw (-0.14,0.38) -- (-0.52,0.92);
\draw (-0.27,0.29) -- (-0.65,0.82);
\draw (-0.206,0.332) -- (-0.595,0.88);
\draw (-0.432,0.648) -- (-0.14,0.62);
\draw (-0.432,0.648) -- (-0.5,0.35);
\draw (-0.8,1.2) [red, fill=red!30] circle (0.4cm);
\draw (-0.8,2) node {\footnotesize{$1$}};
\draw (0.26,0.3) -- (0.65,0.82);
\draw (0.8,1.2) circle (0.4cm);
\draw (0.8,2) node {\footnotesize{$18$}};

\end{tikzpicture} & $34$ \\ \hline	

\begin{tikzpicture}[scale=0.70]
\draw (0.4,0) -- (0.8,0);
\draw (2.8,0) -- (3.2,0);
\draw (0,0) circle (0.4cm);
\draw (0,-0.8) node {\footnotesize{$54$}};
\draw (1.6,0) -- (2,0);
\draw (2.4,0) circle (0.4cm);
\draw (1.2,-0.8) node {\footnotesize{$45$}};
\draw (3.6,0) circle (0.4cm);
\draw (3.6,-0.8) node {\footnotesize{$15$}};
\draw (1.2,1.2) [red,fill=red!30] circle (0.4cm);
\draw (1.2,2) node {\footnotesize{$2$}};
\draw(1.2,0) circle (0.4cm);
\draw (2.4,-0.8) node {\footnotesize{$30$}};
\draw (-2.4,0) circle (0.4cm);
\draw (-2.4,-0.8) node {\footnotesize{$18$}};
\draw (-0.4,0) -- (-0.8,0);
\draw (-1.2,0) circle (0.4cm);
\draw (-1.2,-0.8) node {\footnotesize{$36$}};
\draw (-1.6,0) -- (-2,0);
\draw (0,0.4) -- (0,0.8);
\draw (0,1.2) circle (0.4cm);
\draw (0,2) node {\footnotesize{$27$}};
\draw (1.12,0.4) -- (1.12,0.8);
\draw (1.28,0.4) -- (1.28,0.8);
\draw (1.2,0.4) -- (1.2,0.8);
\draw (1,0.5) -- (1.2,0.7);
\draw (1.4,0.5) -- (1.2,0.7);

\end{tikzpicture} & $41$ \\ \hline

\begin{tikzpicture}[scale=0.70]
\draw (0.4,0) -- (0.8,0);
\draw (2.8,0) -- (3.2,0);
\draw (0,0) circle (0.4cm);
\draw (0,-0.8) node {\footnotesize{$18$}};
\draw (1.6,0) -- (2,0);
\draw (2.4,0) circle (0.4cm);
\draw (1.2,-0.8) node {\footnotesize{$15$}};
\draw (3.6,0) circle (0.4cm);
\draw (3.6,-0.8) node {\footnotesize{$6$}};
\draw (2.4,1.2) [red,fill=red!30] circle (0.4cm);
\draw (2.4,2) node {\footnotesize{$1$}};
\draw(1.2,0) circle (0.4cm);
\draw (2.4,-0.8) node {\footnotesize{$12$}};
\draw (-2.4,0) circle (0.4cm);
\draw (-2.4,-0.8) node {\footnotesize{$6$}};
\draw (-0.4,0) -- (-0.8,0);
\draw (-1.2,0) circle (0.4cm);
\draw (-1.2,-0.8) node {\footnotesize{$12$}};
\draw (-1.6,0) -- (-2,0);
\draw (0,0.4) -- (0,0.8);
\draw (0,1.2) circle (0.4cm);
\draw (0,2) node {\footnotesize{$9$}};
\draw (2.32,0.4) -- (2.32,0.8);
\draw (2.48,0.4) -- (2.48,0.8);
\draw (2.4,0.4) -- (2.4,0.8);
\draw (2.4,0.7) -- (2.6,0.5);
\draw (2.4,0.7) -- (2.2,0.5);

\end{tikzpicture} & $10$ \\ \hline

\begin{tikzpicture}[scale=0.70]	
\draw (0.4,0) -- (0.8,0);
\draw (2.8,0) -- (3.2,0);
\draw (0,0) circle (0.4cm);
\draw (0,-0.8) node {\footnotesize{$18$}};
\draw (1.6,0) -- (2,0);
\draw (2.4,0) circle (0.4cm);
\draw (1.2,-0.8) node {\footnotesize{$15$}};
\draw (3.6,0) circle (0.4cm);
\draw (3.6,-0.8) node {\footnotesize{$9$}};
\draw (3.6,1.2) [red,fill=red!30] circle (0.4cm);
\draw (3.6,2) node {\footnotesize{$2$}};
\draw(1.2,0) circle (0.4cm);
\draw (2.4,-0.8) node {\footnotesize{$12$}};
\draw (-2.4,0) circle (0.4cm);
\draw (-2.4,-0.8) node {\footnotesize{$6$}};
\draw (-0.4,0) -- (-0.8,0);
\draw (-1.2,0) circle (0.4cm);
\draw (-1.2,-0.8) node {\footnotesize{$12$}};
\draw (-1.6,0) -- (-2,0);
\draw (0,0.4) -- (0,0.8);
\draw (0,1.2) circle (0.4cm);
\draw (0,2) node {\footnotesize{$9$}};
\draw (3.52,0.4) -- (3.52,0.8);
\draw (3.68,0.4) -- (3.68,0.8);
\draw (3.6,0.4) -- (3.6,0.8);
\draw (3.6,0.7) -- (3.4,0.5);
\draw (3.6,0.7) -- (3.8,0.5);

\end{tikzpicture} & $5$ \\ \hline
	\end{tabular}
	\caption{Exotic minimally unbalanced quivers of $E_7$-type with inward triple laced edge.}
	\label{tab:E7seriesExoticTripleIn}
\end{table}

\clearpage
\newpage

\subsubsection{Exotic Minimally Unbalanced Quivers with $G$ of Type $E_8$}

Finally, we present all exotic minimally unbalanced quivers with $G=E_8$. Tables \ref{tab:E8seriesExoticDoubleOut} and \ref{tab:E8seriesExoticDoubleIn} contain quivers with the unbalanced node connected by a double laced edge pointing outwards and inwards from the unbalanced node, respectively. Tables \ref{tab:E8seriesExoticTripleOut} and \ref{tab:E8seriesExoticTripleIn} collect quivers with a triple laced edge directed outwards and inwards from the unbalanced node, respectively.

\begin{table}
	\centering
	\begin{tabular}{|c|c|}
	\hline
	Quiver & Excess \\ \hline
	
\begin{tikzpicture}[scale=0.70]
\draw (0.4,0) -- (0.8,0);
\draw (0,0) circle (0.4cm);
\draw (0,-0.8) node {\footnotesize{$8$}};
\draw (1.6,0) -- (2,0);
\draw (2.4,0) circle (0.4cm);
\draw (1.2,-0.8) node {\footnotesize{$6$}};
\draw(1.2,0) circle (0.4cm);
\draw (2.4,-0.8) node {\footnotesize{$4$}};
\draw (-2.4,0) circle (0.4cm);
\draw (-2.4,-0.8) node {\footnotesize{$7$}};
\draw (-2.8,0) -- (-3.2,0);
\draw (-3.6,0) circle (0.4cm);
\draw (-3.6,-0.8) node {\footnotesize{$4$}};
\draw (2.8,0) -- (3.2,0);
\draw (3.6,0) circle (0.4cm);
\draw (3.6,-0.8) node {\footnotesize{$2$}};
\draw (-0.4,0) -- (-0.8,0);
\draw (-1.2,0) circle (0.4cm);
\draw (-1.2,-0.8) node {\footnotesize{$10$}};
\draw (-1.6,0) -- (-2,0);
\draw (-1.2,0.4) -- (-1.2,0.8);
\draw (-1.2,1.2) circle (0.4cm);
\draw (-1.2,2) node {\footnotesize{$5$}};
\draw (-3.6,1.2) [red,fill=red!30] circle  (0.4cm);
\draw (-3.6,2) node {\footnotesize{$1$}};
\draw (-3.52,0.4) -- (-3.52,0.8);
\draw (-3.68,0.4) -- (-3.68,0.8);
\draw (-3.4,0.7) -- (-3.6,0.5);
\draw (-3.8,0.7) -- (-3.6,0.5);

\end{tikzpicture} & $6$\\ \hline	

\begin{tikzpicture}[scale=0.70]
\draw (0.4,0) -- (0.8,0);
\draw (0,0) circle (0.4cm);
\draw (-3.6,-0.8) node {\footnotesize{$7$}};
\draw (-2.4,-0.8) node {\footnotesize{$14$}};
\draw (-1.2,-0.8) node {\footnotesize{$20$}};
\draw (0,-0.8) node     {\footnotesize{$10$}};
\draw (1.2,-0.8) node {\footnotesize{$12$}};
\draw (2.4,-0.8) node {\footnotesize{$8$}};
\draw (3.6,-0.8) node {\footnotesize{$4$}};
\draw (-1.2,2) node {\footnotesize{$10$}};
\draw (1.6,0) -- (2,0);
\draw (2.4,0) circle (0.4cm);
\draw(1.2,0) circle (0.4cm);
\draw (-2.4,0) circle (0.4cm);
\draw (-2.8,0) -- (-3.2,0);
\draw (-3.6,0) circle (0.4cm);
\draw (2.8,0) -- (3.2,0);
\draw (3.6,0) circle (0.4cm);
\draw (-0.4,0) -- (-0.8,0);
\draw (-1.2,0) circle (0.4cm);
\draw (-1.6,0) -- (-2,0);
\draw (-1.2,0.4) -- (-1.2,0.8);
\draw (-1.2,1.2) circle (0.4cm);
\draw (-2.4,1.2) [red,fill=red!30] circle  (0.4cm);
\draw (-2.4,2) node {\footnotesize{$1$}};
\draw (-2.32,0.4) -- (-2.32,0.8);
\draw (-2.48,0.4) -- (-2.48,0.8);
\draw (-2.2,0.7) -- (-2.4,0.5);
\draw (-2.6,0.7) -- (-2.4,0.5);

\end{tikzpicture} & $26$\\ \hline	

\begin{tikzpicture}[scale=0.70]
\draw (0.4,0) -- (0.8,0);
\draw (0,0) circle (0.4cm);
\draw (-3.6,-0.8) node {\footnotesize{$10$}};
\draw (-2.4,-0.8) node {\footnotesize{$20$}};
\draw (-1.2,-0.8) node {\footnotesize{$30$}};
\draw (0,-0.8) node     {\footnotesize{$24$}};
\draw (1.2,-0.8) node {\footnotesize{$18$}};
\draw (2.4,-0.8) node {\footnotesize{$12$}};
\draw (3.6,-0.8) node {\footnotesize{$6$}};
\draw (1.6,0) -- (2,0);
\draw (2.4,0) circle (0.4cm);
\draw(1.2,0) circle (0.4cm);
\draw (-2.4,0) circle (0.4cm);
\draw (-2.8,0) -- (-3.2,0);
\draw (-3.6,0) circle (0.4cm);
\draw (2.8,0) -- (3.2,0);
\draw (3.6,0) circle (0.4cm);
\draw (-0.4,0) -- (-0.8,0);
\draw (-1.6,0) -- (-2,0);
\draw (-1.2,0) circle (0.4cm);
\draw (-2,1.2) [red,fill=red!30] circle  (0.4cm);
\draw (-2,2) node {\footnotesize{$1$}};
\draw (-1.56,0.56) -- (-1.84,0.56);
\draw (-1.56,0.56) -- (-1.488,0.8);
\draw (-1.34,0.38) -- (-1.72,0.92);
\draw (-1.46,0.3) -- (-1.85,0.82);
\draw (-0.4,1.2) circle (0.4cm);
\draw (-0.4,2) node {\footnotesize{$15$}};
\draw (-1,0.334) -- (-0.6,0.865);

\end{tikzpicture} & $58$\\ \hline	

\begin{tikzpicture}[scale=0.70]
\draw (0.4,0) -- (0.8,0);
\draw (0,0) circle (0.4cm);
\draw (-3.6,-0.8) node {\footnotesize{$5$}};
\draw (-2.4,-0.8) node {\footnotesize{$10$}};
\draw (-1.2,-0.8) node {\footnotesize{$15$}};
\draw (0,-0.8) node     {\footnotesize{$12$}};
\draw (1.2,-0.8) node {\footnotesize{$9$}};
\draw (2.4,-0.8) node {\footnotesize{$6$}};
\draw (3.6,-0.8) node {\footnotesize{$3$}};
\draw (-0.4,1.2) node {\footnotesize{$8$}};
\draw (1.6,0) -- (2,0);
\draw (2.4,0) circle (0.4cm);
\draw(1.2,0) circle (0.4cm);
\draw (-2.4,0) circle (0.4cm);
\draw (-2.8,0) -- (-3.2,0);
\draw (-3.6,0) circle (0.4cm);
\draw (2.8,0) -- (3.2,0);
\draw (3.6,0) circle (0.4cm);
\draw (-0.4,0) -- (-0.8,0);
\draw (-1.2,0) circle (0.4cm);
\draw (-1.6,0) -- (-2,0);
\draw (-1.2,0.4) -- (-1.2,0.8);
\draw (-1.2,1.2) circle (0.4cm);
\draw (-1.2,2.4) [red,fill=red!30] circle  (0.4cm);
\draw (-0.4,2.4) node {\footnotesize{$1$}};
\draw (-1.12,1.6) -- (-1.12,2);
\draw (-1.28,1.6) -- (-1.28,2);
\draw (-1,1.9) -- (-1.2,1.7);
\draw (-1.4,1.9) -- (-1.2,1.7);
\draw (0,3) node {\footnotesize{$$}};

\end{tikzpicture} & $14$ \\ \hline	

\begin{tikzpicture}[scale=0.70]
\draw (0.4,0) -- (0.8,0);
\draw (0,0) circle (0.4cm);
\draw (-3.6,-0.8) node {\footnotesize{$8$}};
\draw (-2.4,-0.8) node {\footnotesize{$16$}};
\draw (-1.2,-0.8) node {\footnotesize{$24$}};
\draw (0,-0.8) node     {\footnotesize{$20$}};
\draw (1.2,-0.8) node {\footnotesize{$15$}};
\draw (2.4,-0.8) node {\footnotesize{$10$}};
\draw (3.6,-0.8) node {\footnotesize{$5$}};
\draw (-1.2,2) node {\footnotesize{$12$}};
\draw (1.6,0) -- (2,0);
\draw (2.4,0) circle (0.4cm);
\draw(1.2,0) circle (0.4cm);
\draw (-2.4,0) circle (0.4cm);
\draw (-2.8,0) -- (-3.2,0);
\draw (-3.6,0) circle (0.4cm);
\draw (2.8,0) -- (3.2,0);
\draw (3.6,0) circle (0.4cm);
\draw (-0.4,0) -- (-0.8,0);
\draw (-1.2,0) circle (0.4cm);
\draw (-1.6,0) -- (-2,0);
\draw (-1.2,0.4) -- (-1.2,0.8);
\draw (-1.2,1.2) circle (0.4cm);
\draw (0,1.2) [red,fill=red!30] circle  (0.4cm);
\draw (0,2) node {\footnotesize{$1$}};
\draw (-0.08,0.4) -- (-0.08,0.8);
\draw (0.08,0.4) -- (0.08,0.8);
\draw (-0.2,0.7) -- (0,0.5);
\draw (0.2,0.7) -- (0,0.5);

\end{tikzpicture} & $38$ \\ \hline

\begin{tikzpicture}[scale=0.70]
\draw (0.4,0) -- (0.8,0);
\draw (0,0) circle (0.4cm);
\draw (-3.6,-0.8) node {\footnotesize{$6$}};
\draw (-2.4,-0.8) node {\footnotesize{$12$}};
\draw (-1.2,-0.8) node {\footnotesize{$18$}};
\draw (0,-0.8) node     {\footnotesize{$15$}};
\draw (1.2,-0.8) node {\footnotesize{$12$}};
\draw (2.4,-0.8) node {\footnotesize{$8$}};
\draw (3.6,-0.8) node {\footnotesize{$4$}};
\draw (-1.2,2) node {\footnotesize{$9$}};
\draw (1.6,0) -- (2,0);
\draw (2.4,0) circle (0.4cm);
\draw(1.2,0) circle (0.4cm);
\draw (-2.4,0) circle (0.4cm);
\draw (-2.8,0) -- (-3.2,0);
\draw (-3.6,0) circle (0.4cm);
\draw (2.8,0) -- (3.2,0);
\draw (3.6,0) circle (0.4cm);
\draw (-0.4,0) -- (-0.8,0);
\draw (-1.2,0) circle (0.4cm);
\draw (-1.6,0) -- (-2,0);
\draw (-1.2,0.4) -- (-1.2,0.8);
\draw (-1.2,1.2) circle (0.4cm);
\draw (1.2,1.2) [red,fill=red!30] circle  (0.4cm);
\draw (1.2,2) node {\footnotesize{$1$}};
\draw (1.12,0.4) -- (1.12,0.8);
\draw (1.28,0.4) -- (1.28,0.8);
\draw (1,0.7) -- (1.2,0.5);
\draw (1.4,0.7) -- (1.2,0.5);

\end{tikzpicture} & $22$ \\ \hline

\begin{tikzpicture}[scale=0.70]	
\draw (0.4,0) -- (0.8,0);
\draw (0,0) circle (0.4cm);
\draw (-3.6,-0.8) node {\footnotesize{$4$}};
\draw (-2.4,-0.8) node {\footnotesize{$8$}};
\draw (-1.2,-0.8) node {\footnotesize{$12$}};
\draw (0,-0.8) node     {\footnotesize{$10$}};
\draw (1.2,-0.8) node {\footnotesize{$8$}};
\draw (2.4,-0.8) node {\footnotesize{$6$}};
\draw (3.6,-0.8) node {\footnotesize{$3$}};
\draw (-1.2,2) node {\footnotesize{$6$}};
\draw (1.6,0) -- (2,0);
\draw (2.4,0) circle (0.4cm);
\draw(1.2,0) circle (0.4cm);
\draw (-2.4,0) circle (0.4cm);
\draw (-2.8,0) -- (-3.2,0);
\draw (-3.6,0) circle (0.4cm);
\draw (2.8,0) -- (3.2,0);
\draw (3.6,0) circle (0.4cm);
\draw (-0.4,0) -- (-0.8,0);
\draw (-1.2,0) circle (0.4cm);
\draw (-1.6,0) -- (-2,0);
\draw (-1.2,0.4) -- (-1.2,0.8);
\draw (-1.2,1.2) circle (0.4cm);
\draw (2.4,1.2) [red,fill=red!30] circle  (0.4cm);
\draw (2.4,2) node {\footnotesize{$1$}};
\draw (2.32,0.4) -- (2.32,0.8);
\draw (2.48,0.4) -- (2.48,0.8);
\draw (2.4,0.5) -- (2.6,0.7);
\draw (2.4,0.5) -- (2.2,0.7);
\end{tikzpicture} & $10$ \\ \hline

\begin{tikzpicture}[scale=0.70]	
\draw (0.4,0) -- (0.8,0);
\draw (0,0) circle (0.4cm);
\draw (-3.6,-0.8) node {\footnotesize{$2$}};
\draw (-2.4,-0.8) node {\footnotesize{$4$}};
\draw (-1.2,-0.8) node {\footnotesize{$6$}};
\draw (0,-0.8) node     {\footnotesize{$5$}};
\draw (1.2,-0.8) node {\footnotesize{$4$}};
\draw (2.4,-0.8) node {\footnotesize{$3$}};
\draw (3.6,-0.8) node {\footnotesize{$2$}};
\draw (-1.2,2) node {\footnotesize{$3$}};
\draw (1.6,0) -- (2,0);
\draw (2.4,0) circle (0.4cm);
\draw(1.2,0) circle (0.4cm);
\draw (-2.4,0) circle (0.4cm);
\draw (-2.8,0) -- (-3.2,0);
\draw (-3.6,0) circle (0.4cm);
\draw (2.8,0) -- (3.2,0);
\draw (3.6,0) circle (0.4cm);
\draw (-0.4,0) -- (-0.8,0);
\draw (-1.2,0) circle (0.4cm);
\draw (-1.6,0) -- (-2,0);
\draw (-1.2,0.4) -- (-1.2,0.8);
\draw (-1.2,1.2) circle (0.4cm);
\draw (3.6,1.2) [red,fill=red!30] circle  (0.4cm);
\draw (3.6,2) node {\footnotesize{$1$}};
\draw (3.52,0.4) -- (3.52,0.8);
\draw (3.68,0.4) -- (3.68,0.8);
\draw (3.6,0.5) -- (3.4,0.7);
\draw (3.6,0.5) -- (3.8,0.7);

\end{tikzpicture} & $2$ \\ \hline
	\end{tabular}
	\caption{Exotic minimally unbalanced quivers of $E_8$-type with outward double laced edge.}
	\label{tab:E8seriesExoticDoubleOut}
\end{table}


\begin{table}
	\centering
	\begin{tabular}{|c|c|}
	\hline
Quiver & Excess \\ \hline
	
\begin{tikzpicture}[scale=0.70]
\draw (0.4,0) -- (0.8,0);
\draw (0,0) circle (0.4cm);
\draw (0,-0.8) node {\footnotesize{$16$}};
\draw (1.6,0) -- (2,0);
\draw (2.4,0) circle (0.4cm);
\draw (1.2,-0.8) node {\footnotesize{$12$}};
\draw(1.2,0) circle (0.4cm);
\draw (2.4,-0.8) node {\footnotesize{$8$}};
\draw (-2.4,0) circle (0.4cm);
\draw (-2.4,-0.8) node {\footnotesize{$14$}};
\draw (-2.8,0) -- (-3.2,0);
\draw (-3.6,0) circle (0.4cm);
\draw (-3.6,-0.8) node {\footnotesize{$8$}};
\draw (2.8,0) -- (3.2,0);
\draw (3.6,0) circle (0.4cm);
\draw (3.6,-0.8) node {\footnotesize{$4$}};
\draw (-0.4,0) -- (-0.8,0);
\draw (-1.2,0) circle (0.4cm);
\draw (-1.2,-0.8) node {\footnotesize{$20$}};
\draw (-1.6,0) -- (-2,0);
\draw (-1.2,0.4) -- (-1.2,0.8);
\draw (-1.2,1.2) circle (0.4cm);
\draw (-1.2,2) node {\footnotesize{$10$}};
\draw (-3.6,1.2) [red,fill=red!30] circle  (0.4cm);
\draw (-3.6,2) node {\footnotesize{$1$}};
\draw (-3.52,0.4) -- (-3.52,0.8);
\draw (-3.68,0.4) -- (-3.68,0.8);
\draw (-3.4,0.5) -- (-3.6,0.7);
\draw (-3.8,0.5) -- (-3.6,0.7);

\end{tikzpicture} & $6$\\ \hline	
	
\begin{tikzpicture}[scale=0.70]
\draw (0.4,0) -- (0.8,0);
\draw (0,0) circle (0.4cm);
\draw (-3.6,-0.8) node {\footnotesize{$14$}};
\draw (-2.4,-0.8) node {\footnotesize{$28$}};
\draw (-1.2,-0.8) node {\footnotesize{$40$}};
\draw (0,-0.8) node     {\footnotesize{$32$}};
\draw (1.2,-0.8) node {\footnotesize{$24$}};
\draw (2.4,-0.8) node {\footnotesize{$16$}};
\draw (3.6,-0.8) node {\footnotesize{$8$}};
\draw (-1.2,2) node {\footnotesize{$20$}};
\draw (1.6,0) -- (2,0);
\draw (2.4,0) circle (0.4cm);
\draw(1.2,0) circle (0.4cm);
\draw (-2.4,0) circle (0.4cm);
\draw (-2.8,0) -- (-3.2,0);
\draw (-3.6,0) circle (0.4cm);
\draw (2.8,0) -- (3.2,0);
\draw (3.6,0) circle (0.4cm);
\draw (-0.4,0) -- (-0.8,0);
\draw (-1.2,0) circle (0.4cm);
\draw (-1.6,0) -- (-2,0);
\draw (-1.2,0.4) -- (-1.2,0.8);
\draw (-1.2,1.2) circle (0.4cm);
\draw (-2.4,1.2) [red,fill=red!30] circle  (0.4cm);
\draw (-2.4,2) node {\footnotesize{$1$}};
\draw (-2.32,0.4) -- (-2.32,0.8);
\draw (-2.48,0.4) -- (-2.48,0.8);
\draw (-2.2,0.5) -- (-2.4,0.7);
\draw (-2.6,0.5) -- (-2.4,0.7);

\end{tikzpicture} & $26$\\ \hline	
	
\begin{tikzpicture}[scale=0.70]
\draw (0.4,0) -- (0.8,0);
\draw (0,0) circle (0.4cm);
\draw (-3.6,-0.8) node {\footnotesize{$20$}};
\draw (-2.4,-0.8) node {\footnotesize{$40$}};
\draw (-1.2,-0.8) node {\footnotesize{$60$}};
\draw (0,-0.8) node     {\footnotesize{$48$}};
\draw (1.2,-0.8) node {\footnotesize{$36$}};
\draw (2.4,-0.8) node {\footnotesize{$24$}};
\draw (3.6,-0.8) node {\footnotesize{$12$}};
\draw (1.6,0) -- (2,0);
\draw (2.4,0) circle (0.4cm);
\draw(1.2,0) circle (0.4cm);
\draw (-2.4,0) circle (0.4cm);
\draw (-2.8,0) -- (-3.2,0);
\draw (-3.6,0) circle (0.4cm);
\draw (2.8,0) -- (3.2,0);
\draw (3.6,0) circle (0.4cm);
\draw (-0.4,0) -- (-0.8,0);
\draw (-1.6,0) -- (-2,0);
\draw (-1.2,0) circle (0.4cm);
\draw (-2,1.2) [red,fill=red!30] circle  (0.4cm);
\draw (-2,2) node {\footnotesize{$1$}};
\draw (-1.34,0.38) -- (-1.71,0.91);
\draw (-1.465,0.29) -- (-1.85,0.82);
\draw (-1.632,0.648) -- (-1.34,0.62);
\draw (-1.632,0.648) -- (-1.7,0.35);
\draw (-0.4,1.2) circle (0.4cm);
\draw (-0.4,2) node {\footnotesize{$30$}};
\draw (-1,0.334) -- (-0.6,0.865);

\end{tikzpicture} & $58$\\ \hline	
	
\begin{tikzpicture}[scale=0.70]
\draw (0.4,0) -- (0.8,0);
\draw (0,0) circle (0.4cm);
\draw (-3.6,-0.8) node {\footnotesize{$10$}};
\draw (-2.4,-0.8) node {\footnotesize{$20$}};
\draw (-1.2,-0.8) node {\footnotesize{$30$}};
\draw (0,-0.8) node     {\footnotesize{$24$}};
\draw (1.2,-0.8) node {\footnotesize{$18$}};
\draw (2.4,-0.8) node {\footnotesize{$12$}};
\draw (3.6,-0.8) node {\footnotesize{$6$}};
\draw (-0.4,1.2) node {\footnotesize{$16$}};
\draw (1.6,0) -- (2,0);
\draw (2.4,0) circle (0.4cm);
\draw(1.2,0) circle (0.4cm);
\draw (-2.4,0) circle (0.4cm);
\draw (-2.8,0) -- (-3.2,0);
\draw (-3.6,0) circle (0.4cm);
\draw (2.8,0) -- (3.2,0);
\draw (3.6,0) circle (0.4cm);
\draw (-0.4,0) -- (-0.8,0);
\draw (-1.2,0) circle (0.4cm);
\draw (-1.6,0) -- (-2,0);
\draw (-1.2,0.4) -- (-1.2,0.8);
\draw (-1.2,1.2) circle (0.4cm);
\draw (-1.2,2.4) [red,fill=red!30] circle  (0.4cm);
\draw (-0.4,2.4) node {\footnotesize{$1$}};
\draw (-1.12,1.6) -- (-1.12,2);
\draw (-1.28,1.6) -- (-1.28,2);
\draw (-1,1.7) -- (-1.2,1.9);
\draw (-1.4,1.7) -- (-1.2,1.9);
\draw (0,3) node {\footnotesize{$$}};

\end{tikzpicture} & $14$ \\ \hline	

\begin{tikzpicture}[scale=0.70]
\draw (0.4,0) -- (0.8,0);
\draw (0,0) circle (0.4cm);
\draw (-3.6,-0.8) node {\footnotesize{$16$}};
\draw (-2.4,-0.8) node {\footnotesize{$32$}};
\draw (-1.2,-0.8) node {\footnotesize{$48$}};
\draw (0,-0.8) node     {\footnotesize{$40$}};
\draw (1.2,-0.8) node {\footnotesize{$30$}};
\draw (2.4,-0.8) node {\footnotesize{$20$}};
\draw (3.6,-0.8) node {\footnotesize{$10$}};
\draw (-1.2,2) node {\footnotesize{$24$}};
\draw (1.6,0) -- (2,0);
\draw (2.4,0) circle (0.4cm);
\draw(1.2,0) circle (0.4cm);
\draw (-2.4,0) circle (0.4cm);
\draw (-2.8,0) -- (-3.2,0);
\draw (-3.6,0) circle (0.4cm);
\draw (2.8,0) -- (3.2,0);
\draw (3.6,0) circle (0.4cm);
\draw (-0.4,0) -- (-0.8,0);
\draw (-1.2,0) circle (0.4cm);
\draw (-1.6,0) -- (-2,0);
\draw (-1.2,0.4) -- (-1.2,0.8);
\draw (-1.2,1.2) circle (0.4cm);
\draw (0,1.2) [red,fill=red!30] circle  (0.4cm);
\draw (0,2) node {\footnotesize{$1$}};
\draw (-0.08,0.4) -- (-0.08,0.8);
\draw (0.08,0.4) -- (0.08,0.8);
\draw (-0.2,0.5) -- (0,0.7);
\draw (0.2,0.5) -- (0,0.7);

\end{tikzpicture} & $38$ \\ \hline

\begin{tikzpicture}[scale=0.70]
\draw (0.4,0) -- (0.8,0);
\draw (0,0) circle (0.4cm);
\draw (-3.6,-0.8) node {\footnotesize{$12$}};
\draw (-2.4,-0.8) node {\footnotesize{$24$}};
\draw (-1.2,-0.8) node {\footnotesize{$36$}};
\draw (0,-0.8) node     {\footnotesize{$30$}};
\draw (1.2,-0.8) node {\footnotesize{$24$}};
\draw (2.4,-0.8) node {\footnotesize{$16$}};
\draw (3.6,-0.8) node {\footnotesize{$8$}};
\draw (-1.2,2) node {\footnotesize{$18$}};
\draw (1.6,0) -- (2,0);
\draw (2.4,0) circle (0.4cm);
\draw(1.2,0) circle (0.4cm);
\draw (-2.4,0) circle (0.4cm);
\draw (-2.8,0) -- (-3.2,0);
\draw (-3.6,0) circle (0.4cm);
\draw (2.8,0) -- (3.2,0);
\draw (3.6,0) circle (0.4cm);
\draw (-0.4,0) -- (-0.8,0);
\draw (-1.2,0) circle (0.4cm);
\draw (-1.6,0) -- (-2,0);
\draw (-1.2,0.4) -- (-1.2,0.8);
\draw (-1.2,1.2) circle (0.4cm);
\draw (1.2,1.2) [red,fill=red!30] circle  (0.4cm);
\draw (1.2,2) node {\footnotesize{$1$}};
\draw (1.12,0.4) -- (1.12,0.8);
\draw (1.28,0.4) -- (1.28,0.8);
\draw (1,0.5) -- (1.2,0.7);
\draw (1.4,0.5) -- (1.2,0.7);

\end{tikzpicture} & $22$ \\ \hline

\begin{tikzpicture}[scale=0.70]	
\draw (0.4,0) -- (0.8,0);
\draw (0,0) circle (0.4cm);
\draw (-3.6,-0.8) node {\footnotesize{$8$}};
\draw (-2.4,-0.8) node {\footnotesize{$16$}};
\draw (-1.2,-0.8) node {\footnotesize{$24$}};
\draw (0,-0.8) node     {\footnotesize{$20$}};
\draw (1.2,-0.8) node {\footnotesize{$16$}};
\draw (2.4,-0.8) node {\footnotesize{$12$}};
\draw (3.6,-0.8) node {\footnotesize{$6$}};
\draw (-1.2,2) node {\footnotesize{$12$}};
\draw (1.6,0) -- (2,0);
\draw (2.4,0) circle (0.4cm);
\draw(1.2,0) circle (0.4cm);
\draw (-2.4,0) circle (0.4cm);
\draw (-2.8,0) -- (-3.2,0);
\draw (-3.6,0) circle (0.4cm);
\draw (2.8,0) -- (3.2,0);
\draw (3.6,0) circle (0.4cm);
\draw (-0.4,0) -- (-0.8,0);
\draw (-1.2,0) circle (0.4cm);
\draw (-1.6,0) -- (-2,0);
\draw (-1.2,0.4) -- (-1.2,0.8);
\draw (-1.2,1.2) circle (0.4cm);
\draw (2.4,1.2) [red,fill=red!30] circle  (0.4cm);
\draw (2.4,2) node {\footnotesize{$1$}};
\draw (2.32,0.4) -- (2.32,0.8);
\draw (2.48,0.4) -- (2.48,0.8);
\draw (2.4,0.7) -- (2.6,0.5);
\draw (2.4,0.7) -- (2.2,0.5);

\end{tikzpicture} & $10$ \\ \hline

\begin{tikzpicture}[scale=0.70]
\draw (0.4,0) -- (0.8,0);
\draw (0,0) circle (0.4cm);
\draw (-3.6,-0.8) node {\footnotesize{$4$}};
\draw (-2.4,-0.8) node {\footnotesize{$8$}};
\draw (-1.2,-0.8) node {\footnotesize{$12$}};
\draw (0,-0.8) node     {\footnotesize{$10$}};
\draw (1.2,-0.8) node {\footnotesize{$8$}};
\draw (2.4,-0.8) node {\footnotesize{$6$}};
\draw (3.6,-0.8) node {\footnotesize{$4$}};
\draw (-1.2,2) node {\footnotesize{$6$}};
\draw (1.6,0) -- (2,0);
\draw (2.4,0) circle (0.4cm);
\draw(1.2,0) circle (0.4cm);
\draw (-2.4,0) circle (0.4cm);
\draw (-2.8,0) -- (-3.2,0);
\draw (-3.6,0) circle (0.4cm);
\draw (2.8,0) -- (3.2,0);
\draw (3.6,0) circle (0.4cm);
\draw (-0.4,0) -- (-0.8,0);
\draw (-1.2,0) circle (0.4cm);
\draw (-1.6,0) -- (-2,0);
\draw (-1.2,0.4) -- (-1.2,0.8);
\draw (-1.2,1.2) circle (0.4cm);
\draw (3.6,1.2) [red,fill=red!30] circle  (0.4cm);
\draw (3.6,2) node {\footnotesize{$1$}};
\draw (3.52,0.4) -- (3.52,0.8);
\draw (3.68,0.4) -- (3.68,0.8);
\draw (3.6,0.7) -- (3.4,0.5);
\draw (3.6,0.7) -- (3.8,0.5);

\end{tikzpicture} & $2$ \\ \hline
	\end{tabular}
	\caption{Exotic minimally unbalanced quivers of $E_8$-type with inward double laced edge.}
	\label{tab:E8seriesExoticDoubleIn}
\end{table}


\begin{table}
	\centering
	\begin{tabular}{|c|c|}
	\hline
	Quiver & Excess \\ \hline
	
\begin{tikzpicture}[scale=0.70]
\draw (0.4,0) -- (0.8,0);
\draw (0,0) circle (0.4cm);
\draw (0,-0.8) node {\footnotesize{$8$}};
\draw (1.6,0) -- (2,0);
\draw (2.4,0) circle (0.4cm);
\draw (1.2,-0.8) node {\footnotesize{$6$}};
\draw(1.2,0) circle (0.4cm);
\draw (2.4,-0.8) node {\footnotesize{$4$}};
\draw (-2.4,0) circle (0.4cm);
\draw (-2.4,-0.8) node {\footnotesize{$7$}};
\draw (-2.8,0) -- (-3.2,0);
\draw (-3.6,0) circle (0.4cm);
\draw (-3.6,-0.8) node {\footnotesize{$4$}};
\draw (2.8,0) -- (3.2,0);
\draw (3.6,0) circle (0.4cm);
\draw (3.6,-0.8) node {\footnotesize{$2$}};
\draw (-0.4,0) -- (-0.8,0);
\draw (-1.2,0) circle (0.4cm);
\draw (-1.2,-0.8) node {\footnotesize{$10$}};
\draw (-1.6,0) -- (-2,0);
\draw (-1.2,0.4) -- (-1.2,0.8);
\draw (-1.2,1.2) circle (0.4cm);
\draw (-1.2,2) node {\footnotesize{$5$}};
\draw (-3.6,1.2) [red,fill=red!30] circle  (0.4cm);
\draw (-3.6,2) node {\footnotesize{$1$}};
\draw (-3.52,0.4) -- (-3.52,0.8);
\draw (-3.68,0.4) -- (-3.68,0.8);
\draw (-3.6,0.4) -- (-3.6,0.8);
\draw (-3.4,0.7) -- (-3.6,0.5);
\draw (-3.8,0.7) -- (-3.6,0.5);

\end{tikzpicture} & $10$\\ \hline	

\begin{tikzpicture}[scale=0.70]
\draw (0.4,0) -- (0.8,0);
\draw (0,0) circle (0.4cm);
\draw (-3.6,-0.8) node {\footnotesize{$7$}};
\draw (-2.4,-0.8) node {\footnotesize{$14$}};
\draw (-1.2,-0.8) node {\footnotesize{$20$}};
\draw (0,-0.8) node     {\footnotesize{$16$}};
\draw (1.2,-0.8) node {\footnotesize{$12$}};
\draw (2.4,-0.8) node {\footnotesize{$8$}};
\draw (3.6,-0.8) node {\footnotesize{$4$}};
\draw (-1.2,2) node {\footnotesize{$10$}};
\draw (1.6,0) -- (2,0);
\draw (2.4,0) circle (0.4cm);
\draw(1.2,0) circle (0.4cm);
\draw (-2.4,0) circle (0.4cm);
\draw (-2.8,0) -- (-3.2,0);
\draw (-3.6,0) circle (0.4cm);
\draw (2.8,0) -- (3.2,0);
\draw (3.6,0) circle (0.4cm);
\draw (-0.4,0) -- (-0.8,0);
\draw (-1.2,0) circle (0.4cm);
\draw (-1.6,0) -- (-2,0);
\draw (-1.2,0.4) -- (-1.2,0.8);
\draw (-1.2,1.2) circle (0.4cm);
\draw (-2.4,1.2) [red,fill=red!30] circle  (0.4cm);
\draw (-2.4,2) node {\footnotesize{$1$}};
\draw (-2.32,0.4) -- (-2.32,0.8);
\draw (-2.48,0.4) -- (-2.48,0.8);
\draw (-2.4,0.4) -- (-2.4,0.8);
\draw (-2.2,0.7) -- (-2.4,0.5);
\draw (-2.6,0.7) -- (-2.4,0.5);

\end{tikzpicture} & $40$\\ \hline	
	
\begin{tikzpicture}[scale=0.70]
\draw (0.4,0) -- (0.8,0);
\draw (0,0) circle (0.4cm);
\draw (-3.6,-0.8) node {\footnotesize{$10$}};
\draw (-2.4,-0.8) node {\footnotesize{$20$}};
\draw (-1.2,-0.8) node {\footnotesize{$30$}};
\draw (0,-0.8) node     {\footnotesize{$24$}};
\draw (1.2,-0.8) node {\footnotesize{$18$}};
\draw (2.4,-0.8) node {\footnotesize{$12$}};
\draw (3.6,-0.8) node {\footnotesize{$6$}};
\draw (1.6,0) -- (2,0);
\draw (2.4,0) circle (0.4cm);
\draw(1.2,0) circle (0.4cm);
\draw (-2.4,0) circle (0.4cm);
\draw (-2.8,0) -- (-3.2,0);
\draw (-3.6,0) circle (0.4cm);
\draw (2.8,0) -- (3.2,0);
\draw (3.6,0) circle (0.4cm);
\draw (-0.4,0) -- (-0.8,0);
\draw (-1.6,0) -- (-2,0);
\draw (-1.2,0) circle (0.4cm);
\draw (-2,1.2) [red,fill=red!30] circle  (0.4cm);
\draw (-2,2) node {\footnotesize{$1$}};
\draw (-1.56,0.56) -- (-1.84,0.56);
\draw (-1.56,0.56) -- (-1.488,0.8);
\draw (-1.34,0.38) -- (-1.71,0.91);
\draw (-1.4,0.34) -- (-1.785,0.865);
\draw (-1.465,0.29) -- (-1.85,0.82);
\draw (-0.4,1.2) circle (0.4cm);
\draw (-0.4,2) node {\footnotesize{$15$}};
\draw (-1,0.334) -- (-0.6,0.865);

\end{tikzpicture} & $88$\\ \hline	
	
\begin{tikzpicture}[scale=0.70]
\draw (0.4,0) -- (0.8,0);
\draw (0,0) circle (0.4cm);
\draw (-3.6,-0.8) node {\footnotesize{$5$}};
\draw (-2.4,-0.8) node {\footnotesize{$10$}};
\draw (-1.2,-0.8) node {\footnotesize{$15$}};
\draw (0,-0.8) node     {\footnotesize{$12$}};
\draw (1.2,-0.8) node {\footnotesize{$9$}};
\draw (2.4,-0.8) node {\footnotesize{$6$}};
\draw (3.6,-0.8) node {\footnotesize{$3$}};
\draw (-0.4,1.2) node {\footnotesize{$8$}};
\draw (1.6,0) -- (2,0);
\draw (2.4,0) circle (0.4cm);
\draw(1.2,0) circle (0.4cm);
\draw (-2.4,0) circle (0.4cm);
\draw (-2.8,0) -- (-3.2,0);
\draw (-3.6,0) circle (0.4cm);
\draw (2.8,0) -- (3.2,0);
\draw (3.6,0) circle (0.4cm);
\draw (-0.4,0) -- (-0.8,0);
\draw (-1.2,0) circle (0.4cm);
\draw (-1.6,0) -- (-2,0);
\draw (-1.2,0.4) -- (-1.2,0.8);
\draw (-1.2,1.2) circle (0.4cm);
\draw (-1.2,2.4) [red,fill=red!30] circle  (0.4cm);
\draw (-0.4,2.4) node {\footnotesize{$1$}};
\draw (-1.12,1.6) -- (-1.12,2);
\draw (-1.28,1.6) -- (-1.28,2);
\draw (-1.2,1.6) -- (-1.2,2);
\draw (-1,1.9) -- (-1.2,1.7);
\draw (-1.4,1.9) -- (-1.2,1.7);
\draw (0,3) node {\footnotesize{$$}};

\end{tikzpicture} & $22$ \\ \hline	

\begin{tikzpicture}[scale=0.70]
\draw (0.4,0) -- (0.8,0);
\draw (0,0) circle (0.4cm);
\draw (-3.6,-0.8) node {\footnotesize{$8$}};
\draw (-2.4,-0.8) node {\footnotesize{$16$}};
\draw (-1.2,-0.8) node {\footnotesize{$24$}};
\draw (0,-0.8) node     {\footnotesize{$20$}};
\draw (1.2,-0.8) node {\footnotesize{$15$}};
\draw (2.4,-0.8) node {\footnotesize{$10$}};
\draw (3.6,-0.8) node {\footnotesize{$5$}};
\draw (-1.2,2) node {\footnotesize{$12$}};
\draw (1.6,0) -- (2,0);
\draw (2.4,0) circle (0.4cm);
\draw(1.2,0) circle (0.4cm);
\draw (-2.4,0) circle (0.4cm);
\draw (-2.8,0) -- (-3.2,0);
\draw (-3.6,0) circle (0.4cm);
\draw (2.8,0) -- (3.2,0);
\draw (3.6,0) circle (0.4cm);
\draw (-0.4,0) -- (-0.8,0);
\draw (-1.2,0) circle (0.4cm);
\draw (-1.6,0) -- (-2,0);
\draw (-1.2,0.4) -- (-1.2,0.8);
\draw (-1.2,1.2) circle (0.4cm);
\draw (0,1.2) [red,fill=red!30] circle  (0.4cm);
\draw (0,2) node {\footnotesize{$1$}};
\draw (-0.08,0.4) -- (-0.08,0.8);
\draw (0.08,0.4) -- (0.08,0.8);
\draw (0,0.4) -- (0,0.8);
\draw (-0.2,0.7) -- (0,0.5);
\draw (0.2,0.7) -- (0,0.5);

\end{tikzpicture} & $58$ \\ \hline

\begin{tikzpicture}[scale=0.70]
\draw (0.4,0) -- (0.8,0);
\draw (0,0) circle (0.4cm);
\draw (-3.6,-0.8) node {\footnotesize{$6$}};
\draw (-2.4,-0.8) node {\footnotesize{$12$}};
\draw (-1.2,-0.8) node {\footnotesize{$18$}};
\draw (0,-0.8) node     {\footnotesize{$15$}};
\draw (1.2,-0.8) node {\footnotesize{$12$}};
\draw (2.4,-0.8) node {\footnotesize{$8$}};
\draw (3.6,-0.8) node {\footnotesize{$4$}};
\draw (-1.2,2) node {\footnotesize{$9$}};
\draw (1.6,0) -- (2,0);
\draw (2.4,0) circle (0.4cm);
\draw(1.2,0) circle (0.4cm);
\draw (-2.4,0) circle (0.4cm);
\draw (-2.8,0) -- (-3.2,0);
\draw (-3.6,0) circle (0.4cm);
\draw (2.8,0) -- (3.2,0);
\draw (3.6,0) circle (0.4cm);
\draw (-0.4,0) -- (-0.8,0);
\draw (-1.2,0) circle (0.4cm);
\draw (-1.6,0) -- (-2,0);
\draw (-1.2,0.4) -- (-1.2,0.8);
\draw (-1.2,1.2) circle (0.4cm);
\draw (1.2,1.2) [red,fill=red!30] circle  (0.4cm);
\draw (1.2,2) node {\footnotesize{$1$}};
\draw (1.12,0.4) -- (1.12,0.8);
\draw (1.28,0.4) -- (1.28,0.8);
\draw (1.2,0.4) -- (1.2,0.8);
\draw (1,0.7) -- (1.2,0.5);
\draw (1.4,0.7) -- (1.2,0.5);

\end{tikzpicture} & $34$ \\ \hline

\begin{tikzpicture}[scale=0.70]	
\draw (0.4,0) -- (0.8,0);
\draw (0,0) circle (0.4cm);
\draw (-3.6,-0.8) node {\footnotesize{$4$}};
\draw (-2.4,-0.8) node {\footnotesize{$8$}};
\draw (-1.2,-0.8) node {\footnotesize{$12$}};
\draw (0,-0.8) node     {\footnotesize{$10$}};
\draw (1.2,-0.8) node {\footnotesize{$8$}};
\draw (2.4,-0.8) node {\footnotesize{$6$}};
\draw (3.6,-0.8) node {\footnotesize{$3$}};
\draw (-1.2,2) node {\footnotesize{$6$}};
\draw (1.6,0) -- (2,0);
\draw (2.4,0) circle (0.4cm);
\draw(1.2,0) circle (0.4cm);
\draw (-2.4,0) circle (0.4cm);
\draw (-2.8,0) -- (-3.2,0);
\draw (-3.6,0) circle (0.4cm);
\draw (2.8,0) -- (3.2,0);
\draw (3.6,0) circle (0.4cm);
\draw (-0.4,0) -- (-0.8,0);
\draw (-1.2,0) circle (0.4cm);
\draw (-1.6,0) -- (-2,0);
\draw (-1.2,0.4) -- (-1.2,0.8);
\draw (-1.2,1.2) circle (0.4cm);
\draw (2.4,1.2) [red,fill=red!30] circle  (0.4cm);
\draw (2.4,2) node {\footnotesize{$1$}};
\draw (2.32,0.4) -- (2.32,0.8);
\draw (2.48,0.4) -- (2.48,0.8);
\draw (2.4,0.4) -- (2.4,0.8);
\draw (2.4,0.5) -- (2.6,0.7);
\draw (2.4,0.5) -- (2.2,0.7);

\end{tikzpicture} & $16$ \\ \hline

\begin{tikzpicture}[scale=0.70]	
\draw (0.4,0) -- (0.8,0);
\draw (0,0) circle (0.4cm);
\draw (-3.6,-0.8) node {\footnotesize{$2$}};
\draw (-2.4,-0.8) node {\footnotesize{$4$}};
\draw (-1.2,-0.8) node {\footnotesize{$6$}};
\draw (0,-0.8) node     {\footnotesize{$5$}};
\draw (1.2,-0.8) node {\footnotesize{$4$}};
\draw (2.4,-0.8) node {\footnotesize{$3$}};
\draw (3.6,-0.8) node {\footnotesize{$2$}};
\draw (-1.2,2) node {\footnotesize{$3$}};
\draw (1.6,0) -- (2,0);
\draw (2.4,0) circle (0.4cm);
\draw(1.2,0) circle (0.4cm);
\draw (-2.4,0) circle (0.4cm);
\draw (-2.8,0) -- (-3.2,0);
\draw (-3.6,0) circle (0.4cm);
\draw (2.8,0) -- (3.2,0);
\draw (3.6,0) circle (0.4cm);
\draw (-0.4,0) -- (-0.8,0);
\draw (-1.2,0) circle (0.4cm);
\draw (-1.6,0) -- (-2,0);
\draw (-1.2,0.4) -- (-1.2,0.8);
\draw (-1.2,1.2) circle (0.4cm);
\draw (3.6,1.2) [red,fill=red!30] circle  (0.4cm);
\draw (3.6,2) node {\footnotesize{$1$}};
\draw (3.52,0.4) -- (3.52,0.8);
\draw (3.68,0.4) -- (3.68,0.8);
\draw (3.6,0.4) -- (3.6,0.8);
\draw (3.6,0.5) -- (3.4,0.7);
\draw (3.6,0.5) -- (3.8,0.7);

\end{tikzpicture} & $4$ \\ \hline
	\end{tabular}
	\caption{Exotic minimally unbalanced quivers of $E_8$-type with outward triple laced edge.}
	\label{tab:E8seriesExoticTripleOut}
\end{table}


\begin{table}
	\centering
	\begin{tabular}{|c|c|}
	\hline
	Quiver & Excess \\ \hline
	
\begin{tikzpicture}[scale=0.70]
\draw (0.4,0) -- (0.8,0);
\draw (0,0) circle (0.4cm);
\draw (0,-0.8) node {\footnotesize{$24$}};
\draw (1.6,0) -- (2,0);
\draw (2.4,0) circle (0.4cm);
\draw (1.2,-0.8) node {\footnotesize{$18$}};
\draw(1.2,0) circle (0.4cm);
\draw (2.4,-0.8) node {\footnotesize{$12$}};
\draw (-2.4,0) circle (0.4cm);
\draw (-2.4,-0.8) node {\footnotesize{$21$}};
\draw (-2.8,0) -- (-3.2,0);
\draw (-3.6,0) circle (0.4cm);
\draw (-3.6,-0.8) node {\footnotesize{$12$}};
\draw (2.8,0) -- (3.2,0);
\draw (3.6,0) circle (0.4cm);
\draw (3.6,-0.8) node {\footnotesize{$6$}};
\draw (-0.4,0) -- (-0.8,0);
\draw (-1.2,0) circle (0.4cm);
\draw (-1.2,-0.8) node {\footnotesize{$30$}};
\draw (-1.6,0) -- (-2,0);
\draw (-1.2,0.4) -- (-1.2,0.8);
\draw (-1.2,1.2) circle (0.4cm);
\draw (-1.2,2) node {\footnotesize{$15$}};
\draw (-3.6,1.2) [red,fill=red!30] circle  (0.4cm);
\draw (-3.6,2) node {\footnotesize{$1$}};
\draw (-3.52,0.4) -- (-3.52,0.8);
\draw (-3.68,0.4) -- (-3.68,0.8);
\draw (-3.6,0.4) -- (-3.6,0.8);
\draw (-3.4,0.5) -- (-3.6,0.7);
\draw (-3.8,0.5) -- (-3.6,0.7);

\end{tikzpicture} & $10$\\ \hline	
	
\begin{tikzpicture}[scale=0.70]
\draw (0.4,0) -- (0.8,0);
\draw (0,0) circle (0.4cm);
\draw (-3.6,-0.8) node {\footnotesize{$21$}};
\draw (-2.4,-0.8) node {\footnotesize{$42$}};
\draw (-1.2,-0.8) node {\footnotesize{$60$}};
\draw (0,-0.8) node     {\footnotesize{$48$}};
\draw (1.2,-0.8) node {\footnotesize{$36$}};
\draw (2.4,-0.8) node {\footnotesize{$24$}};
\draw (3.6,-0.8) node {\footnotesize{$12$}};
\draw (-1.2,2) node {\footnotesize{$30$}};
\draw (1.6,0) -- (2,0);
\draw (2.4,0) circle (0.4cm);
\draw(1.2,0) circle (0.4cm);
\draw (-2.4,0) circle (0.4cm);
\draw (-2.8,0) -- (-3.2,0);
\draw (-3.6,0) circle (0.4cm);
\draw (2.8,0) -- (3.2,0);
\draw (3.6,0) circle (0.4cm);
\draw (-0.4,0) -- (-0.8,0);
\draw (-1.2,0) circle (0.4cm);
\draw (-1.6,0) -- (-2,0);
\draw (-1.2,0.4) -- (-1.2,0.8);
\draw (-1.2,1.2) circle (0.4cm);
\draw (-2.4,1.2) [red,fill=red!30] circle  (0.4cm);
\draw (-2.4,2) node {\footnotesize{$1$}};
\draw (-2.32,0.4) -- (-2.32,0.8);
\draw (-2.48,0.4) -- (-2.48,0.8);
\draw (-2.4,0.4) -- (-2.4,0.8);
\draw (-2.2,0.5) -- (-2.4,0.7);
\draw (-2.6,0.5) -- (-2.4,0.7);

\end{tikzpicture} & $40$\\ \hline	
	
\begin{tikzpicture}[scale=0.70]
\draw (0.4,0) -- (0.8,0);
\draw (0,0) circle (0.4cm);
\draw (-3.6,-0.8) node {\footnotesize{$30$}};
\draw (-2.4,-0.8) node {\footnotesize{$60$}};
\draw (-1.2,-0.8) node {\footnotesize{$90$}};
\draw (0,-0.8) node     {\footnotesize{$72$}};
\draw (1.2,-0.8) node {\footnotesize{$54$}};
\draw (2.4,-0.8) node {\footnotesize{$36$}};
\draw (3.6,-0.8) node {\footnotesize{$18$}};
\draw (1.6,0) -- (2,0);
\draw (2.4,0) circle (0.4cm);
\draw(1.2,0) circle (0.4cm);
\draw (-2.4,0) circle (0.4cm);
\draw (-2.8,0) -- (-3.2,0);
\draw (-3.6,0) circle (0.4cm);
\draw (2.8,0) -- (3.2,0);
\draw (3.6,0) circle (0.4cm);
\draw (-0.4,0) -- (-0.8,0);
\draw (-1.6,0) -- (-2,0);
\draw (-1.2,0) circle (0.4cm);
\draw (-2,1.2) [red,fill=red!30] circle  (0.4cm);
\draw (-2,2) node {\footnotesize{$1$}};
\draw (-1.34,0.38) -- (-1.728,0.91);
\draw (-1.41,0.35) -- (-1.79,0.86);
\draw (-1.465,0.29) -- (-1.85,0.82);
\draw (-1.632,0.648) -- (-1.34,0.62);
\draw (-1.632,0.648) -- (-1.7,0.35);
\draw (-0.4,1.2) circle (0.4cm);
\draw (-0.4,2) node {\footnotesize{$45$}};
\draw (-1,0.334) -- (-0.6,0.865);

\end{tikzpicture} & $58$\\ \hline	
	
\begin{tikzpicture}[scale=0.70]
\draw (0.4,0) -- (0.8,0);
\draw (0,0) circle (0.4cm);
\draw (-3.6,-0.8) node {\footnotesize{$15$}};
\draw (-2.4,-0.8) node {\footnotesize{$30$}};
\draw (-1.2,-0.8) node {\footnotesize{$45$}};
\draw (0,-0.8) node     {\footnotesize{$36$}};
\draw (1.2,-0.8) node {\footnotesize{$27$}};
\draw (2.4,-0.8) node {\footnotesize{$18$}};
\draw (3.6,-0.8) node {\footnotesize{$9$}};
\draw (-0.3,1.2) node {\footnotesize{$24$}};
\draw (1.6,0) -- (2,0);
\draw (2.4,0) circle (0.4cm);
\draw(1.2,0) circle (0.4cm);
\draw (-2.4,0) circle (0.4cm);
\draw (-2.8,0) -- (-3.2,0);
\draw (-3.6,0) circle (0.4cm);
\draw (2.8,0) -- (3.2,0);
\draw (3.6,0) circle (0.4cm);
\draw (-0.4,0) -- (-0.8,0);
\draw (-1.2,0) circle (0.4cm);
\draw (-1.6,0) -- (-2,0);
\draw (-1.2,0.4) -- (-1.2,0.8);
\draw (-1.2,1.2) circle (0.4cm);
\draw (-1.2,2.4) [red,fill=red!30] circle  (0.4cm);
\draw (-0.4,2.4) node {\footnotesize{$1$}};
\draw (-1.12,1.6) -- (-1.12,2);
\draw (-1.28,1.6) -- (-1.28,2);
\draw (-1.2,1.6) -- (-1.2,2);
\draw (-1,1.7) -- (-1.2,1.9);
\draw (-1.4,1.7) -- (-1.2,1.9);
\draw (0,3) node {\footnotesize{$$}};

\end{tikzpicture} & $22$ \\ \hline	

\begin{tikzpicture}[scale=0.70]
\draw (0.4,0) -- (0.8,0);
\draw (0,0) circle (0.4cm);
\draw (-3.6,-0.8) node {\footnotesize{$24$}};
\draw (-2.4,-0.8) node {\footnotesize{$48$}};
\draw (-1.2,-0.8) node {\footnotesize{$72$}};
\draw (0,-0.8) node     {\footnotesize{$60$}};
\draw (1.2,-0.8) node {\footnotesize{$45$}};
\draw (2.4,-0.8) node {\footnotesize{$30$}};
\draw (3.6,-0.8) node {\footnotesize{$15$}};
\draw (-1.2,2) node {\footnotesize{$36$}};
\draw (1.6,0) -- (2,0);
\draw (2.4,0) circle (0.4cm);
\draw(1.2,0) circle (0.4cm);
\draw (-2.4,0) circle (0.4cm);
\draw (-2.8,0) -- (-3.2,0);
\draw (-3.6,0) circle (0.4cm);
\draw (2.8,0) -- (3.2,0);
\draw (3.6,0) circle (0.4cm);
\draw (-0.4,0) -- (-0.8,0);
\draw (-1.2,0) circle (0.4cm);
\draw (-1.6,0) -- (-2,0);
\draw (-1.2,0.4) -- (-1.2,0.8);
\draw (-1.2,1.2) circle (0.4cm);
\draw (0,1.2) [red,fill=red!30] circle  (0.4cm);
\draw (0,2) node {\footnotesize{$1$}};
\draw (-0.08,0.4) -- (-0.08,0.8);
\draw (0.08,0.4) -- (0.08,0.8);
\draw (0,0.4) -- (0,0.8);
\draw (-0.2,0.5) -- (0,0.7);
\draw (0.2,0.5) -- (0,0.7);

\end{tikzpicture} & $58$ \\ \hline

\begin{tikzpicture}[scale=0.70]
\draw (0.4,0) -- (0.8,0);
\draw (0,0) circle (0.4cm);
\draw (-3.6,-0.8) node {\footnotesize{$18$}};
\draw (-2.4,-0.8) node {\footnotesize{$36$}};
\draw (-1.2,-0.8) node {\footnotesize{$54$}};
\draw (0,-0.8) node     {\footnotesize{$45$}};
\draw (1.2,-0.8) node {\footnotesize{$36$}};
\draw (2.4,-0.8) node {\footnotesize{$24$}};
\draw (3.6,-0.8) node {\footnotesize{$12$}};
\draw (-1.2,2) node {\footnotesize{$27$}};
\draw (1.6,0) -- (2,0);
\draw (2.4,0) circle (0.4cm);
\draw(1.2,0) circle (0.4cm);
\draw (-2.4,0) circle (0.4cm);
\draw (-2.8,0) -- (-3.2,0);
\draw (-3.6,0) circle (0.4cm);
\draw (2.8,0) -- (3.2,0);
\draw (3.6,0) circle (0.4cm);
\draw (-0.4,0) -- (-0.8,0);
\draw (-1.2,0) circle (0.4cm);
\draw (-1.6,0) -- (-2,0);
\draw (-1.2,0.4) -- (-1.2,0.8);
\draw (-1.2,1.2) circle (0.4cm);
\draw (1.2,1.2) [red,fill=red!30] circle  (0.4cm);
\draw (1.2,2) node {\footnotesize{$1$}};
\draw (1.12,0.4) -- (1.12,0.8);
\draw (1.28,0.4) -- (1.28,0.8);
\draw (1.2,0.4) -- (1.2,0.8);
\draw (1,0.5) -- (1.2,0.7);
\draw (1.4,0.5) -- (1.2,0.7);

\end{tikzpicture} & $34$ \\ \hline

\begin{tikzpicture}[scale=0.70]	
\draw (0.4,0) -- (0.8,0);
\draw (0,0) circle (0.4cm);
\draw (-3.6,-0.8) node {\footnotesize{$12$}};
\draw (-2.4,-0.8) node {\footnotesize{$24$}};
\draw (-1.2,-0.8) node {\footnotesize{$36$}};
\draw (0,-0.8) node     {\footnotesize{$130$}};
\draw (1.2,-0.8) node {\footnotesize{$24$}};
\draw (2.4,-0.8) node {\footnotesize{$18$}};
\draw (3.6,-0.8) node {\footnotesize{$9$}};
\draw (-1.2,2) node {\footnotesize{$18$}};
\draw (1.6,0) -- (2,0);
\draw (2.4,0) circle (0.4cm);
\draw(1.2,0) circle (0.4cm);
\draw (-2.4,0) circle (0.4cm);
\draw (-2.8,0) -- (-3.2,0);
\draw (-3.6,0) circle (0.4cm);
\draw (2.8,0) -- (3.2,0);
\draw (3.6,0) circle (0.4cm);
\draw (-0.4,0) -- (-0.8,0);
\draw (-1.2,0) circle (0.4cm);
\draw (-1.6,0) -- (-2,0);
\draw (-1.2,0.4) -- (-1.2,0.8);
\draw (-1.2,1.2) circle (0.4cm);
\draw (2.4,1.2) [red,fill=red!30] circle  (0.4cm);
\draw (2.4,2) node {\footnotesize{$1$}};
\draw (2.32,0.4) -- (2.32,0.8);
\draw (2.48,0.4) -- (2.48,0.8);
\draw (2.4,0.4) -- (2.4,0.8);
\draw (2.4,0.7) -- (2.6,0.5);
\draw (2.4,0.7) -- (2.2,0.5);

\end{tikzpicture} & $16$ \\ \hline

\begin{tikzpicture}[scale=0.70]
\draw (0.4,0) -- (0.8,0);
\draw (0,0) circle (0.4cm);
\draw (-3.6,-0.8) node {\footnotesize{$6$}};
\draw (-2.4,-0.8) node {\footnotesize{$12$}};
\draw (-1.2,-0.8) node {\footnotesize{$18$}};
\draw (0,-0.8) node     {\footnotesize{$15$}};
\draw (1.2,-0.8) node {\footnotesize{$12$}};
\draw (2.4,-0.8) node {\footnotesize{$9$}};
\draw (3.6,-0.8) node {\footnotesize{$6$}};
\draw (-1.2,2) node {\footnotesize{$9$}};
\draw (1.6,0) -- (2,0);
\draw (2.4,0) circle (0.4cm);
\draw(1.2,0) circle (0.4cm);
\draw (-2.4,0) circle (0.4cm);
\draw (-2.8,0) -- (-3.2,0);
\draw (-3.6,0) circle (0.4cm);
\draw (2.8,0) -- (3.2,0);
\draw (3.6,0) circle (0.4cm);
\draw (-0.4,0) -- (-0.8,0);
\draw (-1.2,0) circle (0.4cm);
\draw (-1.6,0) -- (-2,0);
\draw (-1.2,0.4) -- (-1.2,0.8);
\draw (-1.2,1.2) circle (0.4cm);
\draw (3.6,1.2) [red,fill=red!30] circle  (0.4cm);
\draw (3.6,2) node {\footnotesize{$1$}};
\draw (3.52,0.4) -- (3.52,0.8);
\draw (3.68,0.4) -- (3.68,0.8);
\draw (3.6,0.4) -- (3.6,0.8);
\draw (3.6,0.7) -- (3.4,0.5);
\draw (3.6,0.7) -- (3.8,0.5);

\end{tikzpicture} & $4$ \\ \hline
	\end{tabular}
	\caption{Exotic minimally unbalanced quivers of $E_8$-type with inward triple laced edge.}
	\label{tab:E8seriesExoticTripleIn}
\end{table}

\clearpage
\newpage


\section{Non-Simply Laced Minimally Unbalanced Quivers with Unbalanced Node connected by a Non-Simply Laced Edge} \label{7}

Let us emphasize the main criteria for the classification of minimally unbalanced quivers:
\begin{itemize}
\item Balanced subset of nodes must form a single Dynkin diagram of finite type
\item The unbalanced node must connect to a single node of the Dynkin diagram with either a simple, double or a triple laced edge\footnote{By relaxing the second criterion one can produce \emph{very exotic} minimally unbalanced quivers like the $A$-type Ring quivers in Appendix \ref{C}. The motivation for presenting the quivers in Appendix \ref{C} is that the unbalanced node connects to the adjoint nodes which results in a simpler moduli space compared to a generic cases where the unbalanced node connects to any two arbitrary nodes of a Dynkin diagram.}
\end{itemize}
It follows from this criterion that the remaining part of the classification concerns quivers with two non-simply laced edges. Analogically to section \ref{6} we term the following part of the classification \emph{exotic classification of minimally unbalanced quivers of non-simply laced theories}. In the following the quivers are divided into categories based on:
\begin{itemize}
\item Dynkin diagram type of the balanced part of the quiver
\item Type of the non-simply laced edge:
\begin{itemize}
\item Double Edge
\item Triple Edge
\end{itemize}
\item Orientation of the non-simply laced edge that connects the unbalanced node. The orientation is considered with respect to the unbalanced node:
\begin{itemize}
\item Outwards
\item Inwards
\end{itemize}
\item Position of the unbalanced node (with respect to the balanced sub-quiver)
\end{itemize}
We emphasize that all quivers in this section are shown in their basic form such that the ranks are the lowest possible integers. All other quivers that are also minimally unbalanced and have the same global symmetry $G$ on their Coulomb branch can be obtained by multiplying the basic canonical forms by an integer number.

\subsection{Exotic Minimally Unbalanced Quivers with $G$ of Type $B_n$}

Let us start with exotic minimally unbalanced quivers with $G=SO(2n+1)$ and with the unbalanced node connected to the rest of the quiver via a double laced edge. The eight resulting cases are summarized in table \ref{tab:BseriesexoticDouble}. Analogically, table \ref{tab:BseriesexoticTriple} contains all quivers which contain an extra triple laced edge between the unbalanced node and the rest of the quiver.

\begin{table}[h!]
	\centering
	\begin{tabular}{|c|l|c|}
	\hline
	a &  \multicolumn{1}{c|}{Quiver} & Excess \\ \hline
	$\begin{array}{c}
	a<n\\
	a=2m
	\end{array}$ &
\begin{tikzpicture}[scale=0.70]

\draw (0.4,0) -- (0.8,0);
\draw (0,0) circle (0.4cm);
\draw (0,-0.8) node {\footnotesize{$2m$}};
\draw (1.6,0) -- (2,0);
\draw (1.2,0) node {\footnotesize{\dots}};
\draw (-2.4,0) circle (0.4cm);
\draw (-2.4,-0.8) node {\footnotesize{$3$}};
\draw (-2.8,0) -- (-3.2,0);
\draw (-3.6,0) circle (0.4cm);
\draw (-3.6,-0.8) node {\footnotesize{$2$}};
\draw (-4,0) -- (-4.4,0);
\draw (-4.8,0) circle (0.4cm);
\draw (-4.8,-0.8) node {\footnotesize{$1$}};
\draw (-0.08,0.4) -- (-0.08,0.8);
\draw (0.08,0.4) -- (0.08,0.8);
\draw (0,0.5) -- (-0.2,0.7);
\draw (0,0.5) -- (0.2,0.7);
\draw (2.8,0.08) -- (3.4,0.08);
\draw (2.8,-0.08) -- (3.4,-0.08);
\draw (3.2,0) -- (3,0.2);
\draw (3.2,0) -- (3,-0.2);
\draw (2.4,0) circle (0.4cm);
\draw (3.8,0) circle (0.4cm);
\draw (2.4,-0.8) node {\footnotesize{$2m$}};
\draw (3.8,-0.8) node {\footnotesize{$m$}};
\draw (-0.4,0) -- (-0.8,0);
\draw (-1.2,0) node {\footnotesize{\dots}};
\draw (-1.6,0) -- (-2,0);
\draw (0,1.2)[red,fill=red!30] circle (0.4cm);
\draw (0,2) node {\footnotesize{$1$}};
\draw [decorate,decoration={brace,amplitude=6pt}] (0.22,0.5) to (2.4,0.5);
\draw (1.38,1.1) node {\footnotesize{$n-a$}};
\end{tikzpicture} & $2a-2$\\ \hline	
	$\begin{array}{c}
	a<n\\
	a=2m+1
	\end{array}$ &
\begin{tikzpicture}[scale=0.70]

\draw (0.4,0) -- (0.8,0);
\draw (0,0) circle (0.4cm);
\draw (0,-0.8) node {\footnotesize{$2(2m+1)$}};
\draw (1.6,0) -- (2,0);
\draw (1.2,0) node {\footnotesize{\dots}};
\draw (-2.4,0) circle (0.4cm);
\draw (-2.4,-0.8) node {\footnotesize{$6$}};
\draw (-2.8,0) -- (-3.2,0);
\draw (-3.6,0) circle (0.4cm);
\draw (-3.6,-0.8) node {\footnotesize{$4$}};
\draw (-4,0) -- (-4.4,0);
\draw (-4.8,0) circle (0.4cm);
\draw (-4.8,-0.8) node {\footnotesize{$2$}};
\draw (-0.08,0.4) -- (-0.08,0.8);
\draw (0.08,0.4) -- (0.08,0.8);
\draw (0,0.5) -- (-0.2,0.7);
\draw (0,0.5) -- (0.2,0.7);
\draw (2.8,0.08) -- (4,0.08);
\draw (2.8,-0.08) -- (4,-0.08);
\draw (3.6,0) -- (3.4,0.2);
\draw (3.6,0) -- (3.4,-0.2);
\draw (2.4,0) circle (0.4cm);
\draw (4.4,0) circle (0.4cm);
\draw (2.4,-0.8) node {\footnotesize{$2(2m+1)$}};
\draw (4.4,-0.8) node {\footnotesize{$2m+1$}};
\draw (-0.4,0) -- (-0.8,0);
\draw (-1.2,0) node {\footnotesize{\dots}};
\draw (-1.6,0) -- (-2,0);
\draw (0,1.2)[red,fill=red!30] circle (0.4cm);
\draw (0,2) node {\footnotesize{$2$}};
\draw [decorate,decoration={brace,amplitude=6pt}] (0.22,0.5) to (2.4,0.5);
\draw (1.38,1.1) node {\footnotesize{$n-a$}};
\end{tikzpicture} & $4a-4$\\ \hline	

$\begin{array}{c}
	a<n\\
	\end{array}$ &
\begin{tikzpicture}[scale=0.70]

\draw (0.4,0) -- (0.8,0);
\draw (0,0) circle (0.4cm);
\draw (0,-0.8) node {\footnotesize{$2a$}};
\draw (1.6,0) -- (2,0);
\draw (1.2,0) node {\footnotesize{\dots}};
\draw (-2.4,0) circle (0.4cm);
\draw (-2.4,-0.8) node {\footnotesize{$6$}};
\draw (-2.8,0) -- (-3.2,0);
\draw (-3.6,0) circle (0.4cm);
\draw (-3.6,-0.8) node {\footnotesize{$4$}};
\draw (-4,0) -- (-4.4,0);
\draw (-4.8,0) circle (0.4cm);
\draw (-4.8,-0.8) node {\footnotesize{$2$}};
\draw (-0.08,0.4) -- (-0.08,0.8);
\draw (0.08,0.4) -- (0.08,0.8);
\draw (0,0.7) -- (-0.2,0.5);
\draw (0,0.7) -- (0.2,0.5);
\draw (2.8,0.08) -- (3.4,0.08);
\draw (2.8,-0.08) -- (3.4,-0.08);
\draw (3.2,0) -- (3,0.2);
\draw (3.2,0) -- (3,-0.2);
\draw (2.4,0) circle (0.4cm);
\draw (3.8,0) circle (0.4cm);
\draw (2.4,-0.8) node {\footnotesize{$2a$}};
\draw (3.8,-0.8) node {\footnotesize{$a$}};
\draw (-0.4,0) -- (-0.8,0);
\draw (-1.2,0) node {\footnotesize{\dots}};
\draw (-1.6,0) -- (-2,0);
\draw (0,1.2)[red,fill=red!30] circle (0.4cm);
\draw (0,2) node {\footnotesize{$1$}};
\draw [decorate,decoration={brace,amplitude=6pt}] (0.27,0.5) to (2.4,0.5);
\draw (1.38,1.1) node {\footnotesize{$n-a$}};
\end{tikzpicture} & $2a-2$\\ \hline \hline
%

	$\begin{array}{c}
	a=n\\
	a=2m
	\end{array}$ &
\begin{tikzpicture}[scale=0.70]

\draw (0,0) circle (0.4cm);
\draw (0,-0.8) node {\footnotesize{$2m-1$}};
\draw (1.4,-0.8) node {\footnotesize{$m$}};
\draw (1.4,0) circle (0.4cm);
\draw (2.8,-0.8) node {\footnotesize{$1$}};
\draw (-2.4,0)  circle (0.4cm);
\draw (-2.4,-0.8) node {\footnotesize{$3$}};
\draw (-2.8,0) -- (-3.2,0);
\draw (-3.6,0) circle (0.4cm);
\draw (-3.6,-0.8) node {\footnotesize{$2$}};
\draw (-4,0) -- (-4.4,0);
\draw (-4.8,0) circle (0.4cm);
\draw (-4.8,-0.8) node {\footnotesize{$1$}};
\draw (2.8,0) [red,fill=red!30] circle (0.4cm);
\draw (0.4,0.08) -- (1,0.08);
\draw (0.4,-0.08) -- (1,-0.08);
\draw (0.8,0) -- (0.65,0.20);
\draw (0.8,0) -- (0.65,-0.20);
\draw (-0.4,0) -- (-0.8,0);
\draw (1.8,-0.08) -- (2.4,-0.08);
\draw (1.8,0.08) -- (2.4,0.08);
\draw (2,0) -- (2.2,0.2);
\draw (2,0) -- (2.2,-0.2);
\draw (-1.2,0) node {\footnotesize{\dots}};
\draw (-1.6,0) -- (-2,0);
\draw [decorate,decoration={brace,amplitude=6pt}] (-4.7,0.5) to (-0.1,0.5);
\draw (-2.4,1.15) node {\footnotesize{$n-1=$ odd number}};
\end{tikzpicture} & $a-2$\\ \hline	

$\begin{array}{c}
	a=n\\
	a=2m+1
	\end{array}$ &
\begin{tikzpicture}[scale=0.70]

\draw (0,0) circle (0.4cm);
\draw (-0.1,-0.8) node {\footnotesize{$4m$}};
\draw (1.4,-0.8) node {\footnotesize{$2m+1$}};
\draw (1.4,0) circle (0.4cm);
\draw (2.8,-0.8) node {\footnotesize{$2$}};
\draw (-2.4,0)  circle (0.4cm);
\draw (-2.4,-0.8) node {\footnotesize{$6$}};
\draw (-2.8,0) -- (-3.2,0);
\draw (-3.6,0) circle (0.4cm);
\draw (-3.6,-0.8) node {\footnotesize{$4$}};
\draw (-4,0) -- (-4.4,0);
\draw (-4.8,0) circle (0.4cm);
\draw (-4.8,-0.8) node {\footnotesize{$2$}};
\draw (2.8,0) [red,fill=red!30] circle (0.4cm);
\draw (0.4,0.08) -- (1,0.08);
\draw (0.4,-0.08) -- (1,-0.08);
\draw (0.8,0) -- (0.65,0.2);
\draw (0.8,0) -- (0.65,-0.2);
\draw (-0.4,0) -- (-0.8,0);
\draw (1.8,-0.08) -- (2.4,-0.08);
\draw (1.8,0.08) -- (2.4,0.08);
\draw (2,0) -- (2.2,0.2);
\draw (2,0) -- (2.2,-0.2);
\draw (-1.2,0) node {\footnotesize{\dots}};
\draw (-1.6,0) -- (-2,0);
\draw [decorate,decoration={brace,amplitude=6pt}] (-4.7,0.5) to (-0.1,0.5);
\draw (-2.4,1.15) node {\footnotesize{$n-1=$ even number}};
\end{tikzpicture} &$2a-4$\\ \hline

	$\begin{array}{c}
	a=n\\
	\end{array}$ &
\begin{tikzpicture}[scale=0.70]

\draw (0,0) circle (0.4cm);
\draw (0,-0.8) node {\footnotesize{$2a-2$}};
\draw (1.4,-0.8) node {\footnotesize{$a$}};
\draw (1.4,0) circle (0.4cm);
\draw (2.8,-0.8) node {\footnotesize{$1$}};
\draw (-2.4,0)  circle (0.4cm);
\draw (-2.4,-0.8) node {\footnotesize{$6$}};
\draw (-2.8,0) -- (-3.2,0);
\draw (-3.6,0) circle (0.4cm);
\draw (-3.6,-0.8) node {\footnotesize{$4$}};
\draw (-4,0) -- (-4.4,0);
\draw (-4.8,0) circle (0.4cm);
\draw (-4.8,-0.8) node {\footnotesize{$2$}};
\draw (2.8,0) [red,fill=red!30] circle (0.4cm);
\draw (0.4,0.08) -- (1,0.08);
\draw (0.4,-0.08) -- (1,-0.08);
\draw (0.8,0) -- (0.65,0.20);
\draw (0.8,0) -- (0.65,-0.20);
\draw (-0.4,0) -- (-0.8,0);
\draw (1.8,-0.08) -- (2.4,-0.08);
\draw (1.8,0.08) -- (2.4,0.08);
\draw (2.2,0) -- (2,0.2);
\draw (2.2,0) -- (2,-0.2);
\draw (-1.2,0) node {\footnotesize{\dots}};
\draw (-1.6,0) -- (-2,0);
\draw (0,0.8) node {\footnotesize{$$}};
\end{tikzpicture} & $a-2$\\ \hline
		
	\end{tabular}
	\caption{Exotic minimally unbalanced quivers with $G=SO(2n+1)$ and an extra double laced edge.}
	\label{tab:BseriesexoticDouble}
\end{table}


\begin{table}[h!]
	\centering
	\begin{tabular}{|c|l|c|}
	\hline
	a &  \multicolumn{1}{c|}{Quiver} & Excess \\ \hline
	$\begin{array}{c}
	a<n\\
	a=2m
	\end{array}$ &
\begin{tikzpicture}[scale=0.70]

\draw (0.4,0) -- (0.8,0);
\draw (0,0) circle (0.4cm);
\draw (0,-0.8) node {\footnotesize{$2m$}};
\draw (1.6,0) -- (2,0);
\draw (1.2,0) node {\footnotesize{\dots}};
\draw (-2.4,0) circle (0.4cm);
\draw (-2.4,-0.8) node {\footnotesize{$3$}};
\draw (-2.8,0) -- (-3.2,0);
\draw (-3.6,0) circle (0.4cm);
\draw (-3.6,-0.8) node {\footnotesize{$2$}};
\draw (-4,0) -- (-4.4,0);
\draw (-4.8,0) circle (0.4cm);
\draw (-4.8,-0.8) node {\footnotesize{$1$}};
\draw (-0.08,0.4) -- (-0.08,0.8);
\draw (0.08,0.4) -- (0.08,0.8);
\draw (0,0.4) -- (0,0.8);
\draw (0,0.5) -- (-0.2,0.7);
\draw (0,0.5) -- (0.2,0.7);
\draw (2.8,0.08) -- (3.4,0.08);
\draw (2.8,-0.08) -- (3.4,-0.08);
\draw (3.2,0) -- (3,0.2);
\draw (3.2,0) -- (3,-0.2);
\draw (2.4,0) circle (0.4cm);
\draw (3.8,0) circle (0.4cm);
\draw (2.4,-0.8) node {\footnotesize{$2m$}};
\draw (3.8,-0.8) node {\footnotesize{$m$}};
\draw (-0.4,0) -- (-0.8,0);
\draw (-1.2,0) node {\footnotesize{\dots}};
\draw (-1.6,0) -- (-2,0);
\draw (0,1.2)[red,fill=red!30] circle (0.4cm);
\draw (0,2) node {\footnotesize{$1$}};
\draw [decorate,decoration={brace,amplitude=6pt}] (0.22,0.5) to (2.4,0.5);
\draw (1.38,1.1) node {\footnotesize{$n-a$}};
\end{tikzpicture} & $3a-2$\\ \hline	

	$\begin{array}{c}
	a<n\\
	a=2m+1
	\end{array}$ &
\begin{tikzpicture}[scale=0.70]

\draw (0.4,0) -- (0.8,0);
\draw (0,0) circle (0.4cm);
\draw (0,-0.8) node {\footnotesize{$2(2m+1)$}};
\draw (1.6,0) -- (2,0);
\draw (1.2,0) node {\footnotesize{\dots}};
\draw (-2.4,0) circle (0.4cm);
\draw (-2.4,-0.8) node {\footnotesize{$6$}};
\draw (-2.8,0) -- (-3.2,0);
\draw (-3.6,0) circle (0.4cm);
\draw (-3.6,-0.8) node {\footnotesize{$4$}};
\draw (-4,0) -- (-4.4,0);
\draw (-4.8,0) circle (0.4cm);
\draw (-4.8,-0.8) node {\footnotesize{$2$}};
\draw (-0.08,0.4) -- (-0.08,0.8);
\draw (0.08,0.4) -- (0.08,0.8);
\draw (0,0.4) -- (0,0.8);
\draw (0,0.5) -- (-0.2,0.7);
\draw (0,0.5) -- (0.2,0.7);
\draw (2.8,0.08) -- (4,0.08);
\draw (2.8,-0.08) -- (4,-0.08);
\draw (3.6,0) -- (3.4,0.2);
\draw (3.6,0) -- (3.4,-0.2);
\draw (2.4,0) circle (0.4cm);
\draw (4.4,0) circle (0.4cm);
\draw (2.4,-0.8) node {\footnotesize{$2(2m+1)$}};
\draw (4.4,-0.8) node {\footnotesize{$2m+1$}};
\draw (-0.4,0) -- (-0.8,0);
\draw (-1.2,0) node {\footnotesize{\dots}};
\draw (-1.6,0) -- (-2,0);
\draw (0,1.2)[red,fill=red!30] circle (0.4cm);
\draw (0,2) node {\footnotesize{$2$}};
\draw [decorate,decoration={brace,amplitude=6pt}] (0.22,0.5) to (2.4,0.5);
\draw (1.38,1.1) node {\footnotesize{$n-a$}};
\end{tikzpicture} & $6a-4$\\ \hline	

$\begin{array}{c}
	a<n\\
	a=2m
	\end{array}$ &
\begin{tikzpicture}[scale=0.70]

\draw (0.4,0) -- (0.8,0);
\draw (0,0) circle (0.4cm);
\draw (0,-0.8) node {\footnotesize{$6m$}};
\draw (1.6,0) -- (2,0);
\draw (1.2,0) node {\footnotesize{\dots}};
\draw (-2.4,0) circle (0.4cm);
\draw (-2.4,-0.8) node {\footnotesize{$9$}};
\draw (-2.8,0) -- (-3.2,0);
\draw (-3.6,0) circle (0.4cm);
\draw (-3.6,-0.8) node {\footnotesize{$6$}};
\draw (-4,0) -- (-4.4,0);
\draw (-4.8,0) circle (0.4cm);
\draw (-4.8,-0.8) node {\footnotesize{$3$}};
\draw (-0.08,0.4) -- (-0.08,0.8);
\draw (0.08,0.4) -- (0.08,0.8);
\draw (0,0.4) -- (0,0.8);
\draw (0,0.7) -- (-0.2,0.5);
\draw (0,0.7) -- (0.2,0.5);
\draw (2.8,0.08) -- (3.4,0.08);
\draw (2.8,-0.08) -- (3.4,-0.08);
\draw (3.2,0) -- (3,0.2);
\draw (3.2,0) -- (3,-0.2);
\draw (2.4,0) circle (0.4cm);
\draw (3.8,0) circle (0.4cm);
\draw (2.4,-0.8) node {\footnotesize{$6m$}};
\draw (3.8,-0.8) node {\footnotesize{$3m$}};
\draw (-0.4,0) -- (-0.8,0);
\draw (-1.2,0) node {\footnotesize{\dots}};
\draw (-1.6,0) -- (-2,0);
\draw (0,1.2)[red,fill=red!30] circle (0.4cm);
\draw (0,2) node {\footnotesize{$1$}};
\draw [decorate,decoration={brace,amplitude=6pt}] (0.27,0.5) to (2.4,0.5);
\draw (1.38,1.1) node {\footnotesize{$n-a$}};
\end{tikzpicture} & $3a-2$\\ \hline	

	$\begin{array}{c}
	a<n\\
	a=2m+1
	\end{array}$ &
\begin{tikzpicture}[scale=0.70]

\draw (0.4,0) -- (0.8,0);
\draw (0,0) circle (0.4cm);
\draw (0,-0.8) node {\footnotesize{$12m+6$}};
\draw (1.6,0) -- (2,0);
\draw (1.2,0) node {\footnotesize{\dots}};
\draw (-2.4,0) circle (0.4cm);
\draw (-2.4,-0.8) node {\footnotesize{$18$}};
\draw (-2.8,0) -- (-3.2,0);
\draw (-3.6,0) circle (0.4cm);
\draw (-3.6,-0.8) node {\footnotesize{$12$}};
\draw (-4,0) -- (-4.4,0);
\draw (-4.8,0) circle (0.4cm);
\draw (-4.8,-0.8) node {\footnotesize{$6$}};
\draw (-0.08,0.4) -- (-0.08,0.8);
\draw (0.08,0.4) -- (0.08,0.8);
\draw (0,0.4) -- (0,0.8);
\draw (0,0.7) -- (-0.2,0.5);
\draw (0,0.7) -- (0.2,0.5);
\draw (2.8,0.08) -- (4,0.08);
\draw (2.8,-0.08) -- (4,-0.08);
\draw (3.6,0) -- (3.4,0.2);
\draw (3.6,0) -- (3.4,-0.2);
\draw (2.4,0) circle (0.4cm);
\draw (4.4,0) circle (0.4cm);
\draw (2.4,-0.8) node {\footnotesize{$12m+6$}};
\draw (4.4,-0.8) node {\footnotesize{$6m+3$}};
\draw (-0.4,0) -- (-0.8,0);
\draw (-1.2,0) node {\footnotesize{\dots}};
\draw (-1.6,0) -- (-2,0);
\draw (0,1.2)[red,fill=red!30] circle (0.4cm);
\draw (0,2) node {\footnotesize{$2$}};
\draw [decorate,decoration={brace,amplitude=6pt}] (0.27,0.5) to (2.4,0.5);
\draw (1.38,1.1) node {\footnotesize{$n-a$}};
\end{tikzpicture} & $6a-4$\\ \hline \hline

	$\begin{array}{c}
	a=n\\
	a=2m
	\end{array}$ &
\begin{tikzpicture}[scale=0.70]

\draw (0,0) circle (0.4cm);
\draw (0,-0.8) node {\footnotesize{$2m-1$}};
\draw (1.4,-0.8) node {\footnotesize{$m$}};
\draw (1.4,0) circle (0.4cm);
\draw (2.8,-0.8) node {\footnotesize{$1$}};
\draw (-2.4,0)  circle (0.4cm);
\draw (-2.4,-0.8) node {\footnotesize{$3$}};
\draw (-2.8,0) -- (-3.2,0);
\draw (-3.6,0) circle (0.4cm);
\draw (-3.6,-0.8) node {\footnotesize{$2$}};
\draw (-4,0) -- (-4.4,0);
\draw (-4.8,0) circle (0.4cm);
\draw (-4.8,-0.8) node {\footnotesize{$1$}};
\draw (2.8,0) [red,fill=red!30] circle (0.4cm);
\draw (0.4,0.08) -- (1,0.08);
\draw (0.4,-0.08) -- (1,-0.08);
\draw (0.8,0) -- (0.65,0.20);
\draw (0.8,0) -- (0.65,-0.20);
\draw (-0.4,0) -- (-0.8,0);
\draw (1.8,-0.08) -- (2.4,-0.08);
\draw (1.8,0.08) -- (2.4,0.08);
\draw (1.8,0) -- (2.4,0);
\draw (2,0) -- (2.2,0.2);
\draw (2,0) -- (2.2,-0.2);
\draw (-1.2,0) node {\footnotesize{\dots}};
\draw (-1.6,0) -- (-2,0);
\draw [decorate,decoration={brace,amplitude=6pt}] (-4.7,0.5) to (-0.1,0.5);
\draw (-2.4,1.15) node {\footnotesize{$n-1=$ odd number}};
\end{tikzpicture} & $\frac{3a}{2}-2$\\ \hline	

$\begin{array}{c}
	a=n\\
	a=2m+1
	\end{array}$ &
\begin{tikzpicture}[scale=0.70]

\draw (0,0) circle (0.4cm);
\draw (-0.1,-0.8) node {\footnotesize{$4m$}};
\draw (1.4,-0.8) node {\footnotesize{$2m+1$}};
\draw (1.4,0) circle (0.4cm);
\draw (2.8,-0.8) node {\footnotesize{$2$}};
\draw (-2.4,0)  circle (0.4cm);
\draw (-2.4,-0.8) node {\footnotesize{$6$}};
\draw (-2.8,0) -- (-3.2,0);
\draw (-3.6,0) circle (0.4cm);
\draw (-3.6,-0.8) node {\footnotesize{$4$}};
\draw (-4,0) -- (-4.4,0);
\draw (-4.8,0) circle (0.4cm);
\draw (-4.8,-0.8) node {\footnotesize{$2$}};
\draw (2.8,0) [red,fill=red!30] circle (0.4cm);
\draw (0.4,0.08) -- (1,0.08);
\draw (0.4,-0.08) -- (1,-0.08);
\draw (0.8,0) -- (0.65,0.2);
\draw (0.8,0) -- (0.65,-0.2);
\draw (-0.4,0) -- (-0.8,0);
\draw (1.8,-0.08) -- (2.4,-0.08);
\draw (1.8,0.08) -- (2.4,0.08);
\draw (1.8,0) -- (2.4,0);
\draw (2,0) -- (2.2,0.2);
\draw (2,0) -- (2.2,-0.2);
\draw (-1.2,0) node {\footnotesize{\dots}};
\draw (-1.6,0) -- (-2,0);
\draw [decorate,decoration={brace,amplitude=6pt}] (-4.7,0.5) to (-0.1,0.5);
\draw (-2.4,1.15) node {\footnotesize{$n-1=$ even number}};
\end{tikzpicture} &$3a-4$\\ \hline

	$\begin{array}{c}
	a=n\\
	a=m
	\end{array}$ &
\begin{tikzpicture}[scale=0.70]

\draw (0,0) circle (0.4cm);
\draw (0,-0.8) node {\footnotesize{$6m-3$}};
\draw (1.4,-0.8) node {\footnotesize{$3m$}};
\draw (1.4,0) circle (0.4cm);
\draw (2.8,-0.8) node {\footnotesize{$1$}};
\draw (-2.4,0)  circle (0.4cm);
\draw (-2.4,-0.8) node {\footnotesize{$9$}};
\draw (-2.8,0) -- (-3.2,0);
\draw (-3.6,0) circle (0.4cm);
\draw (-3.6,-0.8) node {\footnotesize{$6$}};
\draw (-4,0) -- (-4.4,0);
\draw (-4.8,0) circle (0.4cm);
\draw (-4.8,-0.8) node {\footnotesize{$3$}};
\draw (2.8,0) [red,fill=red!30] circle (0.4cm);
\draw (0.4,0.08) -- (1,0.08);
\draw (0.4,-0.08) -- (1,-0.08);
\draw (0.8,0) -- (0.65,0.20);
\draw (0.8,0) -- (0.65,-0.20);
\draw (-0.4,0) -- (-0.8,0);
\draw (1.8,-0.08) -- (2.4,-0.08);
\draw (1.8,0.08) -- (2.4,0.08);
\draw (1.8,0) -- (2.4,0);
\draw (2.2,0) -- (2,0.2);
\draw (2.2,0) -- (2,-0.2);
\draw (-1.2,0) node {\footnotesize{\dots}};
\draw (-1.6,0) -- (-2,0);
\draw [decorate,decoration={brace,amplitude=6pt}] (-4.7,0.5) to (-0.1,0.5);
\draw (-2.4,1.15) node {\footnotesize{$n-1=$ odd number}};

\end{tikzpicture} & $3a-2$\\ \hline	

	$\begin{array}{c}
	a=n\\
	a=2m+1
	\end{array}$ &
\begin{tikzpicture}[scale=0.70]

\draw (0,0) circle (0.4cm);
\draw (-0.1,-0.8) node {\footnotesize{$12m$}};
\draw (1.4,-0.8) node {\footnotesize{$6m+3$}};
\draw (1.4,0) circle (0.4cm);
\draw (2.8,-0.8) node {\footnotesize{$2$}};
\draw (-2.4,0)  circle (0.4cm);
\draw (-2.4,-0.8) node {\footnotesize{$18$}};
\draw (-2.8,0) -- (-3.2,0);
\draw (-3.6,0) circle (0.4cm);
\draw (-3.6,-0.8) node {\footnotesize{$12$}};
\draw (-4,0) -- (-4.4,0);
\draw (-4.8,0) circle (0.4cm);
\draw (-4.8,-0.8) node {\footnotesize{$6$}};
\draw (2.8,0) [red,fill=red!30] circle (0.4cm);
\draw (0.4,0.08) -- (1,0.08);
\draw (0.4,-0.08) -- (1,-0.08);
\draw (0.8,0) -- (0.65,0.2);
\draw (0.8,0) -- (0.65,-0.2);
\draw (-0.4,0) -- (-0.8,0);
\draw (1.8,-0.08) -- (2.4,-0.08);
\draw (1.8,0.08) -- (2.4,0.08);
\draw (1.8,0) -- (2.4,0);
\draw (2.2,0) -- (2,0.2);
\draw (2.2,0) -- (2,-0.2);
\draw (-1.2,0) node {\footnotesize{\dots}};
\draw (-1.6,0) -- (-2,0);
\draw [decorate,decoration={brace,amplitude=6pt}] (-4.7,0.5) to (-0.1,0.5);
\draw (-2.4,1.15) node {\footnotesize{$n-1=$ even number}};
\end{tikzpicture} &$3a-4$\\ \hline
		
	\end{tabular}
	\caption{Exotic minimally unbalanced quivers with $G=SO(2n+1)$ and an extra triple laced edge.}
	\label{tab:BseriesexoticTriple}
\end{table}

\clearpage

\subsection{Exotic Minimally Unbalanced Quivers with $G$ of Type $C_n$}

Next, we present the four cases of exotic minimally unbalanced quivers with $\gf=C_n$ and the unbalanced node connected via a double laced edge. The results are summarized in table \ref{tab:CseriesexoticDouble}. In table \ref{tab:CseriesexoticTriple} we collect the remaining four quivers where the unbalanced node is connected via a triple laced edge.

\begin{table}
	\centering
	\begin{tabular}{|c|l|c|}
	\hline
	a &  \multicolumn{1}{c|}{Quiver} & Excess \\ \hline
	$\begin{array}{c}
	a<n\\

	\end{array}$ &
\begin{tikzpicture}[scale=0.70]

\draw (0.4,0) -- (0.8,0);
\draw (0,0) circle (0.4cm);
\draw (0,-0.8) node {\footnotesize{$a$}};
\draw (1.6,0) -- (2,0);
\draw (1.2,0) node {\footnotesize{\dots}};
\draw (-2.4,0) circle (0.4cm);
\draw (-2.4,-0.8) node {\footnotesize{$3$}};
\draw (-2.8,0) -- (-3.2,0);
\draw (-3.6,0) circle (0.4cm);
\draw (-3.6,-0.8) node {\footnotesize{$2$}};
\draw (-4,0) -- (-4.4,0);
\draw (-4.8,0) circle (0.4cm);
\draw (-4.8,-0.8) node {\footnotesize{$1$}};
\draw (2.8,0.08) -- (3.4,0.08);
\draw (2.8,-0.08) -- (3.4,-0.08);
\draw (3,0) -- (3.2,0.2);
\draw (3,0) -- (3.2,-0.2);
\draw (2.4,0) circle (0.4cm);
\draw (3.8,0) circle (0.4cm);
\draw (2.4,-0.8) node {\footnotesize{$a$}};
\draw (3.8,-0.8) node {\footnotesize{$a$}};
\draw (-0.4,0) -- (-0.8,0);
\draw (-1.2,0) node {\footnotesize{\dots}};
\draw (-1.6,0) -- (-2,0);
\draw (-0.08,0.4) -- (-0.08,0.8);
\draw (0.08,0.4) -- (0.08,0.8);
\draw (0,0.5) -- (-0.2,0.7);
\draw (0,0.5) -- (0.2,0.7);
\draw (0,1.2) [red,fill=red!30] circle (0.4cm);
\draw (0,2) node {\footnotesize{$1$}};
\draw [decorate,decoration={brace,amplitude=6pt}] (0.22,0.5) to (2.4,0.5);
\draw (1.36,1.1) node {\footnotesize{$n-a$}};
\end{tikzpicture} & $2a-2 $\\ \hline	
	
	$\begin{array}{c}
	a<n\\
	\end{array}$ &
\begin{tikzpicture}[scale=0.70]

\draw (0.4,0) -- (0.8,0);
\draw (0,0) circle (0.4cm);
\draw (0,-0.8) node {\footnotesize{$2a$}};
\draw (1.6,0) -- (2,0);
\draw (1.2,0) node {\footnotesize{\dots}};
\draw (-2.4,0) circle (0.4cm);
\draw (-2.4,-0.8) node {\footnotesize{$6$}};
\draw (-2.8,0) -- (-3.2,0);
\draw (-3.6,0) circle (0.4cm);
\draw (-3.6,-0.8) node {\footnotesize{$4$}};
\draw (-4,0) -- (-4.4,0);
\draw (-4.8,0) circle (0.4cm);
\draw (-4.8,-0.8) node {\footnotesize{$2$}};
\draw (2.8,0.08) -- (3.4,0.08);
\draw (2.8,-0.08) -- (3.4,-0.08);
\draw (3,0) -- (3.2,0.2);
\draw (3,0) -- (3.2,-0.2);
\draw (2.4,0) circle (0.4cm);
\draw (3.8,0) circle (0.4cm);
\draw (2.4,-0.8) node {\footnotesize{$2a$}};
\draw (3.8,-0.8) node {\footnotesize{$2a$}};
\draw (-0.4,0) -- (-0.8,0);
\draw (-1.2,0) node {\footnotesize{\dots}};
\draw (-1.6,0) -- (-2,0);
\draw (-0.08,0.4) -- (-0.08,0.8);
\draw (0.08,0.4) -- (0.08,0.8);
\draw (0,0.7) -- (-0.2,0.5);
\draw (0,0.7) -- (0.2,0.5);
\draw (0,1.2) [red,fill=red!30] circle (0.4cm);
\draw (0,2) node {\footnotesize{$1$}};
\draw [decorate,decoration={brace,amplitude=6pt}] (0.24,0.5) to (2.4,0.5);
\draw (1.36,1.1) node {\footnotesize{$n-a$}};
\end{tikzpicture} & $2a-2 $\\ \hline \hline

	$\begin{array}{c}
	a=n\\
	\end{array}$ &\begin{tikzpicture}[scale=0.70]
\draw (2,0.08) -- (2.8,0.08);
\draw (2,-0.08) -- (2.8,-0.08);
\draw (2.3,0) -- (2.5,-0.2);
\draw (2.3,0) -- (2.5,0.2);
\draw (0,0) circle (0.4cm);
\draw (0,-0.8) node {\footnotesize{$a-1$}};
\draw (1.6,0) circle (0.4cm);
\draw (1.6,-0.8) node {\footnotesize{$a$}};
\draw (-2.4,0) circle (0.4cm);
\draw (-2.4,-0.8) node {\footnotesize{$3$}};
\draw (-2.8,0) -- (-3.2,0);
\draw (-3.6,0) circle (0.4cm);
\draw (-3.6,-0.8) node {\footnotesize{$2$}};
\draw (-4,0) -- (-4.4,0);
\draw (-4.8,0) circle (0.4cm);
\draw (-4.8,-0.8) node {\footnotesize{$1$}};
\draw (3.2,0) [red,fill=red!30] circle (0.4cm);
\draw (3.2,-0.8) node {\footnotesize{$2$}};
\draw (-0.4,0) -- (-0.8,0);
\draw (-1.2,0) node {\footnotesize{\dots}};
\draw (-1.6,0) -- (-2,0);
\draw (0.4,0.08) -- (1.2,0.08);
\draw (0.4,-0.08) -- (1.2,-0.08);
\draw (0.7,0) -- (0.9,0.2);
\draw (0.7,0) -- (0.9,-0.2);
\draw (0,1) node {\footnotesize{$$}};

\end{tikzpicture} & $2a-4 $\\ \hline

	$\begin{array}{c}
	a=n\\
	\end{array}$ &\begin{tikzpicture}[scale=0.70]
\draw (2,0.08) -- (2.8,0.08);
\draw (2,-0.08) -- (2.8,-0.08);
\draw (2.5,0) -- (2.3,-0.2);
\draw (2.5,0) -- (2.3,0.2);
\draw (0,0) circle (0.4cm);
\draw (0,-0.8) node {\footnotesize{$a-1$}};
\draw (1.6,0) circle (0.4cm);
\draw (1.6,-0.8) node {\footnotesize{$a$}};
\draw (-2.4,0) circle (0.4cm);
\draw (-2.4,-0.8) node {\footnotesize{$3$}};
\draw (-2.8,0) -- (-3.2,0);
\draw (-3.6,0) circle (0.4cm);
\draw (-3.6,-0.8) node {\footnotesize{$2$}};
\draw (-4,0) -- (-4.4,0);
\draw (-4.8,0) circle (0.4cm);
\draw (-4.8,-0.8) node {\footnotesize{$1$}};
\draw (3.2,0) [red,fill=red!30] circle (0.4cm);
\draw (3.2,-0.8) node {\footnotesize{$1$}};
\draw (-0.4,0) -- (-0.8,0);
\draw (-1.2,0) node {\footnotesize{\dots}};
\draw (-1.6,0) -- (-2,0);
\draw (0.4,0.08) -- (1.2,0.08);
\draw (0.4,-0.08) -- (1.2,-0.08);
\draw (0.7,0) -- (0.9,0.2);
\draw (0.7,0) -- (0.9,-0.2);
\draw (0,1) node {\footnotesize{$$}};

\end{tikzpicture} & $a-2 $\\ \hline		

	\end{tabular}
	\caption{Exotic minimally unbalanced quivers with $\gf = Sp(2n)$ and an extra double laced edge.}
	\label{tab:CseriesexoticDouble}
\end{table}


\begin{table}
	\centering
	\begin{tabular}{|c|l|c|}
	\hline
	a &  \multicolumn{1}{c|}{Quiver} & Excess \\ \hline
	$\begin{array}{c}
	a<n\\
	a=m
	\end{array}$ &
\begin{tikzpicture}[scale=0.70]

\draw (0.4,0) -- (0.8,0);
\draw (0,0) circle (0.4cm);
\draw (0,-0.8) node {\footnotesize{$m$}};
\draw (1.6,0) -- (2,0);
\draw (1.2,0) node {\footnotesize{\dots}};
\draw (-2.4,0) circle (0.4cm);
\draw (-2.4,-0.8) node {\footnotesize{$3$}};
\draw (-2.8,0) -- (-3.2,0);
\draw (-3.6,0) circle (0.4cm);
\draw (-3.6,-0.8) node {\footnotesize{$2$}};
\draw (-4,0) -- (-4.4,0);
\draw (-4.8,0) circle (0.4cm);
\draw (-4.8,-0.8) node {\footnotesize{$1$}};
\draw (2.8,0.08) -- (3.4,0.08);
\draw (2.8,-0.08) -- (3.4,-0.08);
\draw (3,0) -- (3.2,0.2);
\draw (3,0) -- (3.2,-0.2);
\draw (2.4,0) circle (0.4cm);
\draw (3.8,0) circle (0.4cm);
\draw (2.4,-0.8) node {\footnotesize{$m$}};
\draw (3.8,-0.8) node {\footnotesize{$m$}};
\draw (-0.4,0) -- (-0.8,0);
\draw (-1.2,0) node {\footnotesize{\dots}};
\draw (-1.6,0) -- (-2,0);
\draw (-0.08,0.4) -- (-0.08,0.8);
\draw (0.08,0.4) -- (0.08,0.8);
\draw (0,0.4) -- (0,0.8);
\draw (0,0.5) -- (-0.2,0.7);
\draw (0,0.5) -- (0.2,0.7);
\draw (0,1.2) [red,fill=red!30] circle (0.4cm);
\draw (0,2) node {\footnotesize{$1$}};
\draw [decorate,decoration={brace,amplitude=6pt}] (0.22,0.5) to (2.4,0.5);
\draw (1.36,1.1) node {\footnotesize{$n-m$}};
\end{tikzpicture} & $3a-2 $\\ \hline

	$\begin{array}{c}
	a<n\\
	a=m
	\end{array}$ &
\begin{tikzpicture}[scale=0.70]

\draw (0.4,0) -- (0.8,0);
\draw (0,0) circle (0.4cm);
\draw (0,-0.8) node {\footnotesize{$3m$}};
\draw (1.6,0) -- (2,0);
\draw (1.2,0) node {\footnotesize{\dots}};
\draw (-2.4,0) circle (0.4cm);
\draw (-2.4,-0.8) node {\footnotesize{$9$}};
\draw (-2.8,0) -- (-3.2,0);
\draw (-3.6,0) circle (0.4cm);
\draw (-3.6,-0.8) node {\footnotesize{$6$}};
\draw (-4,0) -- (-4.4,0);
\draw (-4.8,0) circle (0.4cm);
\draw (-4.8,-0.8) node {\footnotesize{$3$}};
\draw (2.8,0.08) -- (3.4,0.08);
\draw (2.8,-0.08) -- (3.4,-0.08);
\draw (3,0) -- (3.2,0.2);
\draw (3,0) -- (3.2,-0.2);
\draw (2.4,0) circle (0.4cm);
\draw (3.8,0) circle (0.4cm);
\draw (2.4,-0.8) node {\footnotesize{$3m$}};
\draw (3.8,-0.8) node {\footnotesize{$3m$}};
\draw (-0.4,0) -- (-0.8,0);
\draw (-1.2,0) node {\footnotesize{\dots}};
\draw (-1.6,0) -- (-2,0);
\draw (-0.08,0.4) -- (-0.08,0.8);
\draw (0.08,0.4) -- (0.08,0.8);
\draw (0,0.4) -- (0,0.8);
\draw (0,0.7) -- (-0.2,0.5);
\draw (0,0.7) -- (0.2,0.5);
\draw (0,1.2) [red,fill=red!30] circle (0.4cm);
\draw (0,2) node {\footnotesize{$1$}};
\draw [decorate,decoration={brace,amplitude=6pt}] (0.24,0.5) to (2.4,0.5);
\draw (1.36,1.1) node {\footnotesize{$n-m$}};
\end{tikzpicture} & $3a-2 $\\ \hline \hline

	$\begin{array}{c}
	a=n\\
	a=m
	\end{array}$ &\begin{tikzpicture}[scale=0.70]
\draw (2,0.08) -- (2.8,0.08);
\draw (2,-0.08) -- (2.8,-0.08);
\draw (2,0) -- (2.8,0);
\draw (2.3,0) -- (2.5,-0.2);
\draw (2.3,0) -- (2.5,0.2);
\draw (0,0) circle (0.4cm);
\draw (0,-0.8) node {\footnotesize{$m-1$}};
\draw (1.6,0) circle (0.4cm);
\draw (1.6,-0.8) node {\footnotesize{$m$}};
\draw (-2.4,0) circle (0.4cm);
\draw (-2.4,-0.8) node {\footnotesize{$3$}};
\draw (-2.8,0) -- (-3.2,0);
\draw (-3.6,0) circle (0.4cm);
\draw (-3.6,-0.8) node {\footnotesize{$2$}};
\draw (-4,0) -- (-4.4,0);
\draw (-4.8,0) circle (0.4cm);
\draw (-4.8,-0.8) node {\footnotesize{$1$}};
\draw (3.2,0) [red,fill=red!30] circle (0.4cm);
\draw (3.2,-0.8) node {\footnotesize{$2$}};
\draw (-0.4,0) -- (-0.8,0);
\draw (-1.2,0) node {\footnotesize{\dots}};
\draw (-1.6,0) -- (-2,0);
\draw (0.4,0.08) -- (1.2,0.08);
\draw (0.4,-0.08) -- (1.2,-0.08);
\draw (0.7,0) -- (0.9,0.2);
\draw (0.7,0) -- (0.9,-0.2);
\draw (0,1) node {\footnotesize{$$}};

\end{tikzpicture} & $3a-4 $\\ \hline

	$\begin{array}{c}
	a=n\\
	a=m
	\end{array}$ &\begin{tikzpicture}[scale=0.70]
\draw (2,0.08) -- (2.8,0.08);
\draw (2,-0.08) -- (2.8,-0.08);
\draw (2,0) -- (2.8,0);
\draw (2.5,0) -- (2.3,-0.2);
\draw (2.5,0) -- (2.3,0.2);
\draw (0,0) circle (0.4cm);
\draw (-0.1,-0.8) node {\footnotesize{$3m-3$}};
\draw (1.6,0) circle (0.4cm);
\draw (1.76,-0.8) node {\footnotesize{$3m$}};
\draw (-2.4,0) circle (0.4cm);
\draw (-2.4,-0.8) node {\footnotesize{$9$}};
\draw (-2.8,0) -- (-3.2,0);
\draw (-3.6,0) circle (0.4cm);
\draw (-3.6,-0.8) node {\footnotesize{$6$}};
\draw (-4,0) -- (-4.4,0);
\draw (-4.8,0) circle (0.4cm);
\draw (-4.8,-0.8) node {\footnotesize{$3$}};
\draw (3.2,0) [red,fill=red!30] circle (0.4cm);
\draw (3.2,-0.8) node {\footnotesize{$2$}};
\draw (-0.4,0) -- (-0.8,0);
\draw (-1.2,0) node {\footnotesize{\dots}};
\draw (-1.6,0) -- (-2,0);
\draw (0.4,0.08) -- (1.2,0.08);
\draw (0.4,-0.08) -- (1.2,-0.08);
\draw (0.7,0) -- (0.9,0.2);
\draw (0.7,0) -- (0.9,-0.2);
\draw (0,1) node {\footnotesize{$$}};

\end{tikzpicture} & $3a-4 $\\ \hline	
	
	\end{tabular}
	\caption{Exotic minimally unbalanced quivers with $\gf = Sp(2n)$ and an extra triple laced edge.}
	\label{tab:CseriesexoticTriple}
\end{table}

\subsection{Exotic Minimally Unbalanced Quivers with $G$ of Type $F_4$}

In this subsection we present all exotic minimally unbalanced quivers with $\gf=F_4$. Quivers with an extra double laced edge are summarized in table \ref{tab:FveryexoticDouble}. All quivers with extra triple laced edge are collected in table \ref{tab:FveryexoticTriple}.

\begin{table}
	\centering
	\begin{tabular}{|c|c|}
	\hline
	Quiver & Excess \\ \hline
	
\begin{tikzpicture}[scale=0.70]
\draw (0,0) circle (0.4cm);
\draw (0,-0.8) node {\footnotesize{$2$}};
\draw (1.2,-0.8) node {\footnotesize{$1$}};
\draw (1.2,0) circle (0.4cm);
\draw (-2.4,0)  circle (0.4cm);
\draw (-2.4,-0.8) node {\footnotesize{$2$}};
\draw (-1.2,-0.8) node {\footnotesize{$3$}};
\draw (-0.8,0.08) -- (-0.4,0.08);
\draw (-0.8,-0.08) -- (-0.4,-0.08);
\draw (-0.7,0.2) -- (-0.5,0);
\draw (-0.7,-0.2) -- (-0.5,0);
\draw (-2.4,1.2) [red,fill=red!30] circle (0.4cm);
\draw (-2.32,0.4) -- (-2.32,0.8);
\draw (-2.48,0.4) -- (-2.48,0.8);
\draw (-2.2,0.7) -- (-2.4,0.5);
\draw (-2.6,0.7) -- (-2.4,0.5);
\draw (-2.4,2) node {\footnotesize{$1$}};
\draw (0.4,0) -- (0.8,0);
\draw (-1.2,0) circle (0.4cm);
\draw (-1.6,0) -- (-2,0);

\end{tikzpicture} & $2$ \\ \hline	

\begin{tikzpicture}[scale=0.70]
\draw (0,0) circle (0.4cm);
\draw (0,-0.8) node {\footnotesize{$4$}};
\draw (1.2,-0.8) node {\footnotesize{$2$}};
\draw (1.2,0) circle (0.4cm);
\draw (-2.4,0)  circle (0.4cm);
\draw (-2.4,-0.8) node {\footnotesize{$3$}};
\draw (-1.2,-0.8) node {\footnotesize{$6$}};
\draw (-1.2,2) node {\footnotesize{$1$}};
\draw (-0.8,0.08) -- (-0.4,0.08);
\draw (-0.8,-0.08) -- (-0.4,-0.08);
\draw (-0.7,0.2) -- (-0.5,0);
\draw (-0.7,-0.2) -- (-0.5,0);
\draw (-1.2,1.2) [red,fill=red!30] circle (0.4cm);
\draw (0.4,0) -- (0.8,0);
\draw (-1.2,0) circle (0.4cm);
\draw (-1.12,0.4) -- (-1.12,0.8);
\draw (-1.28,0.4) -- (-1.28,0.8);
\draw (-1.4,0.7) -- (-1.2,0.5);
\draw (-1,0.7) -- (-1.2,0.5);
\draw (-1.6,0) -- (-2,0);

\end{tikzpicture} & $10$ \\ \hline

\begin{tikzpicture}[scale=0.70]
\draw (0,0) circle (0.4cm);
\draw (0,-0.8) node {\footnotesize{$6$}};
\draw (1.2,-0.8) node {\footnotesize{$3$}};
\draw (1.2,0) circle (0.4cm);
\draw (-2.4,0)  circle (0.4cm);
\draw (-2.4,-0.8) node {\footnotesize{$4$}};
\draw (-1.2,-0.8) node {\footnotesize{$8$}};
\draw (0,2) node {\footnotesize{$1$}};
\draw (-0.8,0.08) -- (-0.4,0.08);
\draw (-0.8,-0.08) -- (-0.4,-0.08);
\draw (-0.7,0.2) -- (-0.5,0);
\draw (-0.7,-0.2) -- (-0.5,0);
\draw (0,1.2) [red,fill=red!30] circle (0.4cm);
\draw (-0.08,0.4) -- (-0.08,0.8);
\draw (0.08,0.4) -- (0.08,0.8);
\draw (-0.2,0.7) -- (0,0.5);
\draw (0.2,0.7) -- (0,0.5);
\draw (-1.2,0) circle (0.4cm);
\draw (-1.6,0) -- (-2,0);
\draw (0.4,0) -- (0.8,0);

\end{tikzpicture} & $10$ \\ \hline

\begin{tikzpicture}[scale=0.70]	
\draw (0,0) circle (0.4cm);
\draw (0,-0.8) node {\footnotesize{$3$}};
\draw (1.2,-0.8) node {\footnotesize{$2$}};
\draw (1.2,0) circle (0.4cm);
\draw (-2.4,0)  circle (0.4cm);
\draw (-2.4,-0.8) node {\footnotesize{$2$}};
\draw (-1.2,-0.8) node {\footnotesize{$4$}};
\draw (-0.8,0.08) -- (-0.4,0.08);
\draw (-0.8,-0.08) -- (-0.4,-0.08);
\draw (-0.7,0.2) -- (-0.5,0);
\draw (-0.7,-0.2) -- (-0.5,0);
\draw (1.2,1.2) [red,fill=red!30] circle (0.4cm);
\draw (1.12,0.4) -- (1.12,0.8);
\draw (1.28,0.4) -- (1.28,0.8);
\draw (1,0.7) -- (1.2,0.5);
\draw (1.4,0.7) -- (1.2,0.5);
\draw (1.2,2) node {\footnotesize{$1$}};
\draw (-1.2,0) circle (0.4cm);
\draw (-1.6,0) -- (-2,0);
\draw (0.4,0) -- (0.8,0);
\end{tikzpicture} & $2$ \\ \hline \hline


\begin{tikzpicture}[scale=0.70]
\draw (0,0) circle (0.4cm);
\draw (0,-0.8) node {\footnotesize{$4$}};
\draw (1.2,-0.8) node {\footnotesize{$2$}};
\draw (1.2,0) circle (0.4cm);
\draw (-2.4,0)  circle (0.4cm);
\draw (-2.4,-0.8) node {\footnotesize{$4$}};
\draw (-1.2,-0.8) node {\footnotesize{$6$}};
\draw (-0.8,0.08) -- (-0.4,0.08);
\draw (-0.8,-0.08) -- (-0.4,-0.08);
\draw (-0.7,0.2) -- (-0.5,0);
\draw (-0.7,-0.2) -- (-0.5,0);
\draw (-2.4,1.2) [red,fill=red!30] circle (0.4cm);
\draw (-2.32,0.4) -- (-2.32,0.8);
\draw (-2.48,0.4) -- (-2.48,0.8);
\draw (-2.2,0.5) -- (-2.4,0.7);
\draw (-2.6,0.5) -- (-2.4,0.7);
\draw (-2.4,2) node {\footnotesize{$1$}};
\draw (0.4,0) -- (0.8,0);
\draw (-1.2,0) circle (0.4cm);
\draw (-1.6,0) -- (-2,0);

\end{tikzpicture} & $2$ \\ \hline	

\begin{tikzpicture}[scale=0.70]
\draw (0,0) circle (0.4cm);
\draw (0,-0.8) node {\footnotesize{$8$}};
\draw (1.2,-0.8) node {\footnotesize{$4$}};
\draw (1.2,0) circle (0.4cm);
\draw (-2.4,0)  circle (0.4cm);
\draw (-2.4,-0.8) node {\footnotesize{$6$}};
\draw (-1.2,-0.8) node {\footnotesize{$12$}};
\draw (-1.2,2) node {\footnotesize{$1$}};
\draw (-0.8,0.08) -- (-0.4,0.08);
\draw (-0.8,-0.08) -- (-0.4,-0.08);
\draw (-0.7,0.2) -- (-0.5,0);
\draw (-0.7,-0.2) -- (-0.5,0);
\draw (-1.2,1.2) [red,fill=red!30] circle (0.4cm);
\draw (0.4,0) -- (0.8,0);
\draw (-1.2,0) circle (0.4cm);
\draw (-1.12,0.4) -- (-1.12,0.8);
\draw (-1.28,0.4) -- (-1.28,0.8);
\draw (-1.4,0.5) -- (-1.2,0.7);
\draw (-1,0.5) -- (-1.2,0.7);
\draw (-1.6,0) -- (-2,0);

\end{tikzpicture} & $10$ \\ \hline

\begin{tikzpicture}[scale=0.70]
\draw (0,0) circle (0.4cm);
\draw (0,-0.8) node {\footnotesize{$12$}};
\draw (1.2,-0.8) node {\footnotesize{$6$}};
\draw (1.2,0) circle (0.4cm);
\draw (-2.4,0)  circle (0.4cm);
\draw (-2.4,-0.8) node {\footnotesize{$8$}};
\draw (-1.2,-0.8) node {\footnotesize{$16$}};
\draw (0,2) node {\footnotesize{$1$}};
\draw (-0.8,0.08) -- (-0.4,0.08);
\draw (-0.8,-0.08) -- (-0.4,-0.08);
\draw (-0.7,0.2) -- (-0.5,0);
\draw (-0.7,-0.2) -- (-0.5,0);
\draw (0,1.2) [red,fill=red!30] circle (0.4cm);
\draw (-0.08,0.4) -- (-0.08,0.8);
\draw (0.08,0.4) -- (0.08,0.8);
\draw (-0.2,0.5) -- (0,0.7);
\draw (0.2,0.5) -- (0,0.7);
\draw (-1.2,0) circle (0.4cm);
\draw (-1.6,0) -- (-2,0);
\draw (0.4,0) -- (0.8,0);

\end{tikzpicture} & $10$ \\ \hline

\begin{tikzpicture}[scale=0.70]	
\draw (0,0) circle (0.4cm);
\draw (0,-0.8) node {\footnotesize{$6$}};
\draw (1.2,-0.8) node {\footnotesize{$4$}};
\draw (1.2,0) circle (0.4cm);
\draw (-2.4,0)  circle (0.4cm);
\draw (-2.4,-0.8) node {\footnotesize{$4$}};
\draw (-1.2,-0.8) node {\footnotesize{$8$}};
\draw (-0.8,0.08) -- (-0.4,0.08);
\draw (-0.8,-0.08) -- (-0.4,-0.08);
\draw (-0.7,0.2) -- (-0.5,0);
\draw (-0.7,-0.2) -- (-0.5,0);
\draw (1.2,1.2) [red,fill=red!30] circle (0.4cm);
\draw (1.12,0.4) -- (1.12,0.8);
\draw (1.28,0.4) -- (1.28,0.8);
\draw (1,0.5) -- (1.2,0.7);
\draw (1.4,0.5) -- (1.2,0.7);
\draw (1.2,2) node {\footnotesize{$1$}};
\draw (-1.2,0) circle (0.4cm);
\draw (-1.6,0) -- (-2,0);
\draw (0.4,0) -- (0.8,0);
\end{tikzpicture} & $2$ \\ \hline

	\end{tabular}
	\caption{Exotic minimally unbalanced quivers with $\gf=F_4$ and an extra double laced edge.}
	\label{tab:FveryexoticDouble}
\end{table}

\begin{table}
	\centering
	\begin{tabular}{|c|c|}
	\hline
	Quiver & Excess \\ \hline
\begin{tikzpicture}[scale=0.70]
\draw (0,0) circle (0.4cm);
\draw (0,-0.8) node {\footnotesize{$2$}};
\draw (1.2,-0.8) node {\footnotesize{$1$}};
\draw (1.2,0) circle (0.4cm);
\draw (-2.4,0)  circle (0.4cm);
\draw (-2.4,-0.8) node {\footnotesize{$2$}};
\draw (-1.2,-0.8) node {\footnotesize{$3$}};
\draw (-0.8,0.08) -- (-0.4,0.08);
\draw (-0.8,-0.08) -- (-0.4,-0.08);
\draw (-0.7,0.2) -- (-0.5,0);
\draw (-0.7,-0.2) -- (-0.5,0);
\draw (-2.4,1.2) [red,fill=red!30] circle (0.4cm);
\draw (-2.32,0.4) -- (-2.32,0.8);
\draw (-2.48,0.4) -- (-2.48,0.8);
\draw (-2.2,0.7) -- (-2.4,0.5);
\draw (-2.6,0.7) -- (-2.4,0.5);
\draw (-2.4,0.4) -- (-2.4,0.8);
\draw (-2.4,2) node {\footnotesize{$1$}};
\draw (0.4,0) -- (0.8,0);
\draw (-1.2,0) circle (0.4cm);
\draw (-1.6,0) -- (-2,0);
\end{tikzpicture} & $4$\\ \hline	

\begin{tikzpicture}[scale=0.70]
\draw (0,0) circle (0.4cm);
\draw (0,-0.8) node {\footnotesize{$4$}};
\draw (1.2,-0.8) node {\footnotesize{$2$}};
\draw (1.2,0) circle (0.4cm);
\draw (-2.4,0)  circle (0.4cm);
\draw (-2.4,-0.8) node {\footnotesize{$3$}};
\draw (-1.2,-0.8) node {\footnotesize{$6$}};
\draw (-1.2,2) node {\footnotesize{$1$}};
\draw (-0.8,0.08) -- (-0.4,0.08);
\draw (-0.8,-0.08) -- (-0.4,-0.08);
\draw (-0.7,0.2) -- (-0.5,0);
\draw (-0.7,-0.2) -- (-0.5,0);
\draw (-1.2,1.2) [red,fill=red!30] circle (0.4cm);
\draw (0.4,0) -- (0.8,0);
\draw (-1.2,0.4) -- (-1.2,0.8);
\draw (-1.2,0) circle (0.4cm);
\draw (-1.12,0.4) -- (-1.12,0.8);
\draw (-1.28,0.4) -- (-1.28,0.8);
\draw (-1.4,0.7) -- (-1.2,0.5);
\draw (-1,0.7) -- (-1.2,0.5);
\draw (-1.6,0) -- (-2,0);
\end{tikzpicture} & $16$\\ \hline	

\begin{tikzpicture}[scale=0.70]
\draw (0,0) circle (0.4cm);
\draw (0,-0.8) node {\footnotesize{$6$}};
\draw (1.2,-0.8) node {\footnotesize{$3$}};
\draw (1.2,0) circle (0.4cm);
\draw (-2.4,0)  circle (0.4cm);
\draw (-2.4,-0.8) node {\footnotesize{$4$}};
\draw (-1.2,-0.8) node {\footnotesize{$8$}};
\draw (0,2) node {\footnotesize{$1$}};
\draw (-0.8,0.08) -- (-0.4,0.08);
\draw (-0.8,-0.08) -- (-0.4,-0.08);
\draw (-0.7,0.2) -- (-0.5,0);
\draw (-0.7,-0.2) -- (-0.5,0);
\draw (0,1.2) [red,fill=red!30] circle (0.4cm);
\draw (-0.08,0.4) -- (-0.08,0.8);
\draw (0.08,0.4) -- (0.08,0.8);
\draw (-0.2,0.7) -- (0,0.5);
\draw (0.2,0.7) -- (0,0.5);
\draw (0,0.4) -- (0,0.8);
\draw (-1.2,0) circle (0.4cm);
\draw (-1.6,0) -- (-2,0);
\draw (0.4,0) -- (0.8,0);
\end{tikzpicture} & $16$\\ \hline	
	
\begin{tikzpicture}[scale=0.70]
\draw (0,0) circle (0.4cm);
\draw (0,-0.8) node {\footnotesize{$3$}};
\draw (1.2,-0.8) node {\footnotesize{$2$}};
\draw (1.2,0) circle (0.4cm);
\draw (-2.4,0)  circle (0.4cm);
\draw (-2.4,-0.8) node {\footnotesize{$2$}};
\draw (-1.2,-0.8) node {\footnotesize{$4$}};
\draw (-0.8,0.08) -- (-0.4,0.08);
\draw (-0.8,-0.08) -- (-0.4,-0.08);
\draw (-0.7,0.2) -- (-0.5,0);
\draw (-0.7,-0.2) -- (-0.5,0);
\draw (1.2,1.2) [red,fill=red!30] circle (0.4cm);
\draw (1.12,0.4) -- (1.12,0.8);
\draw (1.28,0.4) -- (1.28,0.8);
\draw (1,0.7) -- (1.2,0.5);
\draw (1.4,0.7) -- (1.2,0.5);
\draw (1.2,0.4) -- (1.2,0.8);
\draw (1.2,2) node {\footnotesize{$1$}};
\draw (-1.2,0) circle (0.4cm);
\draw (-1.6,0) -- (-2,0);
\draw (0.4,0) -- (0.8,0);
\end{tikzpicture} & $4$ \\ \hline \hline

\begin{tikzpicture}[scale=0.70]
\draw (0,0) circle (0.4cm);
\draw (0,-0.8) node {\footnotesize{$6$}};
\draw (1.2,-0.8) node {\footnotesize{$3$}};
\draw (1.2,0) circle (0.4cm);
\draw (-2.4,0)  circle (0.4cm);
\draw (-2.4,-0.8) node {\footnotesize{$6$}};
\draw (-1.2,-0.8) node {\footnotesize{$9$}};
\draw (-0.8,0.08) -- (-0.4,0.08);
\draw (-0.8,-0.08) -- (-0.4,-0.08);
\draw (-0.7,0.2) -- (-0.5,0);
\draw (-0.7,-0.2) -- (-0.5,0);
\draw (-2.4,1.2) [red,fill=red!30] circle (0.4cm);
\draw (-2.32,0.4) -- (-2.32,0.8);
\draw (-2.48,0.4) -- (-2.48,0.8);
\draw (-2.2,0.5) -- (-2.4,0.7);
\draw (-2.6,0.5) -- (-2.4,0.7);
\draw (-2.4,0.4) -- (-2.4,0.8);
\draw (-2.4,2) node {\footnotesize{$1$}};
\draw (0.4,0) -- (0.8,0);
\draw (-1.2,0) circle (0.4cm);
\draw (-1.6,0) -- (-2,0);
\end{tikzpicture} & $4$\\ \hline	

\begin{tikzpicture}[scale=0.70]
\draw (0,0) circle (0.4cm);
\draw (0,-0.8) node {\footnotesize{$12$}};
\draw (1.2,-0.8) node {\footnotesize{$6$}};
\draw (1.2,0) circle (0.4cm);
\draw (-2.4,0)  circle (0.4cm);
\draw (-2.4,-0.8) node {\footnotesize{$9$}};
\draw (-1.2,-0.8) node {\footnotesize{$18$}};
\draw (-1.2,2) node {\footnotesize{$1$}};
\draw (-0.8,0.08) -- (-0.4,0.08);
\draw (-0.8,-0.08) -- (-0.4,-0.08);
\draw (-0.7,0.2) -- (-0.5,0);
\draw (-0.7,-0.2) -- (-0.5,0);
\draw (-1.2,1.2) [red,fill=red!30] circle (0.4cm);
\draw (0.4,0) -- (0.8,0);
\draw (-1.2,0.4) -- (-1.2,0.8);
\draw (-1.2,0) circle (0.4cm);
\draw (-1.12,0.4) -- (-1.12,0.8);
\draw (-1.28,0.4) -- (-1.28,0.8);
\draw (-1.4,0.5) -- (-1.2,0.7);
\draw (-1,0.5) -- (-1.2,0.7);
\draw (-1.6,0) -- (-2,0);
\end{tikzpicture} & $16$\\ \hline	

\begin{tikzpicture}[scale=0.70]
\draw (0,0) circle (0.4cm);
\draw (0,-0.8) node {\footnotesize{$18$}};
\draw (1.2,-0.8) node {\footnotesize{$9$}};
\draw (1.2,0) circle (0.4cm);
\draw (-2.4,0)  circle (0.4cm);
\draw (-2.4,-0.8) node {\footnotesize{$12$}};
\draw (-1.2,-0.8) node {\footnotesize{$24$}};
\draw (0,2) node {\footnotesize{$1$}};
\draw (-0.8,0.08) -- (-0.4,0.08);
\draw (-0.8,-0.08) -- (-0.4,-0.08);
\draw (-0.7,0.2) -- (-0.5,0);
\draw (-0.7,-0.2) -- (-0.5,0);
\draw (0,1.2) [red,fill=red!30] circle (0.4cm);
\draw (-0.08,0.4) -- (-0.08,0.8);
\draw (0.08,0.4) -- (0.08,0.8);
\draw (-0.2,0.5) -- (0,0.7);
\draw (0.2,0.5) -- (0,0.7);
\draw (0,0.4) -- (0,0.8);
\draw (-1.2,0) circle (0.4cm);
\draw (-1.6,0) -- (-2,0);
\draw (0.4,0) -- (0.8,0);
\end{tikzpicture} & $16$\\ \hline	
	
\begin{tikzpicture}[scale=0.70]
\draw (0,0) circle (0.4cm);
\draw (0,-0.8) node {\footnotesize{$9$}};
\draw (1.2,-0.8) node {\footnotesize{$6$}};
\draw (1.2,0) circle (0.4cm);
\draw (-2.4,0)  circle (0.4cm);
\draw (-2.4,-0.8) node {\footnotesize{$6$}};
\draw (-1.2,-0.8) node {\footnotesize{$12$}};
\draw (-0.8,0.08) -- (-0.4,0.08);
\draw (-0.8,-0.08) -- (-0.4,-0.08);
\draw (-0.7,0.2) -- (-0.5,0);
\draw (-0.7,-0.2) -- (-0.5,0);
\draw (1.2,1.2) [red,fill=red!30] circle (0.4cm);
\draw (1.12,0.4) -- (1.12,0.8);
\draw (1.28,0.4) -- (1.28,0.8);
\draw (1,0.5) -- (1.2,0.7);
\draw (1.4,0.5) -- (1.2,0.7);
\draw (1.2,0.4) -- (1.2,0.8);
\draw (1.2,2) node {\footnotesize{$1$}};
\draw (-1.2,0) circle (0.4cm);
\draw (-1.6,0) -- (-2,0);
\draw (0.4,0) -- (0.8,0);
\end{tikzpicture} & $4$ \\ \hline

	\end{tabular}
	\caption{Exotic minimally unbalanced quivers with $\gf=F_4$ and an extra triple laced edge.}
	\label{tab:FveryexoticTriple}
\end{table}

\clearpage

\subsection{Exotic Minimally Unbalanced Quivers with $G$ of Type $G_2$}

Finally, we collect all exotic minimally unbalanced quivers with $\gf=G_2$ and with the unbalanced node attached via a double or a triple laced edge in tables \ref{tab:Gseriesveryexotic1} and \ref{tab:Gseriesveryexotic1}, respectively. \\

\begin{table}
	\centering
	\begin{tabular}{|c|c|}
	\hline
	Quiver & Excess \\ \hline
\begin{tikzpicture}[scale=0.70]
\draw (0,0) circle (0.4cm);
\draw (0,-0.8) node {\footnotesize{$1$}};
\draw (-1.2,-0.8) node {\footnotesize{$2$}};
\draw (-1.2,1.2) [red,fill=red!30] circle (0.4cm);
\draw (-1.2,2) node {\footnotesize{$1$}};
\draw (-1.12,0.4) -- (-1.12,0.8);
\draw (-1.28,0.4) -- (-1.28,0.8);
\draw (-1,0.7) -- (-1.2,0.5);
\draw (-1.4,0.7) -- (-1.2,0.5);
\draw (-0.8,0.08) -- (-0.4,0.08);
\draw (-0.8,-0.08) -- (-0.4,-0.08);
\draw (-0.7,0.2) -- (-0.5,0);
\draw (-0.7,-0.2) -- (-0.5,0);
\draw (-0.4,0) -- (-0.8,0);
\draw (-1.2,0) circle (0.4cm);
\end{tikzpicture} & $2$\\ \hline	
	
\begin{tikzpicture}[scale=0.70]
\draw (0,0) circle (0.4cm);
\draw (0,-0.8) node {\footnotesize{$2$}};
\draw (-1.2,-0.8) node {\footnotesize{$3$}};
\draw (0,1.2) [red,fill=red!30] circle (0.4cm);
\draw (0,2) node {\footnotesize{$1$}};
\draw (-0.08,0.4) -- (-0.08,0.8);
\draw (0.08,0.4) -- (0.08,0.8);
\draw (-0.2,0.7) -- (0,0.5);
\draw (0.2,0.7) -- (0,0.5);
\draw (-0.8,0.08) -- (-0.4,0.08);
\draw (-0.8,-0.08) -- (-0.4,-0.08);
\draw (-0.7,0.2) -- (-0.5,0);
\draw (-0.7,-0.2) -- (-0.5,0);
\draw (-0.4,0) -- (-0.8,0);
\draw (-1.2,0) circle (0.4cm);
\end{tikzpicture} & $2$ \\ \hline \hline

\begin{tikzpicture}[scale=0.70]
\draw (0,0) circle (0.4cm);
\draw (0,-0.8) node {\footnotesize{$2$}};
\draw (-1.2,-0.8) node {\footnotesize{$4$}};
\draw (-1.2,1.2) [red,fill=red!30] circle (0.4cm);
\draw (-1.2,2) node {\footnotesize{$1$}};
\draw (-1.12,0.4) -- (-1.12,0.8);
\draw (-1.28,0.4) -- (-1.28,0.8);
\draw (-1,0.5) -- (-1.2,0.7);
\draw (-1.4,0.5) -- (-1.2,0.7);
\draw (-0.8,0.08) -- (-0.4,0.08);
\draw (-0.8,-0.08) -- (-0.4,-0.08);
\draw (-0.7,0.2) -- (-0.5,0);
\draw (-0.7,-0.2) -- (-0.5,0);
\draw (-0.4,0) -- (-0.8,0);
\draw (-1.2,0) circle (0.4cm);
\end{tikzpicture} & $2$\\ \hline	
	
\begin{tikzpicture}[scale=0.70]
\draw (0,0) circle (0.4cm);
\draw (0,-0.8) node {\footnotesize{$4$}};
\draw (-1.2,-0.8) node {\footnotesize{$6$}};
\draw (0,1.2) [red,fill=red!30] circle (0.4cm);
\draw (0,2) node {\footnotesize{$1$}};
\draw (-0.08,0.4) -- (-0.08,0.8);
\draw (0.08,0.4) -- (0.08,0.8);
\draw (-0.2,0.5) -- (0,0.7);
\draw (0.2,0.5) -- (0,0.7);
\draw (-0.8,0.08) -- (-0.4,0.08);
\draw (-0.8,-0.08) -- (-0.4,-0.08);
\draw (-0.7,0.2) -- (-0.5,0);
\draw (-0.7,-0.2) -- (-0.5,0);
\draw (-0.4,0) -- (-0.8,0);
\draw (-1.2,0) circle (0.4cm);
\end{tikzpicture} & $2$ \\ \hline
	\end{tabular}
	\caption{Exotic minimally unbalanced quivers with $\gf=G_2$ and an extra double laced edge.}
	\label{tab:Gseriesveryexotic1}
\end{table}

\begin{table}
	\centering
	\begin{tabular}{|c|c|}
	\hline
	Quiver & Excess \\ \hline
	
\begin{tikzpicture}[scale=0.70]
\draw (0,0) circle (0.4cm);
\draw (0,-0.8) node {\footnotesize{$1$}};
\draw (-1.2,-0.8) node {\footnotesize{$2$}};
\draw (-1.2,1.2) [red,fill=red!30] circle (0.4cm);
\draw (-1.2,2) node {\footnotesize{$1$}};
\draw (-1.2,0.4) -- (-1.2,0.8);
\draw (-1.2,0.4) -- (-1.2,0.8);
\draw (-1.12,0.4) -- (-1.12,0.8);
\draw (-1.28,0.4) -- (-1.28,0.8);
\draw (-1,0.7) -- (-1.2,0.5);
\draw (-1.4,0.7) -- (-1.2,0.5);
\draw (-0.8,0.08) -- (-0.4,0.08);
\draw (-0.8,-0.08) -- (-0.4,-0.08);
\draw (-0.7,0.2) -- (-0.5,0);
\draw (-0.7,-0.2) -- (-0.5,0);
\draw (-0.4,0) -- (-0.8,0);
\draw (-1.2,0) circle (0.4cm);
\end{tikzpicture} & $4$\\ \hline	
	
\begin{tikzpicture}[scale=0.70]
\draw (0,0) circle (0.4cm);
\draw (0,-0.8) node {\footnotesize{$2$}};
\draw (-1.2,-0.8) node {\footnotesize{$3$}};
\draw (0,0.4) -- (0,0.8);
\draw (0,1.2) [red,fill=red!30] circle (0.4cm);
\draw (0,2) node {\footnotesize{$1$}};
\draw (-0.08,0.4) -- (-0.08,0.8);
\draw (0.08,0.4) -- (0.08,0.8);
\draw (-0.2,0.7) -- (0,0.5);
\draw (0.2,0.7) -- (0,0.5);
\draw (-0.8,0.08) -- (-0.4,0.08);
\draw (-0.8,-0.08) -- (-0.4,-0.08);
\draw (-0.7,0.2) -- (-0.5,0);
\draw (-0.7,-0.2) -- (-0.5,0);
\draw (-0.4,0) -- (-0.8,0);
\draw (-1.2,0) circle (0.4cm);
\end{tikzpicture} & $4$ \\ \hline \hline

\begin{tikzpicture}[scale=0.70]
\draw (0,0) circle (0.4cm);
\draw (0,-0.8) node {\footnotesize{$3$}};
\draw (-1.2,-0.8) node {\footnotesize{$6$}};
\draw (-1.2,1.2) [red,fill=red!30] circle (0.4cm);
\draw (-1.2,2) node {\footnotesize{$1$}};
\draw (-1.2,0.4) -- (-1.2,0.8);
\draw (-1.2,0.4) -- (-1.2,0.8);
\draw (-1.12,0.4) -- (-1.12,0.8);
\draw (-1.28,0.4) -- (-1.28,0.8);
\draw (-1,0.5) -- (-1.2,0.7);
\draw (-1.4,0.5) -- (-1.2,0.7);
\draw (-0.8,0.08) -- (-0.4,0.08);
\draw (-0.8,-0.08) -- (-0.4,-0.08);
\draw (-0.7,0.2) -- (-0.5,0);
\draw (-0.7,-0.2) -- (-0.5,0);
\draw (-0.4,0) -- (-0.8,0);
\draw (-1.2,0) circle (0.4cm);
\end{tikzpicture} & $4$\\ \hline	
	
\begin{tikzpicture}[scale=0.70]
\draw (0,0) circle (0.4cm);
\draw (0,-0.8) node {\footnotesize{$6$}};
\draw (-1.2,-0.8) node {\footnotesize{$9$}};
\draw (0,0.4) -- (0,0.8);
\draw (0,1.2) [red,fill=red!30] circle (0.4cm);
\draw (0,2) node {\footnotesize{$1$}};
\draw (-0.08,0.4) -- (-0.08,0.8);
\draw (0.08,0.4) -- (0.08,0.8);
\draw (-0.2,0.5) -- (0,0.7);
\draw (0.2,0.5) -- (0,0.7);
\draw (-0.8,0.08) -- (-0.4,0.08);
\draw (-0.8,-0.08) -- (-0.4,-0.08);
\draw (-0.7,0.2) -- (-0.5,0);
\draw (-0.7,-0.2) -- (-0.5,0);
\draw (-0.4,0) -- (-0.8,0);
\draw (-1.2,0) circle (0.4cm);
\end{tikzpicture} & $4$ \\ \hline

	\end{tabular}
	\caption{Exotic minimally unbalanced quivers with $\gf=G_2$ and an extra triple laced edge.}
	\label{tab:Gseriesveryexotic2}
\end{table}

This completes the classification of minimally unbalanced quivers for which the balanced subset of nodes forms a single finite Dynkin diagram. Quivers found in sections \ref{4} and \ref{5} have Coulomb branches \emph{minimally generated} by two kinds of operators:
\begin{itemize}
\item Operators in the adjoint of the isometry group $\gf$ (corresponding to the balanced Dynkin diagram). These operators appear at order $t^2$ in the Hilbert series.
\item Extra operators in the representation that corresponds to the Dynkin node where the unbalanced extra node attaches (and its complex conjugate representation). In the Hilbert series, these appear at order of $t^k$, where $k$ is determined by the excess  (imbalance of the unbalanced node). In particular, $k=2+e$, where $e$ is the excess. 
\end{itemize}
Section \ref{6} and \ref{7} contain exotic minimally unbalanced quivers. The Coulomb branches of these quivers are, strictly speaking, not \emph{minimally generated} but have additional generators at higher orders. Although these spaces correspond to moduli spaces of gauge theories with eight supercharges, this is the first time most of such quivers appear. The full gauge theoretic examination of such theories is yet to be done.


\section{Conclusions and Prospects} \label{8}

In this paper we develop and present a full classification of all minimally unbalanced quiver gauge theories with the global isometry on the Coulomb branch that corresponds to a single finite Dynkin diagram. This concurrently provides a classification of hyperk\"ahler cones which appear as moduli spaces of supersymmetric gauge theories with $8$ supercharges in various dimensions.
 Minimally unbalanced quivers with $A$-type global symmetry form a two parameter family described by $a$ and $b$. Minimally unbalanced theories with global symmetry of $BCD$-type are classified based on the position of the unbalanced node and the total number of balanced nodes. Minimally unbalanced quivers with exceptional global symmetry (i.e. of $EFG$-type) are found for each different case (i.e. for each possible node to which the unbalanced node is attached). In case of $E_6$, the the number of cases reduces to $4$ due to the $\mathbb{Z}_2$ outer automorphism invariance of the $E_6$ Dynkin diagram.\\
 
A complementary result to the work in this paper is the classification of minimally unbalanced quivers with global symmetry of the form:
\be
G_{global} =  G_1 \times G_2,
\ee
    where $G_1$ and $G_2$ are any two Lie groups. Such extended classification is obtained by combining all pairs of minimally unbalanced quivers found in this paper.\footnote{Combination means an attaching of two minimally unbalanced quivers via a common unbalanced node.} The extended classification is available online at \url{https://www.dropbox.com/s/uxi30bgjis1x4u2/AUX_MU.pdf?dl=0} or for download at \url{https://dl.dropboxusercontent.com/s/uxi30bgjis1x4u2/AUX_MU.pdf?dl=0}.\footnote{Alternatively, the extended classification is available as per request by email.} \\

A possible direction for future research is the classification of unbalanced quivers with $N$ unbalanced nodes, where $N>1$. In such scenario, the global symmetry takes the form:
\be
G_{global} = \prod_{i} G_i \times U(1)^{N-1} ,
\ee
where $G_i$ are the groups corresponding to the Dynkin sub-diagrams formed by the subset of balanced nodes. The number of the $U(1)$ Abelian factors in the global symmetry is one less than the number of unbalanced nodes. Quivers with more than one unbalanced node appear in various contexts in the study of $5d$ and $6d$ Higgs branches \cite{FHMZ17,Hanany:2018vph,DGHZ18}.\\

In the classification of this paper, we find a raft of quiver theories that are not studied in any existing literature. This opens a large and possibly fructiferous domain for extensive future investigations. \emph{Quaerite et invenietis ordinem.}

\section*{Acknowledgements}
S.C. is supported by an EPSRC DTP studentship EP/M507878/1. A.H. is supported by STFC Consolidated Grant ST/J0003533/1, and EPSRC Programme Grant EP/K034456/1. A.Z. would like to extend his gratitude to Rudolph Kalveks for enlightening discussions. A.Z. would also like to extend his gratitude to Matus Plch for coming to his aid when implementing NN tools which brought to light unexpected new minimally unbalanced quivers.

\appendix

\section{The Coulomb Branch: Monopole Operators and Global Symmetries} \label{A}

Moduli spaces of $3d \ \mathcal N=4$ quiver gauge theories have two distinct phases known as the \emph{Coulomb branch} (where the gauge group is typically broken to its maximal torus) and \emph{Higgs branch} (where the gauge group of the theory is typically fully broken). The \emph{Coulomb branch} (and also the \emph{Higgs branch}) is a hyperK\"ahler  variety which can be described by its ring of holomorphic functions. The information about the branch is then encoded in a Hilbert Series which succinctly enumerates holomorphic functions in the ring. A one-to-one correspondence has been observed between holomorphic functions in the moduli space and gauge invariant BPS operators in the chiral ring of the quantum field theory. In \cite{CHZ13} an efficient method for counting these operators is proposed, namely the \emph{monopole formula}:

\begin{equation}
H_G(t,z)=\sum_{m \in \Gamma _{\hat{G}} / \mathcal{W}_{\hat{G}}} z^{J(m)} t^{\Delta(m)} P_{G} (t,m)
\end{equation}
where $G$ is the gauge group of the theory and $m$ is the magnetic charge (see \cite{GNO76}) which takes its value in the lattice:
\begin{align}
\Gamma_{\hat {G}}:=(\Gamma_{G})^*
\end{align}
$(\Gamma_{G})^*$ is the lattice dual to the weight lattice of $G$. It defines a new weight lattice of a new group $\hat G$, which is considered the GNO dual of $G$ \cite{GNO76}. $\mathcal{W}_{\hat{G}}$ is the Weyl group of $\hat G$. $J(m)$ denotes the topological charge counted by the $z$ fugacity. The dressing factor $P_G$ is a generating function for Casimir invariants of the unbroken gauge group. 
\\
$\Delta(m)$ is the conformal dimension which coincides with the R-charge of the monopole operators. We quote the result for the conformal dimension, as it was obtained using radial quantization in \cite{Borokhov:2002ib}:

\begin{equation}
\Delta(m)=-\sum_{\alpha \in \Delta_+} \mid\alpha(m)\mid +\frac{1}{2} \sum_{i=1} ^n \sum_{\rho_i \in R_i} \mid\rho_i (m)\mid
\end{equation}

The two terms of the conformal dimension formula account for vector multiplets and hypermultiplet contributions, respectively. $\Delta_+$ is the set of positive roots of the gauge group. Hypermultiplets transform in representations $R_i$ with weights $\rho_i$. \\

An approach that utilizes division of weight lattice into fans was introduced in \cite{HS16}. For more detailed exposition of the monopole formula, see \cite{CHZ13}. In order to treat non-simply laced quivers, a modification of the hypermultiplet contribution of the conformal dimension introduced in \cite{CFHM14} takes the following form:

\begin{equation}
\frac{1}{2}  \mid\rho_i (m)\mid \rightarrow \frac{1}{2} \sum_{j=1} ^{N_1} \sum_{k=1} ^{N_{2}} \mid \lambda m_j ^{(1)} - m_k ^{(2)}\mid
\end{equation}

where $\rho_i$ is the irrep corresponding to the hypermultiplets assigned to the edge between two nodes $U(N_1)$ and $U(N_2)$. $\lambda =1$ recovers the formula for the quiver when the edge is simple, $\lambda =2$ is used for a double laced edge, and finally, $\lambda =3$ for a triple laced edge. The direction of the edge points from $N_1$ to $N_2$. $m^{(1)}$ and $m^{(2)}$ denote the magnetic fluxes for $U(N_1)$ and $U(N_2)$, respectively. For completeness, we show the function which enumerates the Casimir invariants of residual gauge group of $U(N)$ that is left unbroken by the configuration of magnetic charges:

\begin{equation}
P_{U(N)} (t;m) =\prod_{k=1} ^N \frac{1}{(1-t^{2k})^{{\lambda}(k)(m)}} .
\end{equation}

${\lambda}(k)({m})$ encodes the various configurations of the gauge symmetry braking in form of a partition. As an example, for a $U(2)$ gauge symmetry and magnetic charges ${m} = (m_1,m_2)$ the dressing factor is:

\begin{equation}
P_{U(2)} (t;m_1,m_2) =
\begin{cases}
               \frac{1}{(1-t)(1-t^{2})} \quad if\quad m_1=m_2 \\
          \frac{1}{(1-t)(1-t)}\quad  if\quad m_1 \neq m_2 .
            \end{cases} 
\end{equation}

In order to use the monopole formula, there are certain restrictions for the conformal dimension which translate into the balancing of the quiver nodes. Firstly, for ADE quivers, the excess (or balance) of a particular $U(N_i)$ gauge node is defined as \cite{HK16}:

\begin{equation}
Excess_{ADE}(i) = \sum_{j\in \: adjecent \: nodes} N_j - 2N_i .
\end{equation}

A quiver is said to be fully balanced if the excess of all its nodes is zero. If one or more nodes in the quiver have positive excess the quiver is said to be positively balanced. In case of a quiver with a single node with excess of $1$ or larger, we term the quiver \emph{minimally unbalanced}.\footnote{This differs from the notation introduced in \cite{HK16} where the authors used the term for all quivers with one or more nodes of excess $1$ or greater.} The present work only concerns balanced and minimally unbalanced quivers. For balanced or minimally unbalanced theories the conformal dimension satisfies $\Delta(m)>0$ for all $m\in \Gamma _{\hat G}$ which guarantees that the monopole formula can be applied to calculate the Coulomb branch of the moduli space.\footnote{In fact, there are special cases of balanced quivers with moduli spaces that are not hyperK\"ahler varieties, hence the monopole cannot be applied. Thus, it seems that balance is necessary but not sufficient condition for a quiver to be well behaved and treatable by the currently known methods.} \\

The global symmetry of the Coulomb branch is determined by the operators with $\Delta=1$. From the quiver one can quickly write a set of operators with $\Delta=1$ such that they correspond to the roots of the Dynkin diagram formed by nodes that are balanced. Extra operators with $\Delta=1$ might exist, which would enhance the global symmetry. In the previous pages we are restricted to quivers where only one node is unbalanced, and the remaining nodes form the Dynkin diagram of either a classical or an exceptional Lie algebra.\\

The Higgs branch of $5$d theories at infinite coupling is given by the Coulomb branch of a $3$d quiver. Physically, we can motivate this with a use of 3d mirror symmetry \cite{IS96} and the presence of $8$ supercharges in both theories. Considering a reduction of the $5$d SCFT on a torus leads to a $3$d Higgs branch that is unchanged thanks to the amount of supersymmetry. In addition, many $3$d theories have mirror duals for which the Coulomb and Higgs branch are exchanged. In general a dual theory can lack a Lagrangian description, however, it was argued in \cite{BTX10} that specific class of $5$d theories described by intersecting D5, NS5 and (1,1)-branes reduces to $A$-type class $S$ theories compactified on a circle. It was further argued that reducing class $S$ theories on circle to $3$d leads to $3$d SCFTs with Lagrangian mirrors whose shape is a three-legged unitary quiver. For a $SU(n)$ theory with fundamental matter the bound for the number of flavors is: $N_f>2n$.  $5$d SCFTs with enough matter belong to this class. This is a strong motivation for the approach of this paper.



\section{Very Exotic Minimally Unbalanced (Ring) Quivers with $G$ of Type $A_n$} \label{C}

Carrying out the calculation for a part of the classification, the neural network (NN), working solely with graph theoretical knowledge, produced some peculiar quivers. Among them are quivers included as a caveat in this section. In particular, we include $A$-type ring quivers with an unbalanced node connected to the adjoint nodes of the balanced chain by two non-simply laced edges. There are three cases based on whether there are two double edges, two triple edges, or one double and one triple edge. In all cases the non-simply laced edges point outwards with respect to the unbalanced node. The results are collected in table \ref{tab:Aexoticrings} with the excess shown in a separate column. Note that in all three cases, the unbalanced node connects to the Dynkin nodes corresponding to the adjoint representation of $SU(m+n+2)$.

\begin{table}
	\centering
	\begin{tabular}{|c|c|}
	\hline
	Quiver & Excess \\ \hline
	
\begin{tikzpicture}[scale=0.70]
\draw (0,0) [red,fill=red!30] circle (0.4cm);
\draw (0,-0.8) node {\footnotesize{$1$}};
\draw (1.6,0) -- (2,0);
\draw (1.2,-0.8) node {\footnotesize{$1$}};
\draw (-2.8,0) -- (-3.2,0);
\draw (2.8,0) -- (3.2,0);
\draw (-3.6,0) circle (0.4cm);
\draw (-3.6,-0.8) node {\footnotesize{$1$}};
\draw (2.4,0) node {\footnotesize{\dots}};
\draw (-2.4,0) node {\footnotesize{\dots}};
\draw (3.6,0) circle (0.4cm);
\draw (3.6,-0.8) node {\footnotesize{$1$}};
\draw (-1.2,-0.8) node {\footnotesize{$1$}};
\draw (1.2,0) circle (0.4cm);
\draw (-1.2,0) circle (0.4cm);
\draw (-1.6,0) -- (-2,0);
\draw (0,1.2) circle (0.4cm);
\draw (0,2) node {\footnotesize{$1$}};
\draw (-3.3,0.25) -- (-0.382,1.12);
\draw (3.3,0.25) -- (0.382,1.12);
\draw (-0.4,0.08) -- (-0.8,0.08);
\draw (-0.4,-0.08) -- (-0.8,-0.08);
\draw (-0.7,0) -- (-0.5,0.2);
\draw (-0.7,0) -- (-0.5,-0.2);
\draw (0.4,0.08) -- (0.8,0.08);
\draw (0.4,-0.08) -- (0.8,-0.08);
\draw (0.7,0) -- (0.5,0.2);
\draw (0.7,0) -- (0.5,-0.2);
\draw [decorate,decoration={brace,amplitude=6pt}] (-1.3,-1.2) to (-3.5,-1.2);
\draw (-2.4,-1.7) node {\footnotesize{$m$}};
\draw [decorate,decoration={brace,amplitude=6pt}] (3.5,-1.2) to (1.3,-1.2);
\draw (2.4,-1.7) node {\footnotesize{$n$}};

\end{tikzpicture} & $2$\\ \hline	

\begin{tikzpicture}[scale=0.70]
\draw (0,0) [red,fill=red!30] circle (0.4cm);
\draw (0,-0.8) node {\footnotesize{$1$}};
\draw (1.6,0) -- (2,0);
\draw (1.2,-0.8) node {\footnotesize{$1$}};
\draw (-2.8,0) -- (-3.2,0);
\draw (2.8,0) -- (3.2,0);
\draw (-3.6,0) circle (0.4cm);
\draw (-3.6,-0.8) node {\footnotesize{$1$}};
\draw (2.4,0) node {\footnotesize{\dots}};
\draw (-2.4,0) node {\footnotesize{\dots}};
\draw (3.6,0) circle (0.4cm);
\draw (3.6,-0.8) node {\footnotesize{$1$}};
\draw (-1.2,-0.8) node {\footnotesize{$1$}};
\draw (1.2,0) circle (0.4cm);
\draw (-1.2,0) circle (0.4cm);
\draw (-1.6,0) -- (-2,0);
\draw (0,1.2) circle (0.4cm);
\draw (0,2) node {\footnotesize{$1$}};
\draw (-3.3,0.25) -- (-0.382,1.12);
\draw (3.3,0.25) -- (0.382,1.12);
\draw (-0.4,0.08) -- (-0.8,0.08);
\draw (-0.4,-0.08) -- (-0.8,-0.08);
\draw (-0.4,0) -- (-0.8,0);
\draw (-0.7,0) -- (-0.5,0.2);
\draw (-0.7,0) -- (-0.5,-0.2);
\draw (0.4,0.08) -- (0.8,0.08);
\draw (0.4,-0.08) -- (0.8,-0.08);
\draw (0.4,0) -- (0.8,0);
\draw (0.7,0) -- (0.5,0.2);
\draw (0.7,0) -- (0.5,-0.2);
\draw [decorate,decoration={brace,amplitude=6pt}] (-1.3,-1.2) to (-3.5,-1.2);
\draw (-2.4,-1.7) node {\footnotesize{$m$}};
\draw [decorate,decoration={brace,amplitude=6pt}] (3.5,-1.2) to (1.3,-1.2);
\draw (2.4,-1.7) node {\footnotesize{$n$}};

\end{tikzpicture} & $4$\\ \hline	

\begin{tikzpicture}[scale=0.70]
\draw (0,0) [red,fill=red!30] circle (0.4cm);
\draw (0,-0.8) node {\footnotesize{$1$}};
\draw (1.6,0) -- (2,0);
\draw (1.2,-0.8) node {\footnotesize{$1$}};
\draw (-2.8,0) -- (-3.2,0);
\draw (2.8,0) -- (3.2,0);
\draw (-3.6,0) circle (0.4cm);
\draw (-3.6,-0.8) node {\footnotesize{$1$}};
\draw (2.4,0) node {\footnotesize{\dots}};
\draw (-2.4,0) node {\footnotesize{\dots}};
\draw (3.6,0) circle (0.4cm);
\draw (3.6,-0.8) node {\footnotesize{$1$}};
\draw (-1.2,-0.8) node {\footnotesize{$1$}};
\draw (1.2,0) circle (0.4cm);
\draw (-1.2,0) circle (0.4cm);
\draw (-1.6,0) -- (-2,0);
\draw (0,1.2) circle (0.4cm);
\draw (0,2) node {\footnotesize{$1$}};
\draw (-3.3,0.25) -- (-0.382,1.12);
\draw (3.3,0.25) -- (0.382,1.12);
\draw (-0.4,0.08) -- (-0.8,0.08);
\draw (-0.4,-0.08) -- (-0.8,-0.08);
\draw (-0.4,0) -- (-0.8,0);
\draw (-0.7,0) -- (-0.5,0.2);
\draw (-0.7,0) -- (-0.5,-0.2);
\draw (0.4,0.08) -- (0.8,0.08);
\draw (0.4,-0.08) -- (0.8,-0.08);
\draw (0.7,0) -- (0.5,0.2);
\draw (0.7,0) -- (0.5,-0.2);
\draw [decorate,decoration={brace,amplitude=6pt}] (-1.3,-1.2) to (-3.5,-1.2);
\draw (-2.4,-1.7) node {\footnotesize{$m$}};
\draw [decorate,decoration={brace,amplitude=6pt}] (3.5,-1.2) to (1.3,-1.2);
\draw (2.4,-1.7) node {\footnotesize{$n$}};

\end{tikzpicture} & $3$\\ \hline	
	
	\end{tabular}
	\caption{Classification of very exotic $A$-type minimally unbalanced ring quivers with two non-simply laced edges.}
	\label{tab:Aexoticrings}
\end{table}

\bibliography{main}
\bibliographystyle{JHEP}

\end{document}